\documentclass[11pt,preprint]{aastex}
\usepackage{natbib}
\usepackage{graphicx}
\newcommand{\flux}{erg \,s$^{-1}$ \,cm$^{-2}$}
\newcommand{\lum}{erg \,s$^{-1}$}
\newcommand{\nh}{$N_{\mathrm H}$}
\newcommand{\km}{\,km\,s$^{-1}$}
\newcommand{\norm}{$10^{-5}$\,photons\,keV$^{-1}$\,cm$^{-2}$\,s$^{-1}$ at 1\,keV}

\begin{document}

\title{The Swift BAT-detected Seyfert 1 Galaxies: X-ray Broadband Properties and Warm Absorbers}
\author{Lisa M. Winter\altaffilmark{1,2}, \\ Sylvain Veilleux\altaffilmark{3}, Barry McKernan\altaffilmark{4,5}, T.R. Kallman\altaffilmark{6}}
\altaffiltext{1}{Center for Astrophysics and Space Astronomy, Department of Astrophysical \& Planetary Sciences, University of Colorado, UCB 391, Boulder, CO 80309, USA}
\altaffiltext{2}{Hubble Fellow}
\altaffiltext{3}{Department of Astronomy, University of Maryland, College Park, MD 20742, USA}
\altaffiltext{4}{Department of Science, Borough of Manhattan Community College, City University of New York, New York, NY 10007, USA}
\altaffiltext{5}{Department of Astrophysics, American Museum of Natural History, New York, NY 10024, USA; Graduate Center, City University of New York, 365 5th Avenue, New York, NY 10016, USA}
\altaffiltext{6}{NASA Goddard Space Flight Center, Greenbelt, MD 20771, USA}

\begin{abstract}

We present results from an analysis of the broad-band, 0.3--195\,keV, X-ray spectra of 48 Seyfert 1--1.5 sources detected in the very hard X-rays with the Swift Burst Alert Telescope (BAT).  This sample is selected in an all-sky survey conducted in the 14--195\,keV band.  Therefore, our sources are largely unbiased towards both obscuration and host galaxy properties.  Our detailed and uniform model fits to Suzaku/BAT and XMM-Newton/BAT spectra include the neutral absorption, direct power-law, reflected emission, soft excess, warm absorption, and narrow \ion{Fe}{1} K$\alpha$ emission properties for the entire sample.
We significantly detect O VII and O VIII
 edges in 52\% of our sample. The strength of these detections are
 strongly correlated with the neutral column density measured in the
 spectrum. Among the strongest detections, X-ray grating and UV
 observations, where available, indicate outflowing material. The
 ionized column densities of sources with O VII and O VIII detections
 are clustered in a narrow range with N$_{\rm warm} \sim 10^{21}$\,cm$^{-2}$, while
 sources without strong detections have column densities of ionized
 gas an order of magnitude lower.  Therefore, we note that sources without strong detections likely have warm ionized outflows present but at low column densities that are not easily probed with current X-ray observations. Sources with strong complex
 absorption have a strong soft excess, which may or may not be due to difficulties in modeling the complex spectra of these sources. Still, the detection of a
 flat $\Gamma \sim 1$ and a strong soft excess may allow us to infer the
 presence of strong absorption in low signal-to-noise AGN spectra.  Additionally, we include a useful correction from the Swift BAT luminosity to bolometric luminosity, based on a comparison of our spectral fitting results with published spectral energy distribution fits from 33 of our sources.


\end{abstract}

\section{Introduction}
The Swift Burst Alert Telescope (BAT) is conducting the first all-sky, very hard X-ray survey in thirty years.  With hundreds of detections in a harder X-ray band than the previous survey by HEAO-1 \citep{1982ApJ...253..485P}, the Swift survey presents an unprecedented sample of active galactic nuclei (AGN) sources.  Due to their detection in the 14--195\,keV, very hard X-ray band, the Swift sources are unbiased to all but the highest levels of obscuration.  Therefore, the Swift BAT-detected AGN are an important sample for determining the global properties of AGN.

In \citet{2009ApJ...690.1322W}, the 0.3--10\,keV X-ray properties for the 153 sources detected in the 9-month Swift BAT catalog \citep{2008ApJ...681..113T} are presented.  The 9-month catalog includes all sources with very hard X-ray detections of F$_{14 - 195 {\rm keV}} > 10^{-11}$\flux.  These sources are local, with an average redshift of 0.03, and are bright IR/optical/UV/X-ray sources.  The X-ray analysis of the 9-month sources uncovered many results of the properties of local AGNs, relying on simple models.  However, since both broader X-ray band and higher signal-to-noise spectra are available for a majority of these sources, more detailed analyses are now possible.

Our main goal in this paper is the detection and characterization of ``warm absorbers'', signatures of ionized gas potentially from outflows, in the Type 1 AGN.  Previous studies searching for warm absorbers were biased -- selecting all the bright sources with archived X-ray or UV data \citep{1997MNRAS.286..513R,1998ApJS..114...73G,1999ApJ...516..750C,2007AJ....134.1061D}. Therefore, the results of these studies may also be biased.  Mass outflows or more specifically AGN feedback mechanisms are important in galaxy formation \citep{1998AA...331L...1S,2003ARA&A..41..117C,2005ARA&A..43..769V} and a potential cause of the well-known relationship between galaxies and their black holes (i.e., the M-$\sigma$ relation first described in \citealt{2000ApJ...539L...9F,2000ApJ...539L..13G}). Thus, it is important to understand the AGN warm absorber properties in an unbiased sample.  Since X-ray warm absorbers are most evident in the soft emission ($< 2$\,keV), where more absorbed sources exhibit low flux levels, we rely on a study of the Type 1 sources detected by Swift's BAT. In this paper, we present the detailed broad-band spectral properties of 48/51 Seyfert 1--1.5 sources, with $| b | > 15$\degr and high signal-to-noise X-ray CCD spectra available, which were detected in the Swift BAT 9-month survey.

Our analysis relies preferentially on X-ray spectra from Suzaku, which provides simultaneous coverage from $\sim 0.2$--$50$\,keV, and time-averaged spectra from Swift's BAT in the 14--195\,keV band.  With a broad bandpass, we obtain tight constraints on the continuum emission, which is vital in determining the warm absorber properties.  Where Suzaku spectra were not available, we use data from XMM-Newton, which covers the 0.3--10\,keV band, with the time-averaged Swift BAT spectra.  We describe our data reduction in \S~\ref{sect-data}.  Our spectral analysis is detailed in \S~\ref{sect-spectra}.  Discussion of our results is found in \S~\ref{sect-discussion}, followed by conclusions in \S~\ref{sect-conclusion}.

\section{Data Reduction}~\label{sect-data}
Archived or proprietary Suzaku data from our guest observer Suzaku program to study warm absorbers in the Swift sample (PI: Winter) were available for 32/51 (63\%)  Seyfert 1--1.5 sources in the 9-month Swift catalog.  For our analysis, we prefer to utilize Suzaku spectra, which simultaneously cover the 0.2--12\,keV band with the four X-ray Imaging Spectrometers (XIS; \citealt{2007PASJ...59S..23K}) and the $\sim 15$--50\,keV band with the Hard X-ray Detector (HXD) PIN instrument \citep{2007PASJ...59S..35T}.  The HXD PIN spectra, which overlap in energy band with the non-simultaneous Swift BAT spectra, allow for constraints on the normalization of the BAT spectra.  Where Suzaku observations are not available, we utilize archived XMM-Newton spectra.  We analyzed XMM-Newton spectra for an additional 16 sources, 31\% of the total sample.  Thus, our total sample consists of 48/51 (94\%) of the Swift BAT-detected Seyfert 1--1.5 sources.  A list of our sample, including basic source properties, is provided in Table~\ref{table-sourcelist}.

For the sources with Suzaku observations, we downloaded data for the longest exposure observation from NASA's High Energy Astrophysics Science Archive Research Center (HEASARC\footnote{NASA's HEASARC is accessible from \url{http://heasarc.gsfc.nasa.gov}.}).  Details of the Suzaku observations are included in Table~\ref{tbl-suzaku}.  To analyze the XIS data, we first extracted an image of the source using {\tt XSELECT} and both the ``$3 \times 3$'' and ``$5 \times 5$'' cleaned event files for each of the available XIS observations. We identified circular source regions, centered on the source, with radii ranging from 120\arcsec~to 240\arcsec. We also identified circular background regions in an area free from bright additional sources, which ranged from 120\arcsec~to 180\arcsec~in radius. These regions were used to extract spectra for the source and background. Response and ancillary response files were created, as indicated in The Suzaku Data Reduction Guide\footnote{The Suzaku Data Reduction Guide is found at \url{http://heasarc.gsfc.nasa.gov/docs/suzaku/analysis/abc}.}, with the Suzaku ftools {\tt xisrmfgen} and {\tt xissimarfgen}.  We combined the XIS front-illuminated (XIS0, XIS3, and XIS2, as available) spectra and response files using the ftool {\tt mathpha}.  The spectra and response files for both the back-illuminated (XIS1) and front-illuminated data were each grouped with the ftool {\tt grppha} and binned to a minimum of 20 counts per bin. 

We also used the extracted spectra from the Suzaku HXD PIN cleaned event files.  We downloaded the appropriate tuned background file for each observation, as supplied by the Suzaku team\footnote{Tuned HXD PIN background files are available from \url{ftp://legacy.gsfc.nasa.gov/suzaku/data/background/pinnxb\_ver2.0\_tuned}.}. Source spectra were extracted using XSELECT and corrected for dead time using the Suzaku ftool {\tt hxddtcor}.  Since the tuned background file does not include the cosmic X-ray background (CXB), we accounted for the background using the typical CXB spectrum of \begin{math} CXB(E) = 9.0\times10^{-9} \times (E/3 {\rm keV})^{-0.29} \times \exp(-E/40 {\rm keV})\,{\rm erg\,cm}^{-2}\,{\rm s}^{-1}\,{\rm str}^{-1}\, {\rm keV}^{-1}\end{math} from \citet{1987PhR...146..215B}. The PIN spectra were binned with a signal-to-noise ratio of 3$\sigma$ or 4$\sigma$, depending on the exposure time, brightness, and flux of the source during observation. Standard response files from the Suzaku CALDB were used, as outlined in Table 7.3 of the Suzaku Data Reduction Guide. 

For the sources without Suzaku observations, we utilized archived XMM-Newton pn spectra, where available.  Observation details are included in Table~\ref{tbl-xmm}.  The XMM-Newton data was processed with the XMM-Newton Science Analysis Software (SAS) version 9.0. Processing of the pn data followed the steps laid out in the XMM-Newton ABC guide\footnote{The XMM-Newton ABC Guide is available at \url{http://heasarc.gsfc.nasa.gov/docs/xmm/abc}.}. The source and background regions were extracted from circular regions with radii ranging from 20\arcsec~to 100\arcsec. The background region was selected as an area free of bright point sources, near the source, and located on the same chip as the source region if possible. We used the stringent selection expression ``FLAG==0'' to disregard bad pixels and events near the edges of the CCD. The pn spectra were extracted in the 0.2 to 15 keV energy band, using only patterns of 0-4 (single- and double-pixel events), with the SAS task {\tt especget}. We then used the task {\tt epatplot} to check for pileup, finding significant pileup in only in the observation of Mrk 290.  Pile-up was corrected by taking smaller source region sizes, as described in the XMM-Newton ABC guide.    Additionally, the pn light curves were inspected for signs of background flaring.  The events files were filtered to remove times of high background count rates (i.e. filtering the count rate with appropriate values to exclude values above the normal level, for instance `RATE $< 50$').  Response and ancillary response files were then created using the SAS tasks {\tt rmfgen} and {\tt arfgen}.  The spectra and response files were grouped and binned to 20\,counts per bin, using {\tt grppha}.

For all of the sources, we used time-averaged Swift BAT spectra.  These spectra were created using observations from the first 22 months of the survey and are described in \citet{2010ApJS..186..378T}.  The spectra consist of eight energy bands and are publicly available, along with the diagonal response file, at the Swift 22-Month Survey website\footnote{The Swift BAT spectra are available for download at \url{http://heasarc.gsfc.nasa.gov/docs/swift/results/bs22mon}.}.

\section{X-ray Spectral Fitting}~\label{sect-spectra}
X-ray emission from AGN is believed to originate from the inner most regions, close to the black hole.  The direct emission is likely from inverse Compton scattering of photons from hot electrons in a corona around the accretion disk.  The spectral shape is similar to a cutoff power-law of the form \begin{math} F(E) \propto E^{- (\Gamma - 1)} \end{math}.  This emission is absorbed and reflected by gas along our line of sight.  Soft X-ray ($< 2$\,keV) features often seen in Type 1 spectra include a soft excess, typically fit with a blackbody peaking near 0.1\,keV, and absorption/emission features from abundant metals (e.g., oxygen, iron, neon).  Finally, the most prominent X-ray emission feature is often an Fe K-$\alpha$ line at 6.41\,keV, which is likely created from direct emission reflected in the accretion disk.

For each of our sources, we fit a basic continuum model to either the Suzaku and Swift or XMM-Newton and Swift spectra.  Using XSPEC v12, the Suzaku XIS1 and PIN spectra were fixed at a constant flux ratio of PIN $= 1.16$ or $1.18 \times $XIS1\footnote{As outlined in the Suzaku Data Reduction Guide.  The value 1.16 corresponds to observations in XIS nominal mode, while 1.18 corresponds to HXD nominal mode.}, while the additional spectra were allowed to vary to the best-fit normalization value.  Our base model is a cutoff power-law, absorbed by the Galactic neutral absorption, $N_{\rm H}$(Gal), recorded in Table~\ref{table-sourcelist}.  To account for absorption, we used the XSPEC model {\tt tbabs}, with the abundances set to the \citet{2000ApJ...542..914W} solar values and using the cross sections of \citet{1996ApJ...465..487V}.  Next, we added an additional neutral absorption model to account for intrinsic absorption in the host galaxy, $N_{\rm H}$.  

In addition to Galactic and intrinsic absorption, we added a blackbody model ({\tt zbbody} in XSPEC) to account for a soft excess, an Fe K-$\alpha$ emission line with a gaussian model, and a reflection model ({\tt pexrav} in XSPEC).  Basic parameters for the blackbody model include the energy (kT in keV) and the flux normalization, in addition to the redshift of the source.  Parameters of the gaussian model include the central energy of the line and the width, both in keV, and the flux normalization.  For each of the spectra, we constrained the energy of the Fe K$\alpha$ line to range from 6.0--6.6\,keV.  Where the energy or width were not well-constrained by the model, we fixed these values to 6.41\,keV and 0.01\,keV, respectively.  

Finally, we include the reflection component using the {\tt pexrav} model \citep{1995MNRAS.273..837M}, we fixed both the power-law index, cutoff energy, and normalization of the reflection component to be the same as the direct component.  We also assume solar abundances and the default inclination angle of $\cos i = 0.45$.  We then allowed the reflection parameter ($R$), which is an indicator of the strength of the reflection and defined as approximately the solid angle of reflected material/$2 \pi$, to vary from $0-5$.  For reference, an isotropic continuum source illuminating a flat accretion disk results in $R \sim 1$ (if the inclination angle is constrained).  Large values of the reflection parameter may be the result of several factors including time variability from reflection in the torus (e.g., during a state change in the source as suggested for NGC 4051 by \citealt{2009PASJ...61S.299T}), light bending (e.g., \citealt{2003MNRAS.340L..28F} and \citealt{2004MNRAS.349.1435M}), not allowing the inclination angle of the black hole system to vary in the spectral fits (we fixed the inclination angle to $\sim 63^{\circ}$), or the reflection component being the sum of reflection from multiple regions.

For each of the models added to the base power-law, we record the change in $\chi^2$ upon adding the model in Table~\ref{tbl-fits}, along with the best-fit $\chi^2$ value.  Best-fit model parameters are included in Tables~\ref{tbl-fits1} and~\ref{tbl-fek}.  The quoted errors correspond to the 90\% confidence level for an additional two degrees of freedom ($\Delta\chi^2 = 2.71$).

After we found a best-fit continuum model, we tested for the presence of warm absorber features by adding absorption edge models for \ion{O}{7} and \ion{O}{8} at 0.73\,keV and 0.87\,keV, respectively.  Absorption edges from \ion{O}{7} and \ion{O}{8} are among the strongest ``warm absorber'' signatures detected in X-ray CCD data.  While analysis of high resolution grating data is needed to confirm outflowing gas, through measurement of velocity shifts in individual absorption lines, inflows have yet to be identified in X-ray observations of AGN.  Therefore, the features are potentially produced from outflowing gas.  To detect possible outflows through the presence of \ion{O}{7} and \ion{O}{8} absorption edges, we added absorption edges ({\tt zedge} in XSPEC) for each of the features, fitting for the energy and optical depth of each edge.  Where the energy was poorly constrained, we fixed the values to the expected values of 0.73\,keV and 0.87\,keV, respectively.  Results of these fits are included in Table~\ref{tbl-warmabs}.  We classified sources as having a strong warm absorber detection where $\Delta\chi^2 \ga 13.39$ upon adding the edge models, which corresponds to $P = 0.01$ for four additional degrees of freedom.  We checked our results on the best-fit \ion{O}{7} optical depths by stepping through $\chi^2$-space with the {\tt steppar} command for 10\% of our sample.  Both methods yield consistent results on the optical depth and significance of the edge feature.

We note that the measured edge depths are the result of fits to blends of the indicated oxygen edges with additional emission and absorption features (e.g., the Fe UTA feature) and are crude measurements of the strength of ionized absorption in the X-ray spectroscopy.  Further details of this analysis are also included in the companion paper \citet{2010ApJ...725L.126W}.  We find that the \ion{O}{7} edge is typically the stronger of the absorption features detected.  In some cases, for instance NGC 4051 and NGC 6814, there is no clear detection of the \ion{O}{8} absorption edge.  The detection threshold for the edge features, as indicated by the $\Delta \chi^2$ parameter, tends to be at about $\tau \sim 0.1$.  For sources with firm \ion{O}{7} detections, but not significant \ion{O}{8} detection, we include the \ion{O}{8} absorption edge parameters as an indicator of the limits on the detection.  There are 5 sources in the warm absorption non-detection category with optical depths of \ion{O}{7} absorption in this range with $\Delta\chi^2$ values 
ranging from $\sim 3 - 13$.  All of these sources are on the border for a warm absorption detection.  Finally, the estimated fluxes from this final model ({\tt zedge}*{\tt zedge}*{\tt ztbabs}*{\tt tbabs}*({\tt zbbody} + {\tt zgauss} + {\tt cutoffpl} + {\tt pexrav})*{\tt constant}) in the 0.3--2\,keV, 2--10\,keV, 15--50\,keV, and 14--195\,keV bands are shown in Table~\ref{tbl-flux}.

\begin{figure}
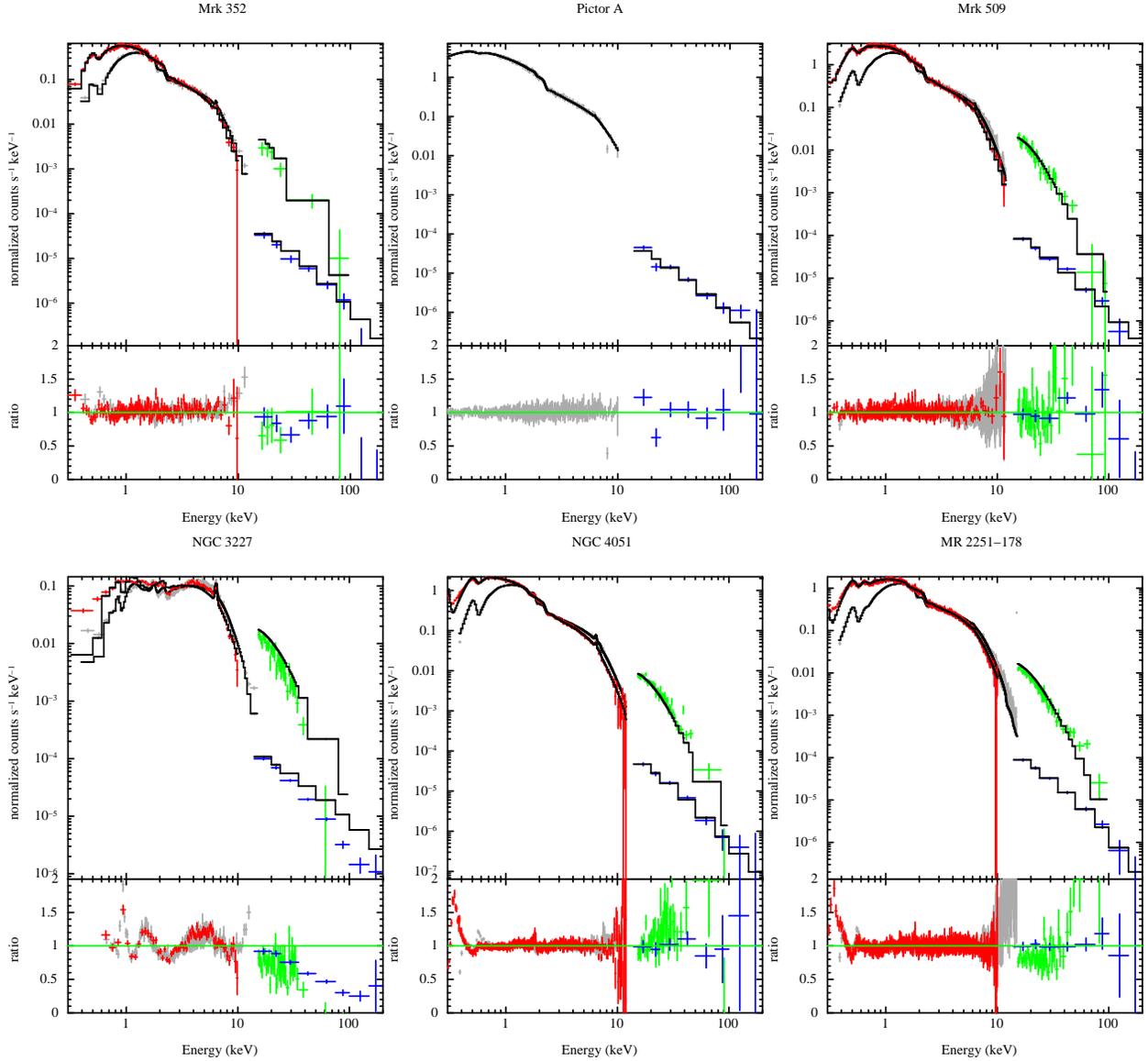

\includegraphics[width=5.5cm]{f1a.ps}
\hspace{-0.2cm}
\includegraphics[width=5.5cm]{f1b.ps}
\hspace{-0.2cm}
\includegraphics[width=5.5cm]{f1c.ps}\\
\includegraphics[width=5.5cm]{f1d.ps}
\hspace{-0.2cm}
\includegraphics[width=5.5cm]{f1e.ps}
\hspace{-0.2cm}
\includegraphics[width=5.5cm]{f1f.ps}
\caption{We plot the best-fit base model for several of our Seyfert 1--1.5 sources, with the ratio of data/model.  The Suzaku XIS data is re-binned to a signal-to-noise ratio of 15--25 for illustrative purposes.  Color-coding of the spectra corresponds to Suzaku XIS front-illuminated/XMM-Newton pn in gray, Suzaku XIS1 in red, Suzaku PIN in green, and Swift BAT data in blue.  The model is shown in black.  Sources in the top row are well-fit (reduced $\chi^2 < 2.0$) with the basemodel ({\tt zedge}*{\tt zedge}*{\tt ztbabs}*({\tt zbbody} + {\tt zgauss} + {\tt cutoffpl} + {\tt pexrav})*{\tt constant}, described in the text), while sources on the bottom row are not well-fit.  Each of the sources in the bottom row has a complex shape in the soft X-rays that is not accounted for with the simple absorption edges modeling \ion{O}{7} and \ion{O}{8}.
\label{fig-basemodel}}
\end{figure}

Examples of spectral fits with our model are shown in Figure~\ref{fig-basemodel}.  The majority of sources have spectra that are well-fit with this model.  Five sources, however, are not well-fit, with reduced $\chi^2 > 2.0$.  These include NGC 3227, NGC 3516, NGC 4051, NGC 4151, and MR 2251-178.  Each of these sources have high signal-to-noise Suzaku spectra with exposures from 79--275\,ks in length and have strong detections of \ion{O}{7} and \ion{O}{8} absorption.  The complexity in the soft X-ray emission is not fully accounted for with the simple addition of the absorption edges.  This is clear in the bottom panels of Figure~\ref{fig-basemodel}, where the ratio of data/model for three of these sources clearly shows structure that is not accounted for with our model.

To determine tighter constraints on warm absorption in the spectra, we used an analytic photo-ionization model to obtain column densities and ionization parameters for absorbers.  Typically, AGN with warm absorbers have two or more absorption components.  For sources with strong warm absorption detections ($\Delta\chi^2 \ga 13.39$ upon adding the absorption edges, as classified in Table~\ref{tbl-warmabs}), we removed the edge models and added two warm absorption model components.  We used the analytic model {\tt warmabs}, which relies on photo-ionization calculations from XSTAR \citep{1982ApJS...50..263K}.  We assume a $T = 10^4$\,K gas illuminated by a $\Gamma = 2$ power-law, solar abundances, and a turbulent velocity of 0\,\km.   While an assumption of the default value of turbulent velocity of 0\,\km~is low (curve of growth estimates lie in the 100--200\,\km~range, which is also comparable to the spectral resolution in the Chandra HETG grating observations; e.g., \citealt{2007MNRAS.379.1359M}), there is little change in the measured parameters (ionized column density and ionization parameter) between a value of 0\,\km and 150\,\km.  For instance, in \citet{2010ApJ...719..737W} we fit the same spectrum of NGC 6860 with the same model, but fit for the turbulent velocity parameter as well as column density and ionization parameter.  Between those results and the values determined in the current analysis, the ionization parameter is the same and the warm ionized column densities are similar (there is a change in the column density of $0.9 - 2.8 \times 10^{20}$\,cm$^{-2}$ between a fixed turbulent velocity at 0\,\km~and a value of around 150\,\km).

In Table~\ref{tbl-warmabs-outflow}, we present the results of these fits.  For sources without strong detections, we added one absorption component and obtained errors on the best-fit parameters, as shown in Table~\ref{tbl-warmabs-nonoutflow}.  Example fits for three sources are shown in Figure~\ref{fig-fitsnooutflow}, including the best-fit base model and one-/two- component analytic model.  Results of this analysis are included in the following discussion.

\section{Discussion}\label{sect-discussion}
We performed detailed, uniform spectral fitting on a sample of 48/51 Seyfert 1--1.5 sources selected in the 14-195\,keV band with the Swift BAT.  In this section, we discuss our findings, based on general spectral properties, including the continuum shape, luminosity, high energy emission lines (e.g., \ion{Fe}{1} K), and soft excess.  Additionally, we specifically highlight results of an analysis of the warm absorption properties and their implications for AGN feedback.

\subsection{Basic Spectral Properties}

\subsubsection{\bf Intrinsic Neutral Column Density and Redshift}\label{sect-nh}
In the X-rays, type 1 AGN typically have low columns of obscuring material in the line of sight to the central source.  Since the emission is largely unobscured, we view directly the central regions surrounding the black hole. In the optical, type 1 AGN exhibit broad permitted emission lines, particularly in H$\alpha$ and H$\beta$.  The distinction between sub-types of Sy 1s relies on the strength of the narrow H$\beta$ emission component, where a Sy 1.5 is intermediate between a Sy 1 and Sy 2 with a clearly distinguishable narrow emission-line component \citep{1977ApJ...215..733O}.  Since optical Sy 1.5 galaxies are in an intermediate state, in a simple unified model of AGN
(e.g., \citealt{1993ARA&A..31..473A}) we expect the Sy 1.5s to have intermediate column densities between the low column Sy 1s and the higher column Sy 2s.

Our sample includes 22 Sy 1s, 9 Sy 1.2s, and 17 Sy 1.5s.  Among the type 1s, there are several broad line radio galaxies (BLRGs), including 1H 0419-577, 3C 120, 3C 382, 3C 390.3, 4C +74.26, 2MASX J21140128+8204483, and MR 2251-178.  Average values for the logarithm of neutral column density (using an upper limit of $10^{19}$\,cm$^{-2}$ for spectra with no clearly detected neutral absorption component) correspond to $19.65 \pm 0.53$ for Sy 1s, $20.12 \pm 0.63$ for Sy 1.2s, and $20.61 \pm 1.14$ for Sy 1.5s.  Broad line radio galaxies, as a subset of Sy 1s, have an average neutral column density measured at $20.26 \pm 0.74$.  Therefore, Sy 1s do have the lowest column densities, while Sy 1.5s are more obscured in the X-rays, on average.  However, if we exclude the four Sy 1.5s with columns higher than $10^{22}$\,cm$^{-2}$, a K-S test shows a smaller probability of the Sy 1s and Sy 1.5s being drawn from different distributions ($D = 0.359$ and $P = 0.264$).  
Further, the optical Sy 1--1.5 sources are, as expected, less obscured in the X-rays than the X-ray obscured sources (with typical values of \nh\,$> 10^{22}$\,cm$^{-2}$, see  \citealt{2009ApJ...690.1322W}).  In Figure~\ref{fig-nhz}, we plot the measured column densities for our sample versus redshift.  The highest redshift sources in our sample are the broad line radio galaxies (black squares), with $\langle z \rangle = 0.068$.  The Sy 1.5s are the lowest redshift sources ($\langle z \rangle = 0.012$), while the Sy 1s have an average redshift of 0.026 and the Sy 1.2s are intermediary with a value of 0.017.  

In Figure~\ref{fig-nhz}, we find that there are no Sy 1s at the lowest redshifts. From the plot, we find that there are 5 Sy 1.5s and 2 Sy 1.2s at $z < 0.008$.  Since Sy 1s are relatively unobscured, we expect that Swift's BAT has no bias against selecting Sy 1s in this redshift range.  The lack of Sy 1.5 detections at higher redshift may indicate that the Sy 1.5s are less luminous, since obscured sources in the BAT sample have lower luminosities than unobscured sources \citep{2009ApJ...690.1322W}.  However, at the low column densities measured from our broad-band X-ray fits (from 0.3--195\,keV), we expect the obscuring material in Sy 1.5s to have little affect at the high energies to which BAT is sensitive.  In the following section, we discuss the derived X-ray luminosities, with respect to optical classifications. 

While we find no Sy 1s at low redshift, it is difficult to make statistical claims on the paucity of low redshift Sy 1s due to the small number of objects in this redshift range.  However, this result is intriguing since the X-ray analyses of the Swift sample reveal that the type 2 sources are less luminous than the type 1 objects \citep{2009ApJ...690.1322W} and the fraction of obscured/unobscured AGN also changes with redshift (e.g., \citealt{koss-thesis}).  We will test this further with optical and X-ray follow-ups of the 600 AGNs detected in the 58-month Swift catalog (Baumgartner et al., submitted).  If the same result of few Sy 1s at low-redshift is found, one possible explanation is that there is a physical difference in the broad line region/accretion state between Sy 1.5s and Sy 1s.  \citet{2009ApJ...701L..91E} posit that in the disk wind model, where the broad line region and torus are part of a wind from the accretion disk, that at low accretion rates the broad line region disappears.  If the Sy 1.5s are accreting at a lower rate (see \S~\ref{lum}), the difference in the appearance of the optical broad lines could be correlated with a different accretion state.  In the current study, there is the possibility, however, of our sources being mis-classified (i.e., between Sy 1 sub-types). This is a concern since the NED classifications we use come from heterogeneous sources.  We will test this further in a follow-up paper examining our uniform optical spectra obtained with the Apache Point Observatory 3.5-m and CTIO SOAR 4-m telescopes.  Using the optical follow-ups, we will carefully classify the optical spectra of our sample to determine whether the NED classifications are correct.  We confirm that the optical follow-ups show broad Balmer lines in all of these sources, however, we will include more detailed analyses in the optical paper (Winter et al., in prep).


\subsubsection{\bf Luminosity, Mass, and Accretion Rate}\label{lum}

The luminosity, black hole mass, and accretion rate are among the most basic properties of an AGN.  Since the Swift BAT bandpass is at high enough energies (14--195\,keV) to be unaffected by all but the highest levels of obscuration, it is a good proxy for the bolometric luminosity of the AGN.  Our measurements of the BAT band luminosity are computed using the 14--195\,keV flux listed in Table~\ref{tbl-flux}.  We find that the BLRGs are the most luminous sources in our sample, while the Sy 1.5s are least luminous.  Average values and standard deviations for $\log$\,L$_{14-195\,{\rm keV}}$ are: $44.8 \pm 0.2$ (BLRGs), $43.8 \pm 0.4$ (Sy 1s), $43.4 \pm 0.6$ (Sy 1.2s), and $43.3 \pm 0.7$ (Sy 1.5s).  The standard deviations demonstrate that there is no statistical difference between the luminosity of Sy 1s, 1.2s, or 1.5s.  However, the BLRGs are significantly more luminous -- approximately 4 times more luminous than the Seyfert sources.

The mass determinations recorded in Table~\ref{tbl-flux} are derived from 2MASS bulge photometry of the host galaxies (see \citealt{2008ApJ...684L..65M} and \citealt{2009ApJ...690.1322W}) and are in good agreement with the independent analysis of 33 of our sources in  \citet{2007MNRAS.381.1235V} and  \citet{2009MNRAS.392.1124V}.  Comparison of the 2MASS-derived masses with both reverberation mapping and H-$\beta$ FWHM derived masses were discussed in \citet{2010ApJ...710..503W}.  We found masses derived from each of these methods to be well-correlated, with $\log M_{2MASS} = (0.91 \pm 0.14) \times \log M_{H\beta} + (1.07 \pm 1.13)$ and the H$\beta$-derived masses equivalent to the reverberation results.  The masses recorded in Table~\ref{tbl-flux} are the 2MASS values corrected to match the more accurate H-$\beta$ derived masses, with the exceptions of NGC 4051, NGC 4593, Mrk 279, and NGC 7469, where we use the reverberation mapping derived masses (in each of these cases there was a significant difference between the masses from the alternate method).  Additionally, we note that the mass of the Sy 1.5 2MASX J11454045-1827149, derived from the 2MASS K-band photometry, is low ($\log M/M_{\sun} = 6.2$), requiring a very high accretion rate ($L_{\rm bol}/L_{\rm Edd} = 3.5$).  We have obtained optical spectroscopy of the H-$\beta$ region of this source from the CTIO SOAR telescope and are in the process of reducing the data.  An alternative mass estimate will be included in Winter et al., in prep.

Comparing the average masses for our sample, we find $\log M/M_{\sun}$ of $8.5 \pm 0.2$ (BLRGs), $7.9 \pm 0.6$ (Sy 1s), $7.9 \pm 0.7$ (Sy 1.2s), and $7.8 \pm 0.7$ (Sy 1.5s).  We find that within the standard deviation the black hole masses are similar between all of our sources.  This is also true for the obscured type 2 sources, shown in \citet{2009ApJ...690.1322W}.


To determine the accretion rate of our sources, we must first determine the bolometric luminosity.  A reliable method for determining the bolometric luminosity is through broad-band fitting of the spectral energy distribution (e.g., \citealt{2002ApJ...579..530W}).  For 33 sources in our sample, bolometric luminosities and accretion rates were obtained through fitting simultaneous optical/UV/X-ray data from XMM-Newton \citep{2007MNRAS.381.1235V} or Swift \citep{2009MNRAS.392.1124V}.  In Figure~\ref{fig-comparebol}, we compare the Swift BAT band luminosities obtained from our spectral fits with the bolometric luminosities calculated from SED fitting in  \citet{2007MNRAS.381.1235V} and  \citet{2009MNRAS.392.1124V}.  We expect that the 14--195\,keV luminosity is the direct unobscured signature from the AGN for our Seyfert 1 sources and our figure confirms this.  Fitting an ordinary least squares line to the data, we find that $\log L_{\rm bol} = (1.1157 \pm 0.1172) \log L_{14-195\,{\rm keV}} + (-4.2280 \pm 5.1376)$.  This fit is very significant, with a correlation coefficient of $R^2 = 0.82$.  Using this relationship, we find that the average bolometric luminosities of our sources are $5.7 \times 10^{45}$\,\lum (BLRGs), $4.4 \times 10^{44}$\,\lum (Sy 1s), $1.6 \times 10^{44}$\,\lum (Sy 1.2s), and $1.2 \times 10^{44}$\,\lum (Sy 1.5s).  The bolometric luminosities are included in Table~\ref{tbl-flux}.  The mass accretion rate, $\dot{M} = \frac{L}{\eta c^2}$, ranges roughly from 1\,M$_{\sun}\,{\rm yr}^{-1}$ for the BLRGs and 0.02\,M$_{\sun}\,{\rm yr}^{-1}$ in the Sy 1.5s (assuming $\eta = 0.1$).

It is useful to parameterize the accretion rate relative to the Eddington limit.  The Eddington ratio is defined as the ratio of the bolometric luminosity to the Eddington luminosity, the luminosity where the gravitational and radiation pressure balance ($L_{\rm Edd} = 1.3 \times 10^{38} \times M/M_{\sun}$\,ergs\,s$^{-1}$), or equivalently the ratio of the mass accretion rate to the Eddington accretion rate ($\dot{M}_{\rm Edd} = \frac{L_{\rm Edd}}{\eta c^2}$).  We include the Eddington ratio for our sample in Table~\ref{tbl-flux}.  We find values of $L_{\rm bol}/L_{Edd}$ corresponding to $0.14 \pm 0.08$ (BLRGs), $0.05 \pm 0.14$ (Sy 1s), $0.03 \pm 0.05$ (Sy 1.2s), and $0.02 \pm 0.09$ (Sy 1.5s).  The main uncertainties in these measurements are in the estimates of the mass.  For one source,   2MASX J11454045-1827149, which does not have a measurement from alternative methods, the Eddington ratio is suspiciously high ($> 1$).  As discussed above, we will determine a mass estimate based on the H-$\beta$ width (Winter et al., in prep).  The general picture, however, emerging from our estimates shows that local AGN selected in the very hard X-rays tend to have accretion rates from $10^{-3} - 0.5$ times the Eddington rate.

The neutral column densities measured in the X-ray spectroscopy are a combination of the effects of absorption features of metals, which impose a soft X-ray cut-off in the spectrum.  This material may be associated with intergalactic material, gas local to the host galaxy, and/or intrinsic to the AGN.  In  \citet{2009ApJ...690.1322W}, we tested whether the obscuration was associated with the accretion rate of the AGN (Figure 10a) and the inclination of the host galaxy (Figure 16).  We found that the bulk of the obscuring material is not associated with the host galaxy (highly inclined hosts have slightly higher neutral column densities, but not enough to account for all of the obscuration measured) and that there is no correlation between the column and accretion rate.  In Figure~\ref{fig-lumlledd}, we test the dependence of the measured column on both the luminosity and Eddington ratio measured from our broad-band spectral fits.  There is no correlation found, which is consistent with the unified model, since our line of sight viewing angle with respect to the torus torus, a potential source of the neutral obscuration, is expected to only be an effect of our line of sight to the AGN (while this is true for low luminosity sources like the BAT-detected AGN, we note that this is not the case for higher luminosity sources).


\subsubsection{\bf Direct and Reflected Continuum}
The direct emission from our sources was modeled as a cutoff power-law.  The power-law index, $\Gamma$, measured values correspond to $1.90 \pm 0.16$ (BLRGs), $1.78 \pm 0.34$ (Sy 1s), $1.84 \pm 0.18$ (Sy 1.2s), and $1.52 \pm 0.61$ (Sy 1.5s).  The low average photon index for the Sy 1.5s is a result of several outliers with $\Gamma \sim 1.0$, including NGC 526A, Mrk 6, NGC 3227, and NGC 3516, with NGC 4151 having $\Gamma << 1.0$. Among these, NGC 3227, NGC 3516, and NGC 4151 are poorly fit with the base model (reduced $\Delta\chi^2 > 2.0$).  Two Sy 1s also have low values: EXO 055620--3820.2 and NGC 3783.  For these sources, we attribute the unusual $\Gamma$ measurements to difficulty modeling the absorbing components. Warm absorber signatures, measured through the \ion{O}{7}/\ion{O}{8} edges (see Table~\ref{tbl-warmabs}), are the strongest in our sample for all but EXO 055620--3820.2.  While we do not measure strong absorbers in EXO 055620--3820.2, analysis of the archived X-ray data by \citet{2009MNRAS.394L...1L} reveals obscuring clouds and complex structure (which we also detect in this observation).  Removing these values, we find that the mean $\Gamma$ value for Sy 1s and Sy 1.5s agree -- $1.90 \pm 0.17$ and $1.90 \pm 0.23$, respectively -- and a K-S test reveals a high probability ($P = 0.947$) that both are drawn from the same distribution.  Further, these average values are consistent with the distribution of $\Gamma$ for the BLRGs, demonstrating that there is no statistical indication of spectral hardening with luminosity/accretion rate in the BAT sample (see Figure~\ref{fig-gammalum}).  

In Figure~\ref{fig-gammalum}, we also plot the average power-law index binned by luminosity/accretion rate (right plots).  We divided the range of luminosities/accretion rates for our sources into equally-spaced bins and computed the average power-law index in each of these bins.  The error-bars indicate the difference between the average value and the maximum/minimum in each bin, with average values of $\Gamma = 1.07$, and are plotted as the logarithm of this value in order to illustrate the average values more clearly.  In  \citet{2009ApJ...690.1322W}, we found that there was no correlation between power-law index and luminosity/accretion rate, consistent with results from previous low-redshift samples \citep{2000ApJ...531...52G}.  Since our study consisted of all of the Swift-BAT detected AGN (e.g., including both absorbed and lower luminosity sources), we concluded that we found no correlation because the properties of the Swift sources were more diverse than the higher-redshift, more luminous sources which showed these correlations \citep{2004ApJ...605...45D,2008AJ....135.1505S}.  We did, however, see correlations for individual sources where we had multiple X-ray observations 
\citep{2008ApJ...674..686W}.  As shown in Figure~\ref{fig-gammalum}, in the current study we find that the photon-index is well-correlated with luminosity and accretion rate, when the values are binned.  We find that: 
\begin{equation}
 \Gamma = (0.16 \pm 0.03) \log L_{\rm bol} + (-5.31 \pm 1.17),
\end{equation}
 with a correlation coefficient of $R^2 = 0.86$, and 
 \begin{equation}
  \Gamma = (0.23 \pm 0.03) \log L_{\rm bol}/L_{\rm Edd} + (2.08 \pm 0.04), 
 \end{equation} with a coefficient of determination (the square of the Pearson linear correlation coefficient) of $R^2 = 0.83$ (the probability for both corresponds to $P < 0.001$ for 48 degrees of freedom).  Our linear correlation between binned $\Gamma$-L$_{\rm bol}$ is consistent with the rest-frame $\Gamma$-(2-10\,keV) luminosity relation found in \citep{2004ApJ...605...45D} ($\Gamma = (0.13 \pm 0.04) L_{2-10\,{\rm keV}} + (-4.1 \pm 1.7)$ for $0.3 \la z \la 0.96$ AGN).  This suggests that the X-ray power-law index and luminosity/accretion rate are related for the Sy 1s and that a similar process is at work as with the higher redshift radio quiet quasars.  However, the fact that a similar relation is not found for the entire Swift BAT-detected sample \citep{2009ApJ...690.1322W} enforces the result that the lowest luminosity sources (absorbed sources whose optical emission line properties classify them as \ion{H}{2} galaxies/composites/LINERs) are in a different accretion state  \citep{2010ApJ...710..503W}.

Cutoff energies of the power-law component are poorly constrained for the majority of our sample (40/48).  For many of the sources, the parameter was best-fit with the upper limit of $E_{cutoff} = 500$\,keV.  Of the sources with constrained cutoffs, EXO 055620--3820.2 and Mrk 79 have cutoffs between 20--60\,keV, and the remaining five sources have cutoffs from the range $\sim 60$--$400$\,keV.  By contrast, \citet{2009MNRAS.399.1293M} find cutoff energies between 50--150\,keV in their sample of Sy 1s, derived from the INTEGRAL survey.  As with our analysis, \citet{2009MNRAS.399.1293M} use broad-band X-ray spectral fits and use a very similar model, including reflection with the {\tt pexrav} model.  Of the overlapping sources between both the INTEGRAL and our Swift study, we find general agreement in the cutoff values, however, our error bars on the cutoff energies are much larger than those in the  \citet{2009MNRAS.399.1293M} analysis.  It is unclear why this is the case, particularly since our study includes, on average, a higher number of degrees of freedom from using the joint Suzaku/BAT or XMM-Newton/BAT data points.
 
 A reflected spectrum is found to be significant in 37/48 (77\%) of the sample, as determined by $\Delta\chi^2$ on adding this parameter (see Table~\ref{tbl-fits}).  We find good agreement with measured reflection parameters between our study and that of  \citet{2009MNRAS.399.1293M}, of which 10 sources overlap.  The average and standard deviation for the measured reflection parameter ($R \sim \Omega/2\pi$) correspond to $1.56 \pm 1.45$ (BLRGs), $2.23 \pm 1.47$ (Sy 1s), $1.96 \pm 1.22$ (Sy 1.2s), and $1.75 \pm 1.61$ (Sy 1.5s).  There is no statistical difference in the reflection parameter between sources of different Sy type.
 
 In Figure~\ref{fig-reflection}, we look for correlations between the reflection parameter and both AGN luminosity and power-law index, $\Gamma$.  Clearly, we find no correlation between $R$ and L$_{\rm bol}$.  A correlation does appear, however, between $R$ and $\Gamma$, where higher $\Gamma$ values coincide with higher reflection parameters (note that $R \sim 0.0$ corresponds to no reflection, while more negative values correspond to larger reflection).  Excluding sources with $\Gamma < 1.5$, which include sources with complex absorbers, we find that $R = (-8.58 \pm 1.84) + (5.74 \pm 0.97) \times \Gamma$.  This correlation is not strong, with a correlation coefficient of $R^2 = 0.49$.
 
 Correlations in $R - \Gamma$ were previously seen in a number of studies, including those of \citet{1999MNRAS.303L..11Z},  \citet{2007ApJ...664..101M},  and \citet{2009MNRAS.399.1293M}.  Since both parameters are linked in the fitting process, this trend may not be physical.  If the correlation is physical, \citet{1999MNRAS.303L..11Z} interpret the $R - \Gamma$ relation as follows.  The reflection parameter is proportional to the angle subtended by the reflector.  This ``cold'' reflector emits soft photons, which irradiate the X-ray source (i.e., the corona) and act as the seed for Compton upscattering.  If $R$ is large, there is more reflecting material and therefore a stronger flux from the soft photons.  This leads to stronger cooling of the plasma, a smaller Comptonization parameter ($y$), and a softer spectrum (or larger $\Gamma$).  \citet{2001ApJ...556..716P} show that with the {\tt pexrav} model larger reflection parameters also correspond to higher corona temperatures ($kT_e \propto E_{cutoff}$) and lower optical depths towards inverse Compton scattering, as well as softer spectra (higher $\Gamma$).  For sources in our sample with well-constrained $R$ and $E_{cutoff}$, there is no correlation between cutoff energy, thereby corona temperature, and reflection parameter/$\Gamma$.

 We conclude that the relationship between the direct and reflected emission is not easy to interpret in our sample.  Particularly, degeneracies between the fitted values from the {\tt pexrav} model make it impossible to determine whether the noted correlation between $R - \Gamma$ is physical.  Sources with low $\Gamma$ tend to have higher N$_H$, low $R$, and a range of luminosities.  Sources with high $\Gamma$ tend to have low N$_H$ and high $R$.  Cutoff energies, which are related to the Comptonization parameter ($y$), corona temperature, and optical depth towards Compton scattering, are not well-constrained for the majority of the sample.  Where they are constrained, there is no correlation with $\Gamma$.  In the future, we will compare the {\tt pexrav} model results with those from alternative reflection models, such as the dusty torus model MYTORUS \citep{2009MNRAS.397.1549M}, but such analysis is beyond the scope of the present paper.
 

\subsubsection{\bf Soft Excess}~\label{softexcess}
An X-ray soft excess, which we modeled with a simple blackbody, is statistically significant in 45/48 (94\%) of our sources.  Of the three sources without a significant blackbody component, the spectrum of Mrk 926 is poorly fit and the spectrum of 2MASX J11454045$-$1827149 has a relatively short exposure time with XMM-Newton (9\,ks).  Further, a soft excess is detected in a long XMM-Newton spectrum of the final source, NGC 5548  \citep{2003MNRAS.341..953P}.  Therefore, a soft excess is likely present in all local AGN.  This contrasts with our previous analyses of the Swift BAT sources, which found a soft excess in $40-50$\% of AGN \citep{2008ApJ...674..686W,2009ApJ...690.1322W}.  The higher detection rate in the current study is likely due to the higher signal-to-noise observations used in this study.  In  \citet{2009ApJ...690.1322W}, we used data from a variety of sources, with the majority of spectra from ASCA and lower signal-to-noise Swift XRT spectra.  This made it difficult to detect fainter blackbody components.

The average and standard deviation for the best-fit blackbody temperature ($kT$) corresponds to $0.11 \pm 0.06$\,keV (BLRGs), $0.11 \pm 0.04$\,keV (Sy 1s), $0.32 \pm 0.60$ (Sy 1.2s), and $0.15 \pm 0.22$ (Sy 1.5s).  The average value for the Sy 1.2s is skewed by one point  (MCG -01-13-025 has a best-fit $kT = 2.0$\,keV, with large error bars) and if this point is disregarded the average and standard deviation ($0.11 \pm 0.04$\,keV) are in line with the other sources in our sample.  Both the measured blackbody temperature and the small amount of scatter in our sample are consistent with measurements of sources in the Lockman Hole from the XMM-Newton survey \citep{2005AA...444...79M}, PG-selected QSOs \citep{2004AA...422...85P, 2005AA...432...15P,2010ApJ...725.1848T}, and our previous analyses of the Swift-selected sources \citep{2008ApJ...674..686W,2009ApJ...690.1322W}.

The origin of the soft excess in AGN is as yet unknown.  It may arise from thermal emission from star formation (particularly as seen in ULIRGs or AGN hosted in galaxies with strong nuclear starbursts), a population of near nuclear X-ray binaries/ULXs  \citep{2010MNRAS.407.2399M}, blurred reflected emission \citep{2003AA...412..317C, 2005MNRAS.358..211R}, or blurred absorption \citep{2004MNRAS.349L...7G}.  In Figure~\ref{fig-bbodytemp}, we test whether a thermal model is plausible.  If the emission is thermal and associated with the black hole, we expect the blackbody temperature to be proportional to $M^{-1/4} L/L_{Edd}^{1/4}$, where $M$ is the black hole mass and $L/L_{Edd}$ is the accretion rate.  We find no correlation between the blackbody temperature and either of these parameters, in agreement with the results of our previous analysis of the X-ray spectra of the Swift BAT-selected AGN \citep{2009ApJ...690.1322W}.  However, we do find a slight correlation between the flux in the power-law and the flux in the blackbody component.  In Figure~\ref{fig-ktnorm}, we plot the blackbody normalization versus the power-law normalization.  The correlation ($\log kT\,{\rm norm} = (-1.64 \pm 0.41) + (1.16 \pm 0.19) \log \Gamma\,{\rm norm}$; excluding the Sy 1.5s, which have more uncertainty in determining the soft excess parameters due to complex absorption) is weak, with $R^2 = 0.28$, but it indicates the same direct correlation shown from our more careful comparison of the luminosity in the blackbody and power-law components in  \citet{2009ApJ...690.1322W} (i.e., $L_{power-law} \propto L_{soft excess}$).  Similarly, this correlation is also seen in the PG QSOs  \citep{2010ApJ...725.1848T}.  The relationship between the power-law and soft excess shows that for the Sy 1s, the soft excess is either created by or affected by the direct AGN emission and not from thermal emission from star formation.  

Further clues to the origin of the soft excess are seen in a comparison with the absorbing gas.  Namely, we find that sources with the largest soft excess also have the highest neutral column densities and the strongest absorber signatures (measured through the optical depth of the O\,VII absorption edge).  These results are clearly shown in Figure~\ref{fig-nhoviiktnorm}.  Binning these relationships, shows that the correlations are very strong.  We find that
\begin{equation}
\log A_{\rm kT} = (1.74 \pm 0.26) \log N_{\rm H} + (-38.77 \pm 5.50),
\end{equation}
with a correlation coefficient of $R^2 = 0.88$.  We also find that
\begin{equation}
\log A_{\rm kT} = (1.62 \pm 0.19) \log \tau_{\rm O\,VII} + (-0.58 \pm 0.35),
\end{equation}
with $R^2 = 0.92$.  Therefore, the soft excess, neutral column density, and warm ionized gas are connected.   Since the soft excess is also correlated with the direct emission, these processes must all be related (i.e., the direct power-law emission, soft excess, absorption).  


\subsubsection{\bf \ion{Fe}{1} K$\alpha$ Emission}
The Fe K$\alpha$ emission feature is typically the most prominent feature in AGN spectra.  Recent work shows that the hard X-ray region surrounding the Fe K$\alpha$ band can be quite complex, with both narrow and broad features from \ion{Fe}{1} K$\alpha$, additional narrow features from, e.g., \ion{Fe}{25} K$\alpha$, \ion{Fe}{1} K$\beta$, and \ion{Ni}{1} K$\alpha$, iron K-edges (7.11\,keV), and absorption features from outflowing highly ionized gas.  We focus only on the narrow \ion{Fe}{1} K$\alpha$ line in this work, and defer further analysis of the hard X-ray emission/absorption to a future study.  We note, however, that many of our sources are included in more in-depth studies, such as the Chandra HETG study in \citet{2004ApJ...604...63Y} and the XMM-Newton study in \citet{2007MNRAS.382..194N}.   

Details of the best-fit parameters for the \ion{Fe}{1} K$\alpha$ emission are included in Table~\ref{tbl-fek}.  For many of our sources (40\% or 19/48), we could not constrain the energy or the width of the emission line.  Instead, we fixed these values to default values of 6.41\,keV and 0.01\,keV, respectively, to determine limits on the equivalent width.  We find that for the sources with well-constrained energies and widths, the average and standard deviation correspond to $E = 6.41 \pm 0.04$\,keV and a range in width from $\sigma = 0.06 - 0.14$\,keV (corresponding to ranges from $FWHM \approx 2780 - 9180$\,km\,s$^{-1}$). Therefore, while the energy of this line is very similar for our sources, we find that the width varies between sources.  However, the average value is still below the resolution of the XIS at 6.4\,keV, which is $\sim 130$\,eV. 
 Still, our best-fit values from the Suzaku and XMM-Newton spectroscopy are in good agreement with Chandra grating results for 14 Seyferts, which found $E = 6.404 \pm 0.005$\,keV and a $FWHM = 2380 \pm 760$\,km\,s$^{-1}$ \citep{2004ApJ...604...63Y}.  
 
Estimates of the $EW$ were made for all of the sources in the sample.  We find average and standard deviations of the equivalent width of the \ion{Fe}{1} K$\alpha$ emission of $108.96 \pm 190.34$\,eV.  Therefore, there is also a large spread in the $EW$, as well as the width, of the emission feature.  The source with the strongest $EW$ line in our sample, with $EW \approx 1.3$\,keV, is EXO 055620-3820.2, whose 2006 observation corresponds to a low flux, possibly Compton-thick phase (see \citealt{2009MNRAS.394L...1L} for a comparison of this observation with previous data showing a high flux state).  Disregarding this outlier, we find $EW = 83.10 \pm 70.08$\,eV.  While past studies found an anti-correlation between the equivalent width and luminosity of AGN (the X-ray Baldwin effect 
\citealt{1993ApJ...413L..15I}), we find no correlation -- in agreement with our earlier results on the Swift sources \citep{2009ApJ...690.1322W}.  In Figure~\ref{fig-fekew}, we show the relationship between the \ion{Fe}{1} K$\alpha$ equivalent width and both the bolometric luminosity and Eddington ratio.  Linear correlations are not seen with either parameter, with correlation coefficients of $R^2 = 0.25$ and $0.19$, respectively.  We previously found this in our \citet{2009ApJ...690.1322W} study, determining that a correlation exists for multiple observations of individual sources or when the luminosity/accretion rate proxy are binned.

To test this further, we looked for evidence of the X-ray Baldwin effect, by binning the data by luminosity and accretion rate.  As shown in Figure~\ref{fig-fekew}, there is no strong correlation between the \ion{Fe}{1} K$\alpha$ equivalent width and the bolometric luminosity.  We find that $EW \propto L_{\rm bol}^{-0.27 \pm 0.08}$, but with a correlation coefficient of $R^2 = 0.36$ ($P \sim 0.1$).  This correlation is consistent, however, with the relationship we found ($EW \propto L^{corr}_{2-10\,{\rm keV}}$) for the entire 9-month sample 
\citep{2009ApJ...690.1322W}.  We find a strong correlation between the \ion{Fe}{1} K$\alpha$ equivalent width and accretion rate.  The linear relationship is parameterized as
\begin{equation}
\log EW = (-0.38 \pm 0.07) \log L_{\rm bol}/L_{\rm Edd} + (1.35 \pm 0.08),
\end{equation}
with $R^2 = 0.80$.  This relationship is consistent with the\\ $EW \propto (L^{corr}_{2-10\,{\rm keV}}/L_{\rm Edd})^{(-0.26 \pm 0.03)}$ relationship found for the entire Swift sample \citep{2009ApJ...690.1322W}.  This confirms that the primary driver for the observed X-ray Baldwin effect is the correlation between the $EW$ and accretion rate, as suggested by \citet{2006ApJ...644..725J}.

\subsection{Warm Absorbers}
Using the base continuum model described in the preceding section, we constrained the properties of potential warm absorber signatures in our sample.  The simplest model for detecting potential outflows in X-ray CCD data is with the addition of absorption edge models fit to the 0.73\,keV \ion{O}{7} and 0.87\,keV \ion{O}{8} features.  We describe details of these fits in \S~\ref{o7}.  We then fit the spectra with more sophisticated models using the {\tt warmabs} model, which determines the ionization state and column density of the warm absorbing gas, described in \S~\ref{warmabs}.  We discuss our conclusions on the warm absorption fits in \S~\ref{subsect-conclusions}.

\subsubsection{\ion{O}{7} and \ion{O}{8} Absorbers}\label{o7}
The results of adding absorption edges to detect \ion{O}{7} and \ion{O}{8} edges are included in Table~\ref{tbl-warmabs}.  Among our sources, 25/48 (52\%) have clear detections of the edge features.  This fraction is higher than our original report of 18/44 (41\%) in \citet{2010ApJ...725L.126W}, since we added analysis of 4 sources with recently available spectra (NGC 985, 2MASX J11454045$-$1827149, NGC 6814, and 2MASX J21140128+8204483) and re-analyzed longer exposures available for 3 additional sources (IC 4329A, NGC 5548, and Mrk 926).  

As in \citet{2010ApJ...725L.126W}, we classify a detection based on $\Delta\chi^2 \ga 13.39$, which corresponds to a probability of $P = 0.01$ for the four additional degrees of freedom added.  The detection rate for sources with Suzaku spectra is 18/34 (53\%), while the detection rate for XMM-Newton spectra is 7/14 (50\%).  We find no relationship between $\Delta\chi^2$, on adding the edges, and the total number of counts in the spectrum (i.e., there are spectra with few counts and strong detections of absorption edges, as well as spectra with many counts and low $\Delta\chi^2$ on adding the absorption edges).  Likewise, there is no relationship between the measured optical depth of the absorption features and the number of counts in the spectrum (e.g., the four sources with the highest $\tau$ in the \ion{O}{7} edge have between $\sim 1.5 - 4.2 \times 10^{5}$\,counts, while spectra in our sample range from total counts of $\sim 5 \times 10^{5} - 2 \times 10^{6}$).  Therefore, we find no evidence for biases in our detection rates or measured edge strengths with the total counts in the spectra.

In the companion paper, \citet{2010ApJ...725L.126W}, we showed that the detection of \ion{O}{7} and \ion{O}{8} is dependent on luminosity, accretion rate, and column density.  In particular, we found that detection rates are higher in sources with larger neutral column densities and lower in the more luminous sources (see Table 2 and Figure 2 in \citealt{2010ApJ...725L.126W}).  Additionally, we find that the strength of the warm absorber features are also dependent upon column density.  This is illustrated in 
Figure~\ref{fig-eddingtonlimit}, where the optical depth in the \ion{O}{7} edge is conveyed by the size of the symbols (i.e., the largest symbols correspond to the highest optical depths).  In the plot, the neutral column densities are from Table~\ref{tbl-fits1}, with the exception of NGC 3783.  Determinations of the neutral column density of NGC 3783 are available from several observations from 2001--2007 utilizing XMM-Newton, Chandra, and Suzaku.  The published measurements range from $\sim 5 \times 10^{21} - 10^{22}$\,cm$^{-2}$ \citep{2004ApJ...602..648R,2007MNRAS.379.1359M,2009PASJ...61.1331M}.  The source is variable in the X-rays and it is unclear whether our low measurement in the 2009 Suzaku observation represents a change in column density from the earlier epochs or is the result of difficulties measuring the column in such a complex spectrum.  In the plot, we adopt N$_{\rm H} = 5 \times 10^{21}$\,cm$^{-2}$ for NGC 3783.

The N$_{\rm H}$-Eddington ratio plot is a useful illustration of the effects of radiation pressure on dusty gas surrounding the AGN (see for instance \citealt{2006MNRAS.373L..16F}, \citealt{2008MNRAS.385L..43F}, and \citealt{2010MNRAS.402.1081V}).  In Figure~\ref{fig-eddingtonlimit}, the blue line represents the {\it effective} Eddington limit, where gravitational pressure balances with radiation pressure on dusty gas with solar abundances for the dust grains, as computed in  \citet{2006MNRAS.373L..16F}.  The radiation pressure on dusty gas is enhanced relative to that of ionized gas; such that the the Eddington limit for dusty gas is lower.  Outflows driven by radiation pressure in a dusty medium are expected when both the Eddington ratio and the neutral column density are high (i.e., where the effective Eddington limit is exceeded).  Sources in this region are believed to be short-lived, as the radiation pressure will eventually drive away the obscuring dusty material, leaving unobscured, low Eddington ratio sources.  

The sources with the strongest absorption features, i.e., highest optical depth for \ion{O}{7} absorption, clearly fit with the Fabian et al., model.  They have high column densities and Eddington ratios near or above the effective Eddington limit for dusty gas.  The estimates of the Eddington ratios for 4/6 of these sources are secure, since the mass estimates are based on reverberation mapping \citep{2004ApJ...613..682P}.  Therefore, a wind driven by radiation pressure on the dusty gas is a likely mechanism for creating the absorption features seen in these sources.  Assuming that the absorption edges trace outflowing ionized gas, the sources with the strongest outflow detections are NGC 3516, NGC 4151, Mrk 6, NGC 3227, NGC 526A, and NGC 3783.  For NGC 3516, NGC 4151, NGC 3227, and NGC 3783, UV and X-ray grating results confirm that the warm ionized gas is indeed outflowing, with measured absorption lines blue-shifted by $\sim 100 - 600$\km~(see, for instance,  \citealt{1999ApJ...516..750C,2001ApJ...555..633C} for analysis of UV Hubble spectroscopy; also note that the low-ionization lines of NGC 3227 are red-shifted, whereas the higher ionization lines are blue-shifted).  Neither Mrk 6 or NGC 526A have available UV spectroscopy, as these are highly reddened sources, and X-ray grating observations of these sources do not have sufficient signal-to-noise to confirm that the gas is outflowing.

Since the strong outflow sources also have complicated spectra, we searched the literature for confirmation that the high neutral column densities we measured were also detected in alternative spectral fits.   We already described above the column density measurements for NGC 3783, which indicate N$_H \ga 5 \times 10^{21}$\,cm$^{-2}$.  \citet{2011ApJ...731...21M} find that there is a variable absorber in Mrk 6, with N$_{\rm H} \sim 10^{21} - 10^{23}$\,cm$^{-2}$ over the past 6 years of available X-ray spectroscopy.  

The ASCA and Chandra analyses of NGC 4151 also reveal variable high column density gas (e.g., \citealt{2010ApJ...714.1497W}).  For this source, we note that the soft X-rays are dominated by emission features, with a weak and heavily absorbed continuum, and that the grating observations reveal that the edges from \ion{O}{7}, \ion{O}{8}, \ion{Ne}{9}, and \ion{Ne}{10} are blended together \citep{2005ApJ...633..693K}.  Simple fits to the Suzaku data of NGC 4151 indicate that there is a strong component of photoionized emission, as in the earlier observations, but a more comprehensive analysis of the complex spectrum is beyond the scope of the current paper.  Despite the complexity of NGC 4151's X-ray spectra, the grating observations clearly indicated a high column density ($\sim 10^{22}$\,cm$^{-2}$) of absorbed gas, consistent with this study, particularly when NGC 4151 is in a lower flux state (as is the case in the Suzaku spectrum presented here).  Since the soft continuum is very weak and dominated by strong emission, we can not constrain the warm absorption properties of NGC 4151.  We can say that it is heavily absorbed and that the complexity of this source is unique in our sample, as no other source shows as weak a continuum or as strong of emission features.

An independent analysis of the XMM-Newton spectrum of NGC 526A agrees with our measured column of $\sim 10^{22}$\,cm$^{-2}$ \citep{2011MNRAS.413.1206B}.  NGC 3516 also is known to exhibit high column density ($\ga 10^{22}$\,cm$^{-2}$) gas in its X-ray spectrum \citep{2008PASJ...60S.277M}.
The XMM-Newton analysis by \citet{2009ApJ...691..922M} of NGC 3227 uncovers a lower neutral component, $9 \times 10^{20}$\,cm$^{-2}$, than our measured value of $\sim 1.6 \times 10^{22}$\,cm$^{-2}$, but since the source's column density has been observed to vary in the past and there is no published analysis of the Suzaku spectrum we analyzed, it is unclear how our results compare.  As a whole, our literature search supports our conclusion that the sources with the strongest outflow detections are associated with higher column densities of gas.

Our spectral fits allow insight into the spectral properties of sources with strong outflows.  In particular, we identify signs of strong absorbers that can be used in lower quality data, for instance, for low flux or higher redshift sources where X-ray grating spectroscopy is infeasible.  Two clear signs of sources with strong outflows are a flat power-law index and a strong soft excess. We find that among the sources with the strongest warm absorber detections, all of these sources have measured values for $\Gamma \approx 1.0$.  Two other sources with $\Gamma < 1.5$ also have complex absorption in their spectra: EXO 055620-3820.2 and NGC 5548.  These low measured values are likely due to difficulties in modeling the spectra, which are absorbed by multiple components of dusty and ionized gas.  Likely, the ``true'' direct emission of these sources is a $\Gamma \sim 1.9$ power-law, as found for the majority of our sources, but the absorption distorts the soft X-ray spectrum, flattening the fit.  Even with the extended hard X-ray spectra from Suzaku HXD and Swift BAT, the intrinsic power-law continuum is difficult to uncover from the fitting process.  While a flat spectrum is not physically representative of the direct emission, it is an easy diagnostic for identifying sources with complex absorbed spectra, both for Seyfert 1s and heavily absorbed Compton-thick spectra, in lower signal-to-noise data.  

Additionally, we find that sources with strongly detected outflows (confirmed through the archived grating and UV data) also have the strongest soft excesses (see also \S~\ref{softexcess}). As with the flat power-law index, a strong soft excess can be an effect of strong absorption features distorting the spectrum.  Alternatively, the soft excess could be related to the warm ionized absorbing gas.  If the soft excess is a feature of the outflow/ionized absorption, this is in agreement with the soft excess as complex absorption model \citep{2004MNRAS.349L...7G,2007MNRAS.374..150S}.


From our spectral fitting of  \ion{O}{7} and \ion{O}{8} absorption edges, we find that the 0.73\,keV \ion{O}{7} edge tends to have the higher optical depth.  For the sources classified as not having strong warm absorber detections, based on $\Delta\chi^2$, only 10 sources have an optical depth of zero in the \ion{O}{7} feature.  This suggests that warm ionized absorbing gas is present in at least 80\% of the sample.  However, it is unclear whether this material is outflowing, as high signal-to-noise UV or X-ray grating observations are needed to measure the velocity shifts in individual absorption lines.  Among the sources with zero optical depth, three sources are BLRGs, four are Sy 1s, two are Sy 1.2s, and one is a Sy 1.5.  All of these sources have low neutral column densities (N$_{\rm H} \la 10^{20}$\,cm$^{-2}$).   In \citet{2010ApJ...725L.126W}, we discussed how the published observations of each of these three BLRGs reveal ionized absorption, with ionization parameters higher or lower than are expected to create the \ion{O}{7} and \ion{O}{8} edges.  It is possible that this is the case for the remaining seven sources, as well, in which case the covering fraction of an outflow is $\Omega \sim 1$.  In the following section, we explore this through a discussion of our analytic warm absorber fits.  With these models, we determined both the column density and ionization parameter of ionized gas in our entire sample, including both the sources with and without clear detections of \ion{O}{7} and \ion{O}{8} absorption.


\subsubsection{Analytic Warm Absorption Models}\label{warmabs}
In \citet{2010ApJ...725L.126W}, we claimed that while the detection rate of \ion{O}{7} and \ion{O}{8} edges was low in the highest luminosity sources (33\%) and high in the low luminosity sources (60\%), more detailed spectral fits reveal the presence of ionized gas that in many cases is either more or less ionized than required to produce strong \ion{O}{7} and \ion{O}{8} features.  As a follow-up to this work, we fit analytic models to the X-ray CCD spectra.  Results of these fits are included in Tables~\ref{tbl-warmabs-outflow} and \ref{tbl-warmabs-nonoutflow}.  The X-ray CCD data does not have the energy resolution to accurately determine the velocity of outflow components.  However, we were able to determine both the column density of warm ionized gas (N$_{\rm warm}$) and the ionization parameter ($\xi =  {\rm L}_{\rm ion}/(n_e R^2)$; where L$_{\rm ion}$ is the ionizing luminosity, $n_e$ is the electron density, and $R$ is the distance of the ionized gas from the central ionizing source) for our entire sample.  

For the sources with clear detections of absorbers through \ion{O}{7} and \ion{O}{8} edges, we fit the spectra with two ionized components. We add only one warm absorber component for the sources without strong detections of the edge features, to determine limits on ionized absorbers present.  For the majority of sources in Table~\ref{tbl-warmabs-outflow}, the two component analytic model improves the fits significantly, as indicated by the $\Delta\chi^2$ values (e.g., out of the 25 sources fit with a two-component model, only 8 show a second component to have low significance with $\Delta\chi^2 < 20$ upon adding the second warm absorber).

\subsubsection*{Caveats on the Analysis and Detection of Ionized Absorbers in CCD Data}
While ideally we would compare our results to those from an X-ray grating analysis, the complication of X-ray variability makes such comparisons difficult.  For instance, while 8 sources overlap between our analysis and the uniform Chandra grating analysis in \citet{2007MNRAS.379.1359M}, a comparison of the observed soft fluxes from our observations and the Chandra observations reveals that 6 of the sources vary considerably.  We find, for example, that NGC 4051 is ten times brighter in the Suzaku observation, while NGC 4593 is four times fainter.  Therefore, there is no consistent way to compare our results unless we know that a source does not vary or we analyze simultaneous grating and CCD data.  In \citet{2010ApJ...719..737W}, such an analysis is presented using the 2009 XMM-Newton CCD and grating spectrum of NGC 6860, along with a Suzaku observation, taken a year earlier, which showed no variability from the XMM observation.  In this analysis, the best-fit neutral and ionized column densities were consistent within the error-bars between the Suzaku CCD, XMM-Newton CCD, and XMM-Newton grating observations.  We find, however, that the ionization parameter, consistent in both CCD analyses ($\log \xi \sim 2.0$), is higher in the grating observation ($\log \xi = 2.4$).  The second warm absorber fit had best-fit values that varied considerably between the CCD and grating measurements.  Since multiple components are present in AGN spectra and the quality of the grating spectrum from NGC 6860 was relatively low signal-to-noise, it is unclear how these results reflect on our entire sample.  However, it is likely that the measured column densities are accurate, particularly for significant warm absorption components.  

It is also important to note that since we do not have velocity information on the detected ionized absorbers, due to the low resolution of the CCD data, we can not confirm that the absorbers originate in an AGN driven outflow (i.e., due to the low velocity resolution, we fixed the velocity of the ionized absorbers to the systemic velocity of the AGN).  We have shown that for the sources with the strongest detections, outflows are likely present, since UV/X-ray grating observations confirm blue-shifted absorption features.  For the remaining sources, the absorbing gas could be intrinsic to the AGN or a feature from ionized gas in our own Galaxy, the host galaxy, or intervening systems (signatures of the warm hot ionized medium).  A Chandra grating study by \citet{2004ApJ...617..232M}, for instance, detected oxygen absorption in half of a sample of 15 type 1 AGN sources that was identified as hot gas from local interstellar structures and potential intergalactic medium features.  In order to test whether the ionized gas is outflowing from the AGN, high signal-to-noise X-ray grating observations and/or ultraviolet spectra are needed to measure the velocity of the absorption features.  In future work, we will present an analysis of the X-ray grating and ultraviolet spectra of our sample.  

\subsubsection*{Results of the Analytic Warm Absorption Spectral Fits}
For the six sources with the strongest \ion{O}{7} and \ion{O}{8} absorption features, we find average column densities of warm ionized gas of a few $10^{21}$\,cm$^{-2}$ in each of the two absorption components fit to the spectra.  All but NGC 3227 have a warm ionized component with an ionization parameter near $\xi \approx 100$\,ergs\,s$^{-1}$ (NGC 3227's highest measured ionization component is $\xi \approx 28$\,ergs\,s$^{-1}$).  Out of the entire sample of sources with detected \ion{O}{7} and \ion{O}{8} edges, we find that the average column and ionization parameters, for the component with the highest warm ionized column density, are $N_{\rm  warm} = (3.8 \pm 9.1) \times 10^{21}$\,cm$^{-2}$ and $\log \xi = 2.04 \pm 1.27$.  We find $N_{\rm warm} = (3.9 \pm 4.1) \times 10^{20}$\,cm$^{-2}$ and $\log \xi = 1.59 \pm 1.69$ for the second component.  These values are in line with the range in column density (N$_{\rm warm} \sim 10^{20} - 10^{23}$\,cm$^{-2}$) and ionization parameter ($\xi \sim 10^{0-4}$\,ergs\,s$^{-1}$) found from studies of X-ray grating observations (see \citealt{2005AA...431..111B} and \citealt{2007MNRAS.379.1359M}).

For the sources without strong detections of \ion{O}{7} and \ion{O}{8}, we find $N_{\rm warm} = (1.8 \pm 2.2) \times 10^{20}$\,cm$^{-2}$ and $\log \xi = 1.09 \pm 2.04$.  Of these, three have significant warm absorption detections with the analytic model, including, 1H 0419-577 (low column and low ionization parameter), EXO 055620-3820.2 (low column and low ionization parameter), and 3C 382 (with low column and a higher ionization parameter).  To test whether the analytic model result of a very low ionization parameter in the spectrum of 1H 0419-577 (the source with the highest significance with the analytic model in Table~\ref{tbl-warmabs-nonoutflow} and lowest measured ionization parameter) was plausible, we alternatively fit the spectrum with an edge model for lowly ionized oxygen (\ion{O}{1}/\ion{O}{2}) at 0.545\,keV.  We found that the lowly ionized edge was very significant, with $\Delta\chi^2 = 73.3$, and find an optical depth of $\tau = 0.15^{+0.03}_{-0.03}$.

Our results, then, show that sources without strong detections of \ion{O}{7} and \ion{O}{8} absorption edges tend to have lower column densities and lower ionization parameters, on average, than sources with strong detections.  In particular, the column densities of warm ionized gas are an order of magnitude lower in the non-detection sources.  The ionization parameters, however, do show a large range of values, including both lowly ionized and highly ionized gas.  

In Figure~\ref{fig-nwarm1}, we compare the results on the properties of the warm ionized gas in sources with strong \ion{O}{7} and \ion{O}{8} absorption edges.  Component 1 is chosen as the model component with the highest column density from Table~\ref{tbl-warmabs-outflow}.  In the top plots, we compare the ionization parameters to the bolometric luminosity.  We find that the ionization parameter, $\xi$, has no dependence on the luminosity of the AGN.  Instead, we find the majority of values clustered near $\xi = 100$\,ergs\,s$^{-1}$ for component 1 and a broader range in $\xi$ for component 2.  This
method is not particularly sensitive to very high $\xi$ ($>3.5$) or very low  $\xi$
($<1.0$)
absorbers, since those absorber components are driven by fits to a small
number of highly ionized lines (e.g. \ion{Si}{14} Lya) or Fe UTAs, which require
grating spectral resolution to fit properly. The Fe UTAs in particular may
show up in the \ion{O}{7} and \ion{O}{8} edge fits.  The most likely scenario to account for the different components is that the absorbing region consists of a warm ionized medium with higher density blobs or filaments embedded within it \citep{2001ApJ...561..684K}.  In this case, the sources without the strong detections of \ion{O}{7} and \ion{O}{8} have similar ionization parameters with the lower density, ionized medium also detected in the sources with 'strong' absorber detections.

\subsubsection{Conclusions on Warm Absorber Properties}\label{subsect-conclusions}
The most comparable previous studies to the work we present on the warm absorber properties in the Swift-detected Seyfert 1s are the ASCA studies of  \citet{1997MNRAS.286..513R} and \citet{1998ApJS..114...73G}.  In \citet{2010ApJ...725L.126W}, we discussed the overlap in our sample with the  \citet{1997MNRAS.286..513R} study.  Out of the 24 sources in the ASCA study, 18 are also in our sample.  Between publication of \citet{2010ApJ...725L.126W} and our present results, we revised the classification of two sources (IC 4329A and NGC 5548) to have warm absorption detections.  Thus, we find good agreement in our classification scheme with the \citet{1997MNRAS.286..513R} study, with 16/18 sources classified accordingly.  Both 3C 382 and 3C 390.3 were classified by  \citet{1997MNRAS.286..513R} as exhibiting \ion{O}{7} and \ion{O}{8} features in the ASCA data, while our analysis shows that these sources do not have significant detections.  Grating spectroscopy of these sources reveal that ionized outflows are present, but at higher ionization parameters 
\citep{2010MNRAS.401L..10T,2009ApJ...700.1473S}.  

Our detection rate for warm absorption features, revealed by the significance of \ion{O}{7} and \ion{O}{8} absorption edge features, is 52\%.  This is in agreement both with the more biased samples presented in the ASCA study of  \citet{1997MNRAS.286..513R} and the UV outflow studies of \citet{1999ApJ...516..750C}, but much lower than the $\sim 70$\% detection rate reported in \citet{1998ApJS..114...73G}.  While our detection rate, assuming that the warm absorbers are produced in outflowing gas, initially suggests a covering fraction of AGN outflows of $\Omega \sim 0.5$, there is clearly more that needs to be considered in this simple picture.

In \citet{2010ApJ...725L.126W}, we pointed out that the detection rate of \ion{O}{7} and \ion{O}{8} edges was related to the luminosity of the AGN.  There are fewer detections in the most luminous ($\sim 30$\%) and higher detection rates for the least luminous ($\sim 60$\%) sources.  Grating observations, however, reveal that the detection rate is higher -- consistent with outflows in at least 80\% of the most luminous sources.  In the current study, we find that the main difference between sources with and without \ion{O}{7} and \ion{O}{8} detections is the measured column density of potentially outflowing ionized gas.  We find that sources with strong detections have an order of magnitude or higher ionized column densities than those without detections.  Therefore, we can not rule out that sources without \ion{O}{7} and \ion{O}{8} detections do not have warm absorbers present.  Our results support the hypothesis that they do have ionized gas, just at lower column densities than are easily detected in the X-ray data, which is most sensitive to absorption with N$_{\rm H} > 10^{20}$\,cm$^{-2}$.  This is still orders of magnitude higher than the ionized gas detected in the UV, which has column densities from N$_{\rm ion} \sim 10^{12} - 10^{14}$\,cm$^{-2}$.

Through this work, we suggest a change in the paradigm of assuming a 50\% covering fraction for AGN warm absorbers (and by extension outflows, since grating observations reveal that this gas is typically outflowing) to $\sim$100\%.  Our analysis of an unbiased sample of type 1 AGNs selected in the very hard X-rays with Swift shows that while the covering fraction is likely unity, there are clear distinctions between the higher column density warm absorbers, which have neutral and ionized columns  $> 10^{20}$\,cm$^{-2}$, and lower column density warm absorbers.  The higher column density sources make up half of the sample.  Among these sources with strong detections of warm absorbers/outflows, we suggest that the strongest outflows are driven by radiation pressure on dusty gas, as claimed by  \citet{2006MNRAS.373L..16F}.

A main goal of AGN outflow study work is to determine how much mass and energy is entrained in the outflow.  While work such as the grating study of \citet{2005AA...431..111B} places estimates on the energy in the outflows (e.g., \citealt{2005AA...431..111B} estimate that outflows account for less than 1\% of the bolometric AGN luminosity), many assumptions, particularly relating to the geometry of the outflow, are necessary to make these estimates (see, for instance, \citealt{2007MNRAS.379.1359M} for a discussion of these assumptions).  Therefore, we do not include estimates of the mass outflow rate or the total energy in the current paper.  In future work, we will follow-up the current study with a uniform analysis of the archived grating observations, available for the majority of our sample but without as uniformly high a signal-to-noise ratio as in the CCD data presented in this work.   Additionally, estimates of the total outflow rate/energy must rely on multi-wavelength data, characterizing the AGN outflow contribution from neutral through the most highly ionized gas (e.g., 
\citealt{2011arXiv1109.2882T} detect Fe K-shell absorption lines in $> 35$\% of their sample of 42 low redshift, radio-quiet AGN), and include studies of the obscured AGN (X-ray type 2) sources, whose UV and soft X-ray ionized absorber properties are difficult to determine.  We will explore both of these lines of research in our future work.

\section{Summary}\label{sect-conclusion}
In this paper, we fit broad-band spectra of 48 Seyfert 1--1.5 sources in the 0.3--195\,keV X-ray band.  We utilized a combination of Suzaku XIS/HXD and Swift BAT or XMM-Newton pn and Swift BAT observations.  Through our spectral fits, we present the full X-ray properties of our sources, including direct emission, reflected emission, Fe K$\alpha$, soft excess, and warm absorption.  

\begin{itemize}
\item{\bf Basic properties, black hole masses, and accretion rates.}
The sources in our sample consist of all of the Seyfert 1--1.5 sources detected with $5\sigma$ significance in the Swift BAT 9-month survey 
\citep{2008ApJ...681..113T}.  The sample consists of 15 Sy 1s, 9 Sy 1.2s, and 17 Sy 1.5s, as well as 7 broad line radio galaxies, which are at a higher redshift than the Seyferts (i.e., $\langle z \rangle = 0.07$ for BLRGs, while $\langle z \rangle = 0.02$ for the Seyferts).  We caution that Sy1/1.2/1.5 classifications can be somewhat fluid, nevertheless we find that on average the neutral column densities of our sources are higher for Seyfert 1.5s than Seyfert 1s, with average values for Seyfert 1.5s ($\langle N_{\rm H} \rangle = 4 \times 10^{20}$\,cm$^{-2}$) 10 times higher than those for Seyfert 1s ($\langle N_{\rm H} \rangle = 4 \times 10^{19}$\,cm$^{-2}$).  We will test this result further with a larger sample from the newest Swift BAT catalog.  The average black hole masses of our sample, estimated from 2MASS bulge magnitudes corrected to match H$\beta$ FWHM estimates from a sub-set of our sample, are $M/M_{\sun} = 7.9 \times 10^7$ for the Seyferts and are slightly higher at $M/M_{\sun} = 3.2 \times 10^8$ for the BLRGs (see \citealt{2010ApJ...710..503W} for more information on mass determinations in the Swift sample).  The Swift BAT-band luminosities range from $\log L_{14-195\,{\rm keV}} = 41.7 - 45.1$, corresponding to bolometric luminosities of $L_{\rm bol} \sim 43.0 - 46.1$ (see \S~\ref{lum} for details on our correction from BAT band to bolometric luminosity).  We estimate an average accretion rate for our sample of $L_{\rm bol}/L_{\rm Edd} \sim 0.025$.


\item{\bf Direct power-law and reflected emission.}
The direct emission for our sources is best-fit with a power-law of the form $\Gamma = 1.90 \pm 0.2$.  The power-law index is correlated
with both the bolometric luminosity and accretion rate, such that $\Gamma \propto L_{\rm bol}^{(0.16 \pm 0.03)}$ and $\Gamma \propto (L_{\rm bol}/L_{\rm Edd})^{(0.23 \pm 0.03)}$.  A population of sources exhibit flat power-law indices of $\Gamma \sim 1$.  The spectra of these sources exhibit complex absorption and include the targets with the strongest detections of \ion{O}{7} and \ion{O}{8} absorption.  This result, low power-law indices as a diagnostic for complex absorbers, is useful for determining complex absorption sources in low signal-to-noise data.  In addition to determining the power-law index for our sample, we attempted to determine the cut-off energy using our broad-band 0.3-195\,keV fits.  For the majority of sources, we could not constrain the cut-off energy.  Of the 9/48 spectra where the cut-off energy was constrained, we find values in the range of $\sim 20-400$\,keV.  This allows us to rule out a rapidly rising gamma-ray contribution to the spectral energy distribution up to 400\,keV and is in agreement with the lack of a gamma-ray detection for Swift-detected Seyferts with Fermi/LAT \citep{2011arXiv1109.2734T}.  We also determined the properties of the reflected emission, using the {\tt pexrav} model.  In our sample, we find reflected emission is significant in 77\% of the spectra.  The reflection parameter, $R$, is correlated with the photon index but not the luminosity of the source.  

\item{\bf Soft excess.}
The soft excess is detected in 94\% of the sample.  In our previous analysis of lower signal-to-noise spectra, we found a detection rate of 40--50\% \citep{2008ApJ...674..686W,2009ApJ...690.1322W}.  The higher signal-to-noise in the present study allowed us to detect fainter soft excesses than our previous study.   The temperature of this component, modeled with a blackbody, is extremely uniform, with $kT = 0.11 \pm 0.04$\,keV.  We find that the luminosity of the soft excess is well-correlated with the luminosity of the power-law emission.  We also find that the strongest soft excesses are seen in sources with the highest neutral column densities (i.e., Sy 1.5s) and the strongest \ion{O}{7} and \ion{O}{8} absorption detections.  In particular, we find that the soft excess normalization is strongly correlated with the neutral column density ($A_{\rm kT} \propto N_{\rm H}^{1.74 \pm 0.26}$) and the strength of the ionized absorber ($A_{\rm kT} \propto \tau_{\rm O\,VII}^{1.62 \pm 0.19}$).  This shows that the soft excess, neutral column density, and ionized gas are all related.

\item{\bf Fe K$\alpha$ emission.}
Additionally, we determined the properties of the neutral Fe\,K$\alpha$ emission in our sample.  We determined the energy, width, and equivalent width of the narrow emission; finding average values of $E = 6.41 \pm 0.04$\,keV and $EW = 83.10 \pm 70.08$\,eV, with widths ranging from $FWHM \sim 2780 - 9180$\km.  The $EW$ is weakly correlated with the bolometric luminosity, but we find a strong correlation with accretion rate ($EW \propto (L_{\rm bol}/L_{\rm Edd})^{(-0.38 \pm 0.07)}$).  This shows that the primary driver of the X-ray Baldwin effect is the correlation with accretion rate.

\item{\bf Warm Absorbers.}
With a uniform base model fit to the broad-band X-ray spectra, we tested for the presence of warm ionized gas using two methods.  In the first, we fit absorption edges to the 0.73\,keV \ion{O}{7} and 0.87\,keV \ion{O}{8} features.  These features were used to uniformly detect ionized gas in CCD data with ASCA \citep{1997MNRAS.286..513R}.  Based on these features, we detect significant absorption in 52\% of our sample, which is in agreement with earlier ASCA studies, although an Fe UTA may complicate edge measurements.  In our companion paper, \citet{2010ApJ...725L.126W}, however, we showed that these detections are correlated with high column densities.  Additionally, we found that the detection rate was lower in higher luminosity sources (30\%) than in low luminosity sources (60\%).  

The sources with the highest optical depths in the absorption features have the highest measured column densities.  They also have strong soft excesses, although it is possible that these high values are a result of degeneracies in the modeling their complex spectra.  Among the strong warm absorber sources, 4/6 were observed to show outflowing gas of $\sim 100 - 600$\,\km\,in UV/X-ray grating analyses, while the final two have no UV/high signal-to-noise X-ray grating data available.  The sources are close to or exceed the effective Eddington limit on dusty gas.  We conclude that for these sources, radiation pressure on dusty gas likely drives their outflows.

As a second method of detecting ionized gas, we fit the spectra with analytical warm absorption models using the XSTAR code, fitting directly for the ionized column density and ionization parameter ($\xi$) (since we used CCD data, we can not determine the velocity).  We find that the sources with strong detections of \ion{O}{7} and \ion{O}{8} absorption edges show multiple absorber components, in agreement with general results from grating studies (e.g., \citealt{2005AA...431..111B} and \citealt{2007MNRAS.379.1359M}),with at least one component of $N_{\rm warm} \ga 10^{21}$\,cm$^{-2}$ and $\xi \sim 100$\,erg\,s$^{-1}$.  A second measured component tends to exhibit lower column densities and a range of ionization parameters.  It is the first, higher column density, absorber to which the measurements of \ion{O}{7} and \ion{O}{8} edges are most sensitive.  Sources without strong \ion{O}{7} and \ion{O}{8} detections also exhibit warm absorber signatures in their spectra, but we find that the column densities of these absorbers are an order of magnitude less ($N_{\rm warm} \sim 10^{20}$\,cm$^{-2}$) than in the higher column component of sources with strong \ion{O}{7} and \ion{O}{8} absorbers. In future work, we will determine the velocities of these absorbers, to confirm outflowing gas, utilizing archived X-ray grating data.  Additionally, we are currently in the process of analyzing optical spectroscopy to determine the properties of less ionized ([\ion{O}{3}]) and neutral gas (\ion{Na}{1}D) and compare the results with the warm absorber properties presented in this study.


\end{itemize}

\acknowledgements
LMW acknowledges CU undergraduate student Tatiana Taylor for assistance with the initial data reductions.
Also, we gratefully acknowledge support from NASA grant NNX09AV60G for Suzaku guest observer observations and NASA grant HST-HF-51263.01-A, through a Hubble Fellowship from the Space Telescope Science Institute, which is operated by the Association of Universities for Research in Astronomy, Incorporated, under NASA contract NAS5-26555.  This work utilizes observations obtained with XMM-Newton, an ESA science mission with instruments and contributions directly funded by ESA Member States and NASA.  This research has made use of data obtained through the High Energy Astrophysics Science Archive Research Center Online Service, provided by the NASA/Goddard Space Flight Center.

{\it Facilities:} \facility{Suzaku()}, \facility{Swift()}, \facility{XMM()}

\bibliography{MyBibtex.bib}

\begin{deluxetable}{llllllc}
\tabletypesize{\scriptsize}
\tablecaption{Swift BAT Seyfert 1--1.5 Source List\label{table-sourcelist}}
\tablewidth{0pt}
\tablehead{
\colhead{Source} & \colhead{RA (h m s)} & 
\colhead{Dec ($\degr$\,$\arcmin$\,$\arcsec$)} & \colhead{$z$\tablenotemark{\dagger}} &
\colhead{\nh(Gal)\tablenotemark{\ddagger}} & \colhead{X-ray Obs.} & \colhead{Type\tablenotemark{\dagger}}
}
\startdata
\object{Mrk 352}                  & 00 59 53.3  & $+$31 49 36.8     & 0.0149  & 0.0559  & Suzaku  & 1    \\
\object{Fairall 9}                & 01 23 45.8  & $-$58 48 20.5  & 0.0470  & 0.0328  & Suzaku  & 1    \\
\object{NGC 526A}                 & 01 23 54.4  & $-$35 03 55.4  & 0.0191  & 0.0218  & XMM     & 1.5  \\
\object{Mrk 1018}                 & 02 06 16.0  & $-$00 17 29.0     & 0.0424  & 0.0258  & Suzaku     & 1.5  \\
\object{Mrk 590}                  & 02 14 33.6  & $-$00 46 00.1     & 0.0264  & 0.0269  & XMM     & 1.2  \\
\object{NGC 931}                  & 02 28 14.5  & $+$31 18 42.1     & 0.0167  & 0.0723  & XMM     & 1.5  \\
\object{NGC 985}			& 02 34 37.8 & $-$08 47 15 & 0.0431 & 0.0290 & Suzaku & 1 \\
\object{ESO 198-024}              & 02 38 19.7  & $-$52 11 32.3  & 0.0455  & 0.0305  & XMM     & 1    \\
\enddata
\tablenotetext{\dagger}{Redshift and optical Seyfert type, as recorded in the NASA/IPAC Extragalactic Database, or from optical spectroscopy in \citet{2010ApJ...710..503W}.}
\tablenotetext{\ddagger}{The Milky Way Galactic column density along the line of sight to the indicated source, in units of $10^{22}$\,cm$^{-2}$, from the \ion{H}{1} measurements of \citet{1990ARAA..28..215D}.}

\end{deluxetable}

\begin{deluxetable}{lllllllll}
\tabletypesize{\scriptsize}
\tablecaption{X-ray Observation Details for Suzaku Spectra\label{tbl-suzaku}}
\tablewidth{0pt}
\tablehead{
\colhead{Source} & \colhead{Observation ID} & 
\colhead{Date} & \colhead{Pointing} &
\colhead{Rate(XIS)} & \colhead{Time(XIS)} & 
\colhead{Rate(HXD)} & \colhead{Time(HXD)} &
\colhead{Ref}
}
\startdata
Mrk 352                 & 704025010  & 2010-01-06  & HXD  & 0.74, 0.98  & 91, 46    & 0.38  & 21   &      \\
Fairall 9               & 702043010  & 2007-06-07  & XIS  & 2.01, 2.78  & 336, 168  & 0.10  & 127  & 1    \\
Mrk 1018 		& 704044010 & 2009-07-03 & HXD & 0.72, 1.02 & 88, 44 & 0.05 & 38 & \\
NGC 985 		& 704042010 & 2009-07-15 & HXD & 0.86, 1.12 & 64, 32 & 0.04 & 24\\
ESO 548-G081            & 704026010  & 2009-08-03  & HXD  & 0.80, 1.00  & 79, 39    & 0.37  & 33   &      \\
1H 0419-577             & 702041010  & 2007-07-25  & XIS  & 1.53, 2.21  & 412, 206  & 0.09  & 143  & 2    \\
3C 120 			& 700001010 &  2006-02-09 & XIS & 4.78 & 42 & 0.11 & 32 & 3\\
Ark 120                 & 702014010  & 2007-04-01  & HXD  & 2.28, 3.09  & 202, 101  & 0.12  & 89   &      \\
\enddata
\\
Observation details include the ID, date, pointing (HXD or XIS nominal), and count rates and exposure times for the XIS (combined front-illuminated (XIS0, XIS3, and where available XIS2), XIS1) and HXD PIN detectors.  The final column indicates references to papers with a previously published analysis of the data.\\
References: (1) \citet{2009ApJ...703.2171S}, (2) \citet{2009ApJ...698...99T}, (3) \citet{2006AN....327.1079R}, (4) \citet{2006AN....327.1087M}, (5) \citet{2009PASJ...61.1331M}, (6) \citet{2010ApJ...709.1230T}, (7) \citet{2007A&A...475..121T}, (8) \citet{2009ApJ...705..496M}, (9) \citet{2010ApJ...710..360K}, (10) \citet{2010ApJ...710.1228L}, 
(11) \citet{2008sf2a.conf..263P}, (12) \citet{2010ApJ...719..700T},
(13) \citet{2010ApJ...719..737W},
(14) \citet{2008MNRAS.388.1037L}, and
(15) \citet{2009MNRAS.394.1487P}
. 
\end{deluxetable}

\begin{deluxetable}{llllll}
\tabletypesize{\scriptsize}
\tablecaption{X-ray Observation Details for XMM-Newton pn Spectra\label{tbl-xmm}}
\tablewidth{0pt}
\tablehead{
\colhead{Source} & \colhead{Observation ID} & 
\colhead{Date} & \colhead{Exposure Time} & 
\colhead{Count Rate} &
\colhead{References}
}
\startdata
NGC 526A           & 0150940101  & 2003-06-21  & 42  & 3.69   & 1  \\
Mrk 590            & 0201020201  & 2004-07-04  & 31  & 3.67   & 2  \\
NGC 931            & 0554990101  & 2009-02-13  & 82  & 10.34  &    \\
ESO 198-024        & 0305370101  & 2006-02-04  & 85  & 5.71   &    \\
MCG -01-13-025     & 0103861401  & 2000-08-30  & 2   & 4.29   & 3  \\
PICTOR A           & 0206390101  & 2005-01-14  & 46  & 5.99   & 4  \\
EXO 055620-3820.2  & 0404260101  & 2006-08-11  & 25  & 0.49   & 5  \\
Mrk 6              & 0061540101  & 2001-03-27  & 26  & 1.29   & 6  \\
\enddata
\\
Observation details include the ID, date, exposure time and count rate for the pn spectrum, and references to papers with a previously published analysis of the data.
References: (1) \citet{2005ApJ...634..161H}, (2) \citet{2006AN....327.1020L}, (3) \citet{2006MNRAS.365..688G}, (4) \citet{2007ApJ...668..203M}, (5) \citet{2009MNRAS.394L...1L}, (6) \citet{2006MNRAS.371..211S}, (7) \citet{2011MNRAS.410.2274Z},
and (8) \citet{2008MNRAS.390.1217M}.
\end{deluxetable}


\begin{deluxetable}{llllll}
\tabletypesize{\scriptsize}
\tablecaption{X-ray Spectral Fits: Base Model\label{tbl-fits}}
\tablewidth{0pt}
\tablehead{
\colhead{Source} & \colhead{$\Delta\chi^2$ N$_{\rm H}$} & 
\colhead{$\Delta\chi^2$ $kT$} & \colhead{$\Delta\chi^2$ Fe K$\alpha$} &
 \colhead{$\Delta\chi^2 R$ } & 
\colhead{$\chi^2/{\rm dof}$} 
}
\startdata
Mrk 352                  & 0      & 521     & 79    & 51     & 1757/1643   \\
Fairall 9                & 0      & 3807    & 2652  & 943    & 4205/2969   \\
NGC 526A                 & 23207  & 1244    & 84    & 47     & 1956/1728   \\
Mrk 1018		& 0 &  426 & 39 & 154 & 1903/1602 \\
Mrk 590                  & 0      & 398     & 199   & 11     & 1157/1180   \\
NGC 931                  & 7262   & 63      & 147   & $-$23  & 4236/1918   \\
NGC 985 			& 0 &  13 & 43 & 1369 & 1795/1548 \\ 
ESO 198-024              & 0      & 566     & 527   & 212    & 1747/1657   \\
\enddata
\\
\vspace{0.5cm}
The change in $\chi^2$ ($\Delta \chi^2 = \chi^2_{base} - \chi^2_{new}$) is shown for the addition of intrinsic neutral absorption (\nh), a blackbody ($kT$), an Fe K$\alpha$ emission line (Fe K$\alpha$), and reflection ($R$, with the {\tt pexrav} model) to the base power-law model.  The $\Delta\chi^2$ and best-fit parameters for the addition of \ion{O}{7} and \ion{O}{8} absorption edges are included in Table 1 of 
\citet{2010ApJ...725L.126W}.  The best-fit $\chi^2/{\rm dof}$ upon the addition of all of the described models in shown in the final column.  For the five sources where our base model does not provide a statistically good fit (reduced $\chi^2 > 2.0$), the $\chi^2$-value is in bold.
\end{deluxetable}

\begin{deluxetable}{lllllll}
\tabletypesize{\scriptsize}
\tablecaption{X-ray Spectral Fits: Base Model\label{tbl-fits1}}
\tablewidth{0pt}
\tablehead{
\colhead{Source} & \colhead{N$_{\rm H}$} & 
\colhead{$kT$} & \colhead{$A_{kT}$} & 
\colhead{$\Gamma$} & \colhead{E$_{\rm cutoff}$ (keV)} &
\colhead{$A_{\Gamma}$}
}
\startdata
Mrk 352                  & (0.2)                   & $0.136^{+0.029}_{-0.032}$ & $4^{+1}_{-0.6}$          & $1.85^{+0.06}_{-0.06}$ & (500)              & $365^{+15}_{-16}$  \\
Fairall 9                & (0.0)                   & $0.142^{+0.004}_{-0.005}$ & $6^{+0.5}_{-0.3}$        & $2.06^{+0.01}_{-0.02}$ & (500)              & $828^{+6}_{-8}$    \\
NGC 526A                 & $65.0^{+4.7}_{-3.2}$    & $0.029^{+0.005}_{-0.004}$ & $25000^{+12000}_{-21000}$ & $1.09^{+0.02}_{-0.02}$ & (15)               & $318^{+8}_{-7}$    \\
Mrk 1018                 & $2.2^{+0.6}_{-0.6}$       & $0.087^{+0.012}_{-0.008}$ & $5^{+2}_{-2}$            & $2.00^{+0.03}_{-0.03}$ & (500)              & $330^{+7}_{-6}$  \\
Mrk 590                  & (0.2)                   & $0.156^{+0.035}_{-0.013}$ & $0.4^{+0.1}_{-0.3}$      & $1.74^{+0.05}_{-0.03}$ & (500)              & $161^{+5}_{-3}$    \\
NGC 931                  & $24.3^{+0.7}_{-0.5}$    & $0.048^{+0.002}_{-0.001}$ & $335^{+48}_{-52}$        & $1.83^{+0.02}_{-0.02}$ & (500)              & $842^{+10}_{-11}$  \\
NGC 985                  & (0.5)    & $0.091^{+0.006}_{-0.007}$ & $10^{+2}_{-2}$        & $1.84^{+0.07}_{-0.04}$ & 163.4              & $385^{+16}_{-10}$  \\
ESO 198-024              & (0.4)                   & $0.086^{+0.013}_{-0.018}$ & $0.5^{+0.4}_{-0.1}$      & $1.86^{+0.02}_{-0.01}$ & (500)              & $287^{+3}_{-2}$    \\
\enddata
\\
\vspace{0.5cm}
Best-fit parameters and errors are given for the intrinsic column density (\nh~in units of $10^{20}$\,cm$^{-2}$), the blackbody model (blackbody temperature, $kT$, in units of keV and the normalization, $A_{kT}$ with units of \norm), and a base cutoff power-law model (with photon index $\Gamma$, cutoff energy, and normalization, $A_{\Gamma}$ with units of \norm).  Upper limits are indicated in parentheses.  For the five sources where errors can not be determined, as the reduced $\chi^2 > 2.0$, the computed values are in italics and the source names are emphasized in bold.  We follow this convention throughout the following tables.

\end{deluxetable}

\begin{deluxetable}{llllc}
\tabletypesize{\scriptsize}
\tablecaption{Fe\,K$\alpha$ and Reflection Parameters\label{tbl-fek}}
\tablewidth{0pt}
\tablehead{
\colhead{Source} & \colhead{Fe\,K E (keV)} & 
\colhead{Fe\,K $\sigma$ (keV)} & \colhead{Fe\,K EW (eV)}  &\colhead{$R$} 
}
\startdata
Mrk 352                  & 6.41                    & 0.01                    & $60.59^{+18.04}_{-20.34}$      & $1.79^{+0.80}_{-1.13}$  \\
Fairall 9                & $6.39^{+0.01}_{-0.01}$  & $0.03^{+0.02}_{-0.03}$  & $83.31^{+7.14}_{-9.14}$        & $4.00^{+0.36}_{-0.26}$  \\
NGC 526A                 & $6.42^{+0.05}_{-0.04}$  & $0.12^{+0.09}_{-0.06}$  & $75.76^{+23.67}_{-19.27}$      & $0.28^{+0.08}_{-0.10}$  \\
Mrk 1018                 & 6.41                    & 0.01                    & $16.37^{+16.99}_{-16.37}$     & $3.47^{+0.56}_{-0.34}$     \\
Mrk 590                  & $6.43^{+0.03}_{-0.03}$  & $0.01^{+0.04}_{-0.01}$  & $74.38^{+22.29}_{-20.75}$      & $3.54^{+0.51}_{-1.07}$  \\
NGC 931                  & $6.38^{+0.02}_{-0.02}$  & $0.06^{+0.03}_{-0.03}$  & $49.10^{+9.38}_{-8.41}$        & $1.85^{+0.25}_{-0.23}$  \\
NGC 985                  & $6.40^{+0.05}_{-0.05}$  & $0.01^{+0.10}$  & $47.86^{+23.27}_{-20.58}$        & $1.57^{+0.41}_{-0.74}$  \\
ESO 198-024              & $6.43^{+0.02}_{-0.02}$  & $0.05^{+0.04}_{-0.05}$  & $56.81^{+14.40}_{-13.59}$      & $2.54^{+0.28}_{-0.33}$  \\
\enddata
\\
\vspace{0.5cm}
Best-fit parameters and errors are given for the Fe K$\alpha$ emission line component, including the energy of the line (in keV), the width of the line, $\sigma$, in keV, and the equivalent width (in eV), along with the reflection parameter ($R = \Omega/2\pi$) from the {\tt pexrav} component.  Where the energy of the emission line was not well-constrained, it was fixed to 6.41\,keV.  Similarly, the width of the line was fixed to 0.01\,keV, where it was poorly constrained.  Values for the reflection component which were poorly constrained are indicated in parentheses.
\end{deluxetable}

\begin{deluxetable}{llllll}
\tablecaption{Warm Absorption through the Detection of O\,VII and O\,VIII Absorption Edges\label{tbl-warmabs}}
\tabletypesize{\scriptsize}
\tablewidth{0pt}
\tablehead{
\colhead{Source} & \colhead{$\Delta\chi^2$} & \colhead{O\,VII} & \colhead{O\,VII} &  \colhead{O\,VIII} & \colhead{O\,VIII} \\
\colhead{} & \colhead{} & \colhead{$E$ (keV)} & \colhead{$\tau$} & \colhead{$E$ (keV)} & \colhead{$\tau$} 
}
\startdata
\multicolumn{6}{c}{\bf Sources With Warm Absorption Detections} \\
\hline
NGC 526A         & 17.99      & $0.70^{+0.01}_{-0.70}$    & $1.354^{+0.243}_{-0.256}$  & 0.87                      & $0.763^{+0.087}_{-0.099}$  \\
NGC 931          & 7008.86    & 0.73                      & $0.453^{+0.027}_{-0.029}$  & $0.87^{+0.01}_{-0.01}$    & $0.096^{+0.016}_{-0.019}$  \\
NGC 985	       & 156.38 & $0.727^{+0.017}_{-0.016}$ & $0.290^{+0.085}_{-0.076}$ & $0.857^{+0.016}_{-0.014}$ & $0.272^{+0.070}_{-0.054}$ \\
Mrk 6           & 94.69      & 0.73                      & $1.907^{+0.312}_{-0.289}$  & 0.87                      & $0.951^{+0.336}_{-0.625}$  \\
Mrk 79           & 134.82     & $0.72^{+0.01}_{-0.01}$    & $0.424^{+0.047}_{-0.044}$  & $0.90^{+0.01}_{-0.01}$    & $0.217^{+0.019}_{-0.042}$  \\
{\bf NGC 3227}         & {\it 363.1}     & {\it 0.73}    & {\it 1.79}  & {\it 0.87}    & {\it 1.54}  \\
{\bf NGC 3516}        & {\it 676}  & {\it 0.73}                      & {\it 2.539}                      & {\it 0.87}                      & {\it 1.084}                      \\
\enddata
\vspace{0.2cm}
\tablenotetext{\dagger}{We note that the soft X-ray spectrum of NGC 4151 is extremely complex and dominated by strong emission features, with a very weak and absorbed continuum.  For this reason, warm absorber properties are not constrained from our fits to the CCD data.}\\
This table includes results of fitting {\tt zedge} models to the O\,VII and O\,VIII absorption edges at 0.73\,keV and 0.87\,keV, respectively.  Type is the optical Seyfert type for each AGN.  The $\Delta\chi^2$ measurement represents the improvement of the fit upon adding the edges to our baseline model.  We record both the measured edge energies (keV) and optical depths ($\tau$).  Where the energies were not well-constrained, we fixed these to the lab value.  For NGC 3227, NGC 3516, NGC 4051, NGC 4151, and MR 2251-178, our baseline model was not sufficient to obtain a good fit to their spectra ($\chi^2/dof > 2$).  Therefore, error bars are not computed for these sources.  Parentheses indicate upper-limits on the measured parameter.
\end{deluxetable}

\begin{deluxetable}{lllllccc}
\tabletypesize{\scriptsize}
\tablecaption{Flux, Luminosity, and Black Hole Mass\label{tbl-flux}}
\tablewidth{0pt}
\tablehead{
\colhead{Source} & \colhead{F$_{0.3-2 {\rm keV}}$} & 
\colhead{F$_{2-10 {\rm keV}}$} & \colhead{F$_{15-50 {\rm keV}}$}  &\colhead{F$_{14-195 {\rm keV}}$} & 
\colhead{L$_{\rm bol}$} &
\colhead{M/M$_{\sun}$} & \colhead{L$_{\rm bol}$/L$_{\rm Edd}$}
}
\startdata
Mrk 352 & -11.0 & -10.9 & -10.7 & -10.4 & 44.1 & 6.8 & 0.147 \\
Fairall 9 & -10.6 & -10.6 & -10.4 & -10.4 & 45.2 & 8.6 & 0.030 \\
NGC 526A & -11.6 & -10.6 & & -10.4 & 44.3 & 7.6 & 0.039 \\
Mrk 1018 & -11.0 & -11.0 & -10.7 & -10.6 & 44.9 & 8.6 & 0.014 \\
Mrk 590 & -11.4 & -11.1 & & -11.0 & 44.0 & 8.6 & 0.002 \\
NGC 931 & -11.0 & -10.5 & & -10.2 & 44.4 & 8.3 & 0.010 \\
NGC 985 & -10.9 & -10.8 & -10.6 & -10.5 & 45.0 & 8.6 & 0.018 \\
ESO 198-024 & -11.1 & -11.0 & & -10.5 & 45.1 & 8.1 & 0.074 \\
\enddata
\begin{center}
{The $\log$ of flux is given in each band using the best-fit model to the broad-band spectral fits.  Units of flux are ergs\,s$^{-1}$\,cm$^{-2}$.  The 15--50\,keV band flux is given using the Suzaku HXD PIN spectrum and is not available for sources without Suzaku spectra.  The $\log$ luminosity is calculated using the 14--195\,keV flux, the redshift of the source, and the correction to bolometric described in \S~\ref{lum} from SED fits, with luminosity in units of ergs\,s$^{-1}$.  We assume $H_0 = 71$\,km\,s$^{-1}$\,Mpc$^{-1}$.  The $\log$ of mass, in solar masses, is computed using the 2MASS K$_s$ bulge flux, taken from \citet{2009ApJ...690.1322W}, corrected using the correlation between the H$\beta$ derived masses and the 2MASS derived masses, shown in \citet{2010ApJ...710..503W}.  Finally, the ratio of bolometric luminosity to Eddington luminosity is presented, with Eddington luminosity computed as $1.3 \times 10^{38} \times (M/M_{\sun})$, in ergs\,s$^{-1}$.

Note: for NGC 4051, NGC 4593, Mrk 279, and NGC 7469, the H-$\beta$ derived masses are significantly higher than the reverberation mapping values.  Therefore, the recorded value, which we use throughout, is from the reverberation mapping method \citep{2004ApJ...613..682P}.
}
\end{center}
\end{deluxetable}

\begin{deluxetable}{lllllll}
\tabletypesize{\scriptsize}
\tablecaption{Results of Spectral Fitting with the Warm Absorber Model for Sources With O\,VII/O\,VIII Detections\label{tbl-warmabs-outflow}}
\tablewidth{0pt}
\tablehead{
\colhead{Source} & \colhead{$\Delta\chi^2_1$} & \colhead{$\log$ \nh(warm$_1$)} & \colhead{$\log \xi_1$} &\colhead{$\Delta\chi^2_2$} & \colhead{$\log$ \nh(warm$_2$)} & \colhead{$\log \xi_2$} \\
\colhead{} & \colhead{} &  \colhead{cm$^{-2}$} & \colhead{ergs\,s$^{-1}$} & \colhead{} &  \colhead{cm$^{-2}$} & \colhead{ergs\,s$^{-1}$}
}
\startdata
NGC 526A & 20.9 & $20.53^{+0.06}_{-0.06}$ & $-3.79^{+3.07}_{-3.84}$ & 12.2 & $20.30^{+0.64}_{-0.56}$ & $1.86^{+0.13}_{-0.01}$\\
NGC 931 & 364.2 & $20.23^{+0.02}_{-0.03}$ & $0.25^{+0.10}_{-0.10}$ & 376.8 & $20.59^{+0.46}_{-0.30}$ & $2.42^{+0.60}_{-0.30}$\\
NGC 985 & 89.3 & (22.66) & $1.93^{+0.10}_{-0.06}$ & 33.0 & $21.21^{+0.09}_{-0.17}$ & $2.54^{+0.08}_{-0.13}$ \\
Mrk 6 & 116.5 & $20.93^{+0.23}_{-0.21}$ & $-0.29^{+0.07}_{-0.48}$ &  94.0 &   $21.70^{+0.02}_{-0.02}$ & $2.04^{+0.04}_{-0.12}$\\
Mrk 79 & 573.6 & $20.91^{+0.22}_{-0.14}$& $2.75^{+0.04}_{-0.13}$& 37.73 & $20.91^{+0.05}_{-0.04}$& $1.99^{+0.03}_{-0.03}$\\
NGC 3227 & 706.6 & $21.19^{+0.01}_{-0.01}$ & $1.44^{+0.01}_{-0.08}$ & 148.3 & $20.89^{+0.01}_{-0.01}$ & $0.42^{+0.05}_{-0.04}$ \\
NGC 3516 & 8591.2& (21.82) & (2.04) & 17732.8 & (20.79) & (1.88) \\
NGC 3783 & 22373.5 & (21.48) & (1.99) & 5410.0 & (21.02) & (2.04)\\
\enddata
\vspace{0.1cm}
\tablenotetext{\dagger}{We note that the soft X-ray spectrum of NGC 4151 is extremely complex and dominated by strong emission features, with a very weak and absorbed continuum.  For this reason, warm absorber properties are not constrained from our fits to the CCD data.}\\
{\footnotesize
The change in $\chi^2$, column density, and ionization parameter is included for each of two components of warm ionized gas, using the analytic {\tt warmabs} model in place of \ion{O}{7} and \ion{O}{8} edges.  We indicate values that were not well-constrained with parentheses.
}
\end{deluxetable}

\begin{deluxetable}{lrcc}
\tablecaption{Results of Spectral Fitting with the Warm Absorber Model for Sources Without O\,VII/O\,VIII Detections \label{tbl-warmabs-nonoutflow}}
\tablewidth{0pt}
\tablehead{
\colhead{Source} & \colhead{$\Delta\chi^2$} & \colhead{$\log$ \nh(warm)} & \colhead{$\log \xi$}  \\
\colhead{} & \colhead{} & \colhead{cm$^{-2}$} & \colhead{ergs\,s$^{-1}$} 
}
\startdata
Mrk 352 & 9.9 & $20.69^{+0.12}_{-0.06}$ & $1.99^{+0.08}_{-0.01}$ \\
Fairall 9 & 11.4 &  $19.02^{+0.21}_{-0.29}$ & $1.74^{+0.07}_{-0.13}$\\
Mrk 1018 & $- 57.2$ & $20.77^{+0.12}_{-0.20}$ & $2.12^{+0.05}_{-0.09}$ \\
Mrk 590 & $-$9.3 & $20.76^{+1.86}_{-0.24}$ & $3.22^{+0.00}_{-0.82}$\\
ESO 198-024 & 3.9 & $18.86^{+0.36}_{-0.47}$ & $-3.76^{+3.70}_{-3.99}$ \\
ESO 548-G081 & 6.0 & $20.42^{+0.82}_{-0.98}$ & $2.92^{+0.10}_{-0.24}$\\
1H 0419-577 & 63.2 & $19.47^{+0.06}_{-0.15}$ & $-3.60^{+1.67}_{-0}$\\
3C 120 & $-$27.7 & ($20.16$) & $2.84^{+0.03}_{-0.09}$\\
\enddata
\vspace{0.1cm}
{\footnotesize
The change in $\chi^2$, column density, and ionization parameter is included for warm ionized gas, using the analytic {\tt warmabs} model in place of \ion{O}{7} and \ion{O}{8} edges.  We indicate values that were not well-constrained with parentheses.
}
\end{deluxetable}


\begin{figure}
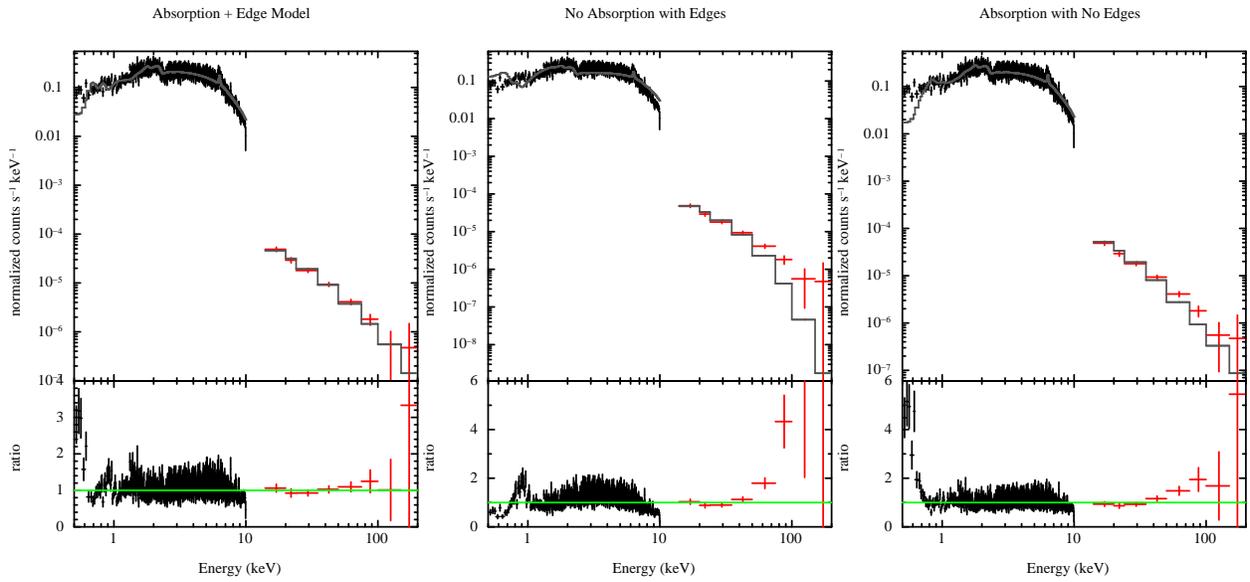

\includegraphics[width=5.5cm]{f2a.ps}
\hspace{-0.2cm}
\includegraphics[width=5.5cm]{f2b.ps}
\hspace{-0.2cm}
\includegraphics[width=5.5cm]{f2c.ps}\\
\caption{We plot example spectral fits demonstrating the presence of both neutral absorbing gas and ionized gas, evidenced through the \ion{O}{7}/\ion{O}{8} edges.  We show fits to the XMM-Newton + Swift BAT spectra of Mrk 6, with both absorption and edges (left), with no absorption but with edges (middle), and with absorption and no edges (right).  The best-fit model includes both absorption and edges, as demonstrated in the ratio of the data to model.
\label{fig-edges}}
\end{figure}

\begin{figure}
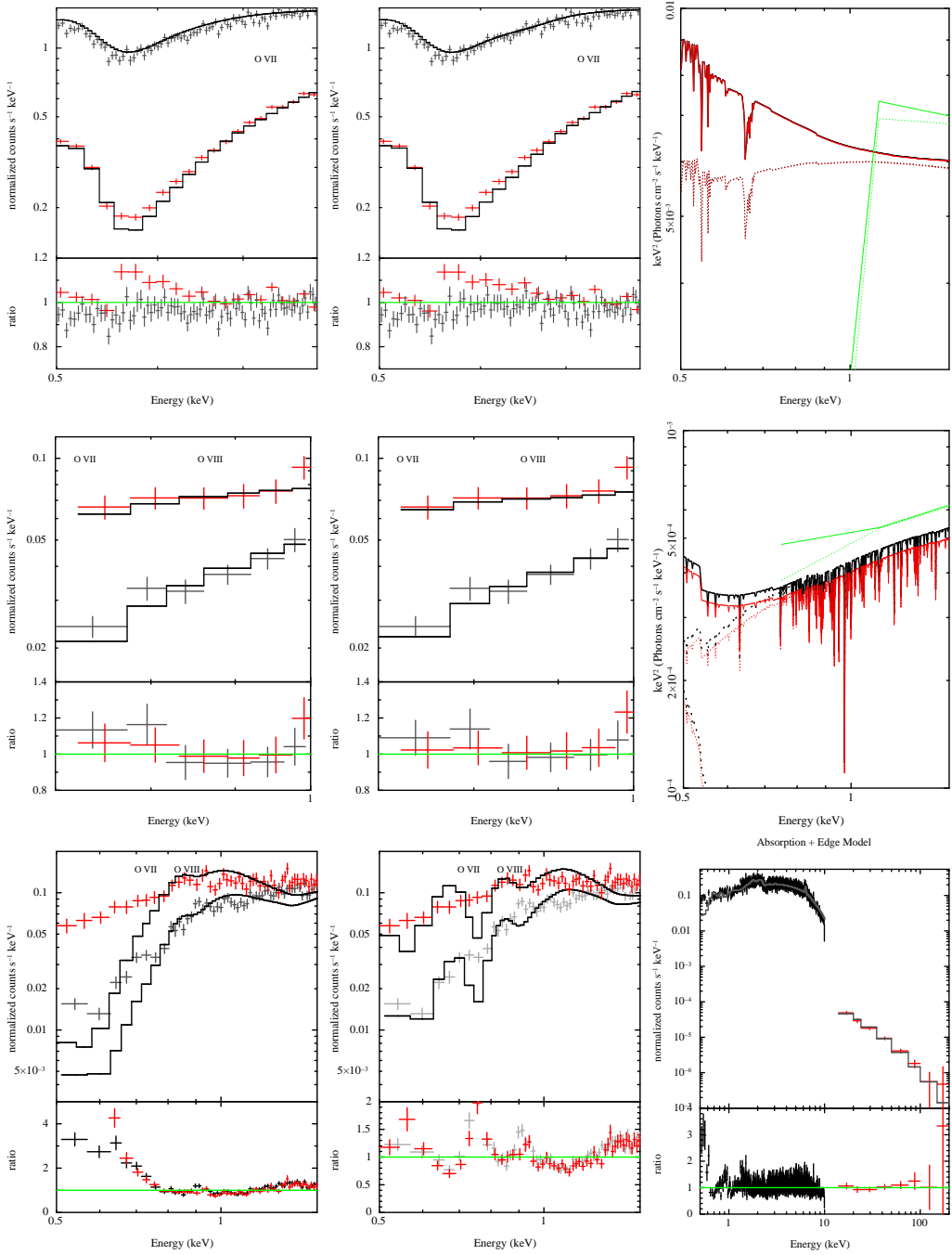

\begin{center}
\includegraphics[width=4.5cm]{f3a.ps}
\includegraphics[width=4.5cm]{f3b.ps}
\includegraphics[width=4.3cm]{f3c.ps}\\

\includegraphics[width=4.5cm]{f3d.ps}
\includegraphics[width=4.5cm]{f3e.ps}
\includegraphics[width=4.3cm]{f3f.ps}\\

\includegraphics[width=4.5cm]{f3g.ps}
\includegraphics[width=4.5cm]{f3h.ps}
\includegraphics[width=4.3cm]{f2a.ps}\\
\end{center}
\caption{Shown are examples of the base model fit (left), without any fits to the warm absorption, analytic one (two for NGC 3227) component warm absorber fits (middle), and the warm absorber model (right) for three representative sources (top 1H 0419-577, middle 2MASX J09043699+5536025, and bottom NGC 3227).  Fits are shown in the 0.5--1.5\,keV band, with data re-binned to signal-to-noise of 10--20 for illustration.  We show the 0.5-0.8\,keV region of the spectra of 1H 0419-577, since this source exhibits lowly ionized gas.  We note that differences in the fits are subtle, due to re-binning of the data for illustration.
\label{fig-fitsnooutflow}}
\end{figure}

\begin{figure}
\plotone{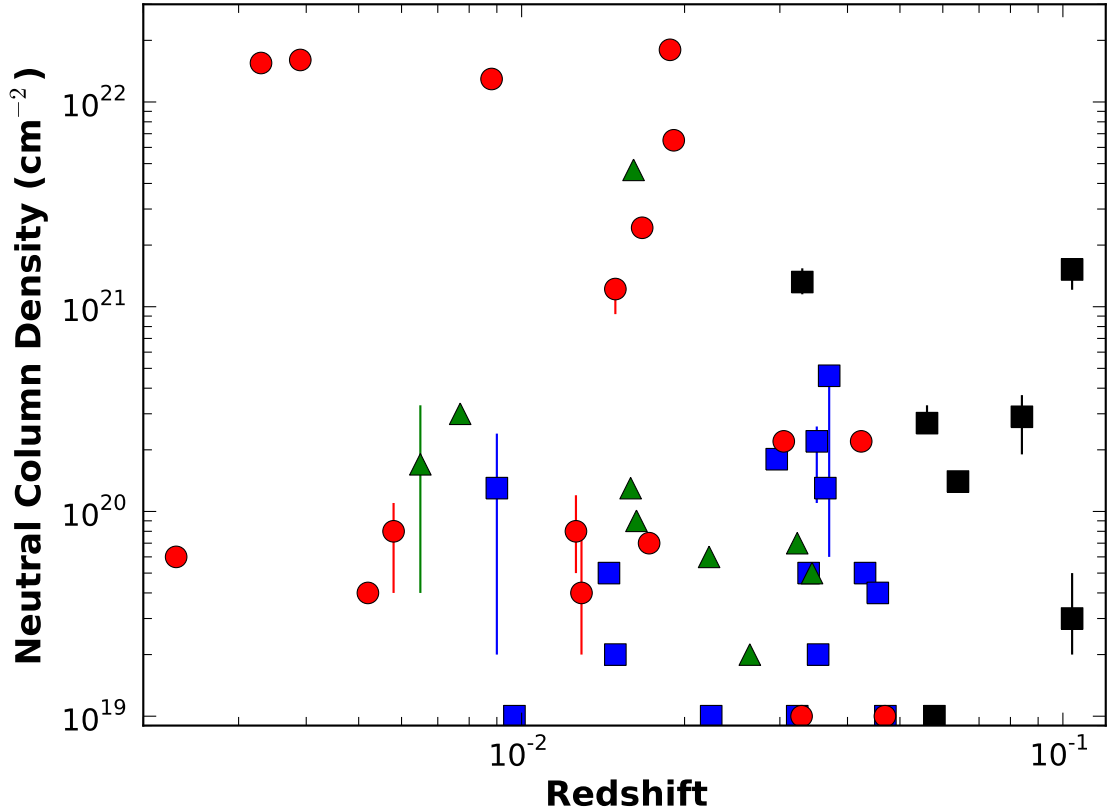}
\caption{We show the best-fit neutral column densities for our sample versus redshift.  The Sy 1 sources, blue squares, are less obscured in the X-rays, on average, than the Sy 1.2s (green triangles) and Sy 1.5s (red circles).  Broad line radio galaxies (black squares) include the highest redshift sources in the sample.  These same symbols are used in the subsequent figures.  There is significant scatter in the properties of the Sy 1.5s, which could be due to inaccuracies in classification (we use the classifications from NED) or the potentially heterogeneous nature of sources classified as Sy 1.5s. 
\label{fig-nhz}}
\end{figure}

\begin{figure}
\plotone{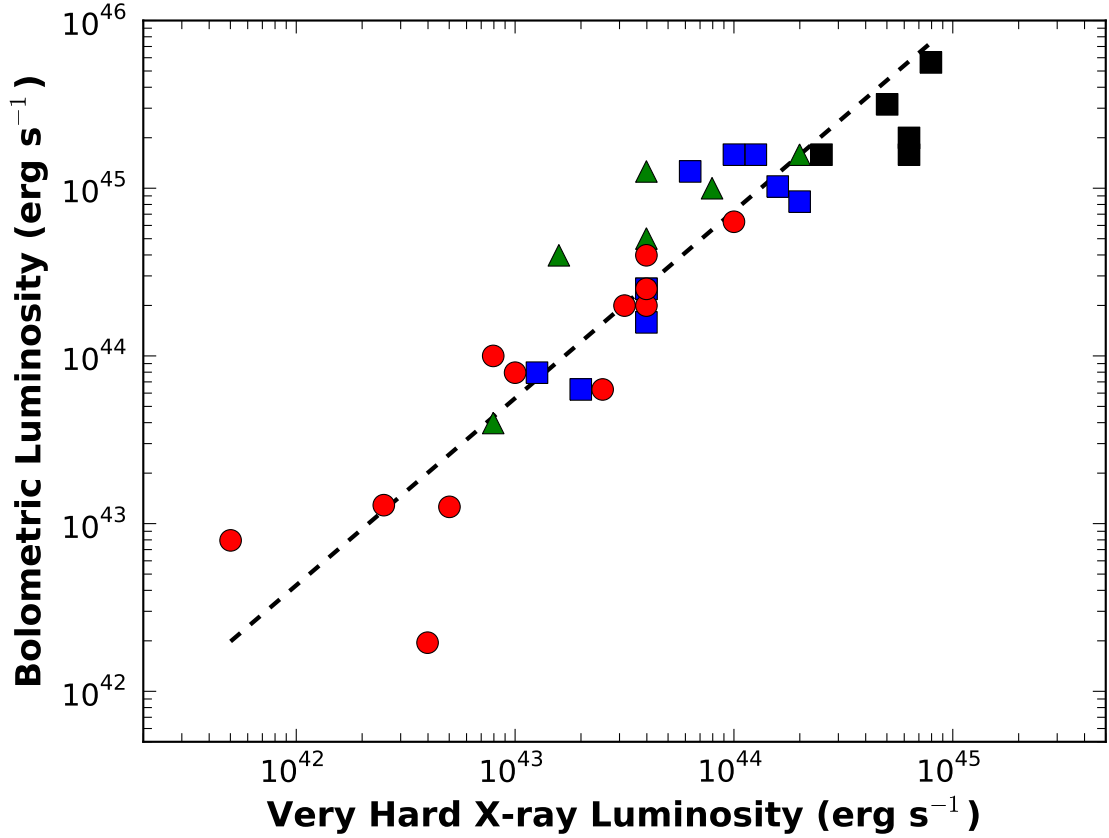}
\caption{We compare the Swift BAT band luminosity derived from our spectral fits with the bolometric luminosity obtained through fits to the SEDs of 33 of our sources \citep{2007MNRAS.381.1235V,2009MNRAS.392.1124V}.  As expected, we find that the Swift BAT luminosity is highly correlated with the bolometric luminosity (with a correlation coefficient of $R^2 = 0.82$).  This allows us to derive a simple correction from the BAT luminosity to bolometric: $\log L_{\rm bol} = (1.1157 \pm 0.117) \log L_{\rm 14-195 keV} + (-4.2280 \pm 5.1376)$ (see text). 
\label{fig-comparebol}}
\end{figure}

\begin{figure}
\begin{center}
\includegraphics[height=8.5cm]{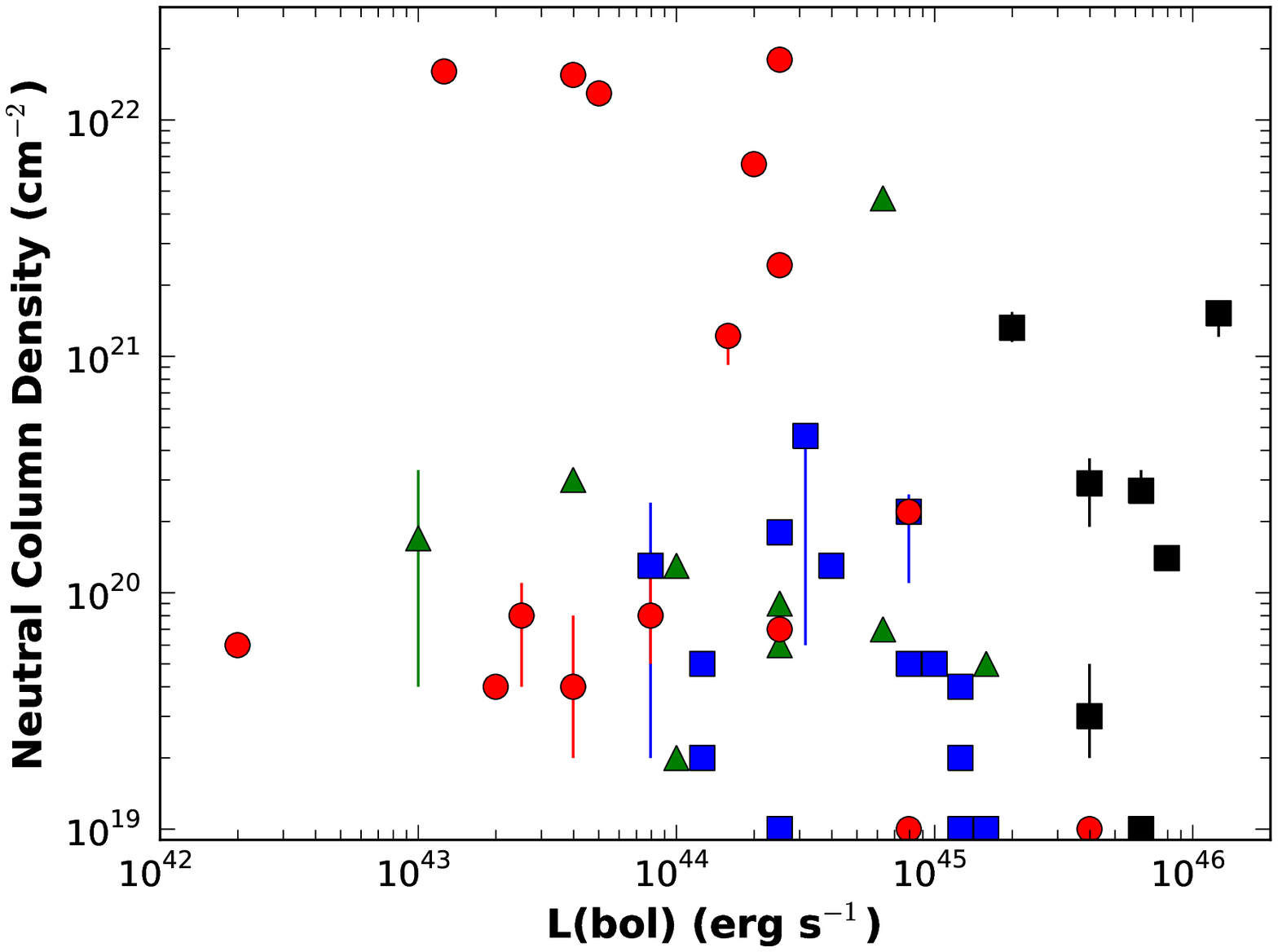}
\includegraphics[height=8.5cm]{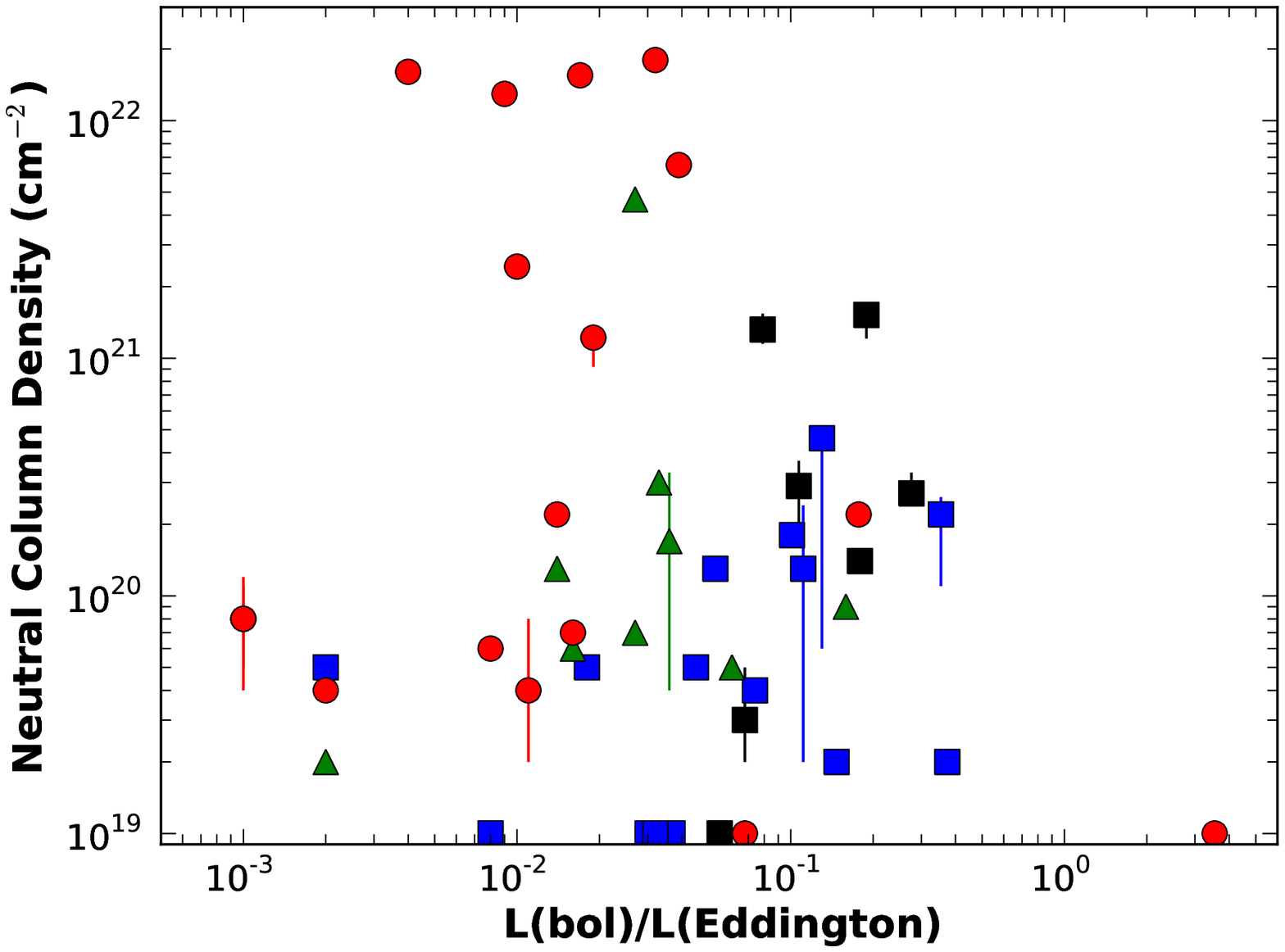}
\end{center}
\caption{We show the relationship between the measured neutral column density and the bolometric luminosity (top) and Eddington ratio (bottom).  No correlation is seen between the column density and either of these quantities.  As in Figure~\ref{fig-nhz}, we find that the properties of the Sy 1.5s (column densities, luminosities, and accretion rates) are not homogeneous.  Either there are errors in the NED classifications or the class of Sy 1.5s includes sources with very different intrinsic properties.
\label{fig-lumlledd}}
\end{figure}

\begin{figure}
\begin{center}
\includegraphics[height=6.25cm]{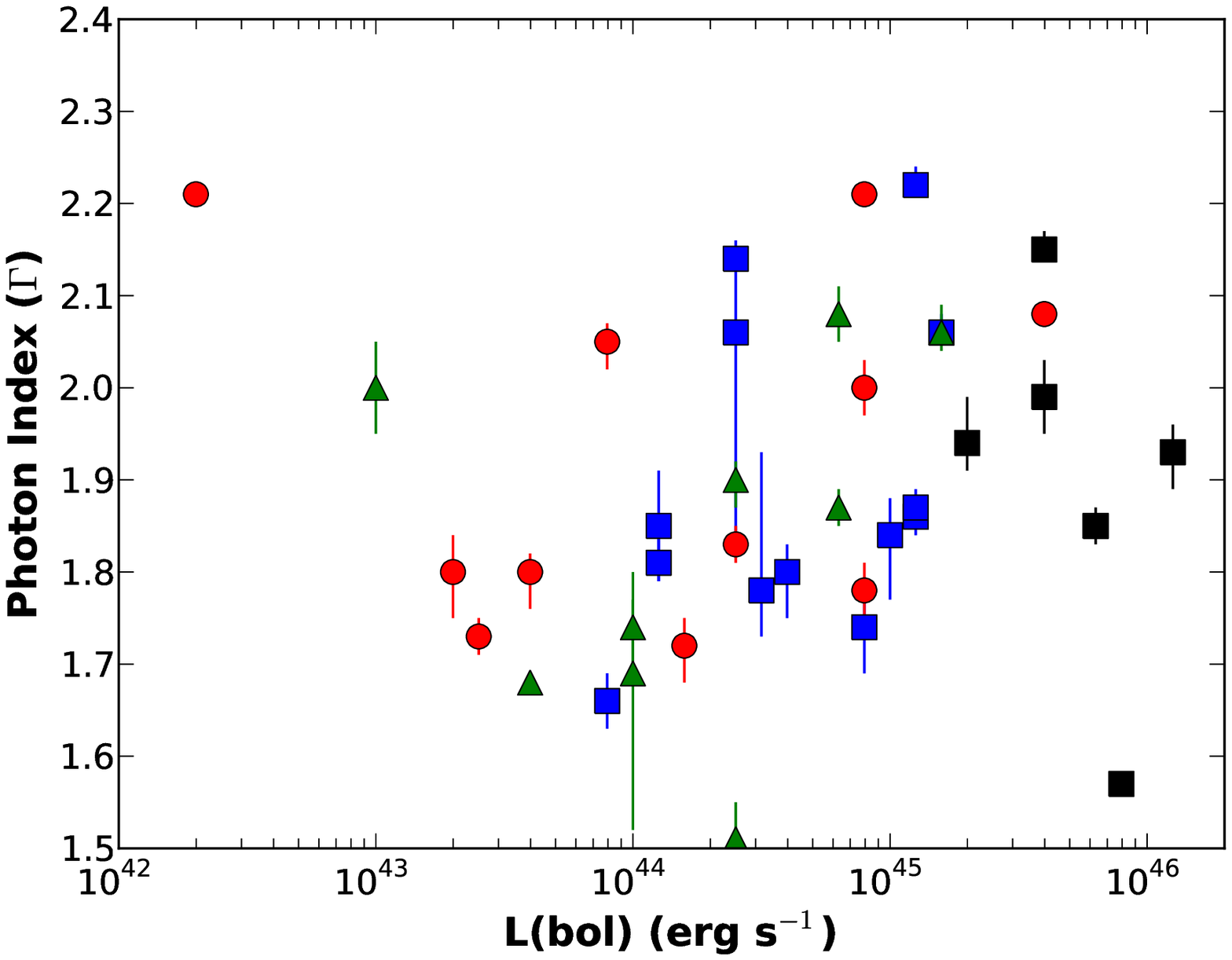}
\hspace{-1cm}
\includegraphics[height=6.25cm]{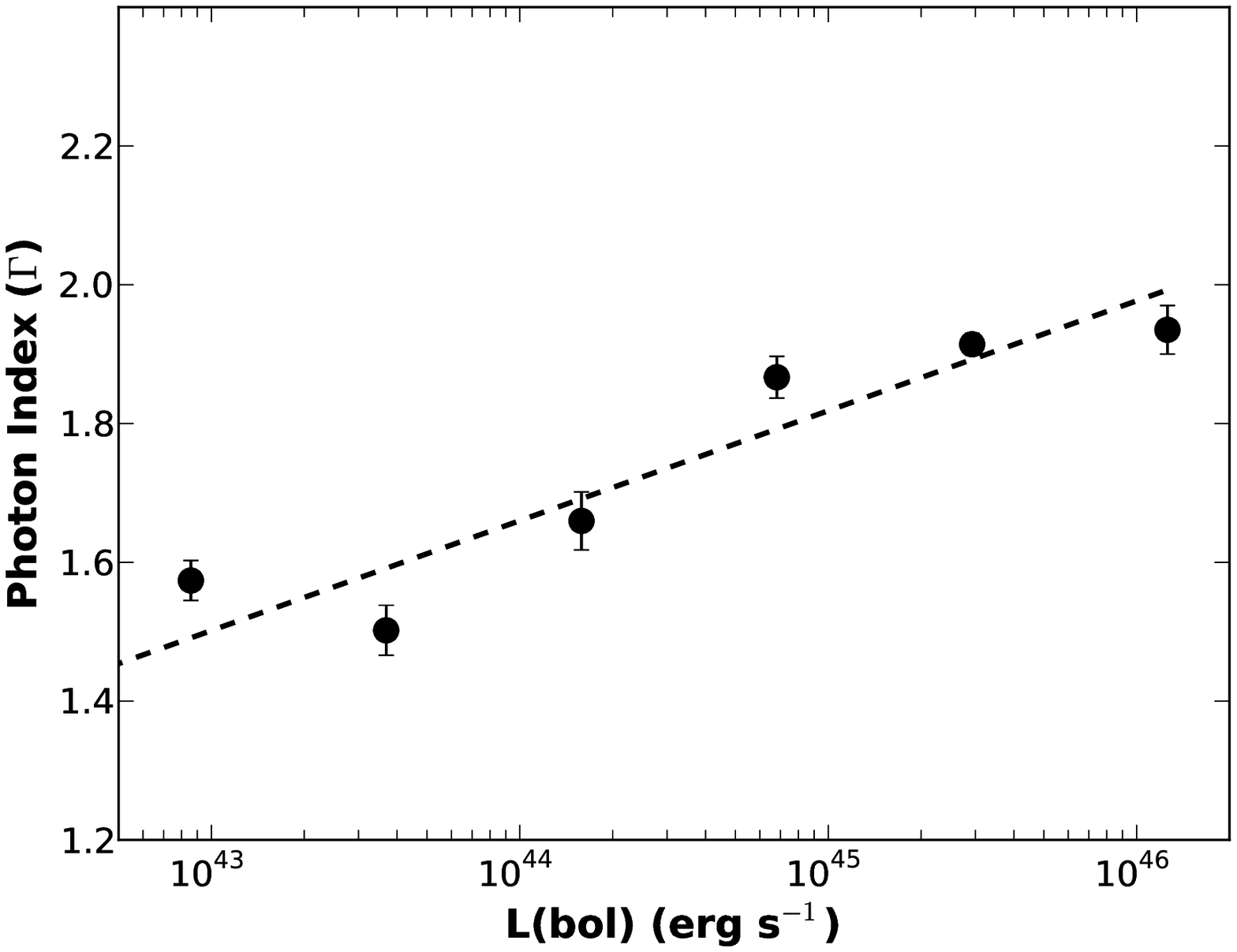}\\

\includegraphics[height=6.25cm]{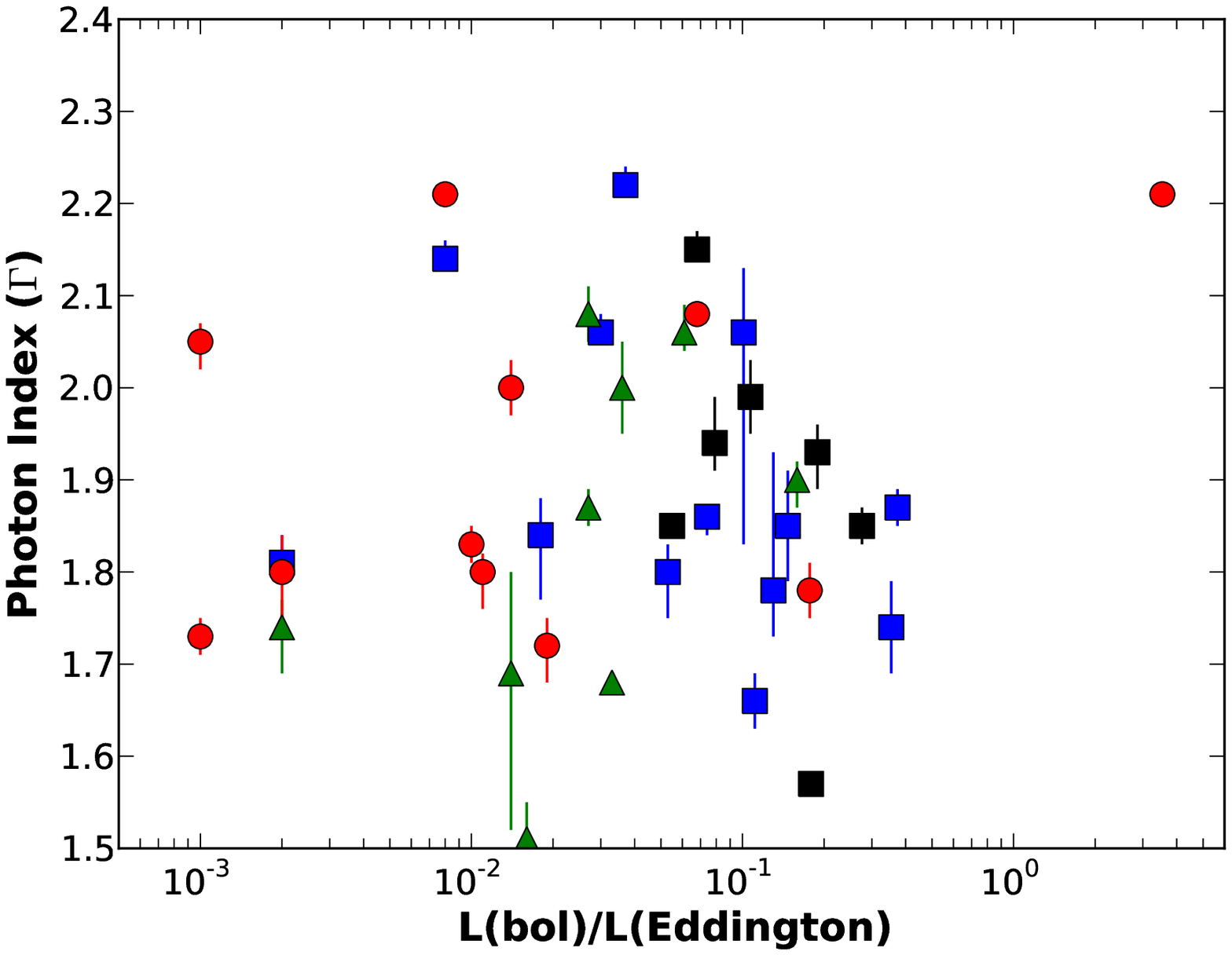}
\hspace{-1cm}
\includegraphics[height=6.25cm]{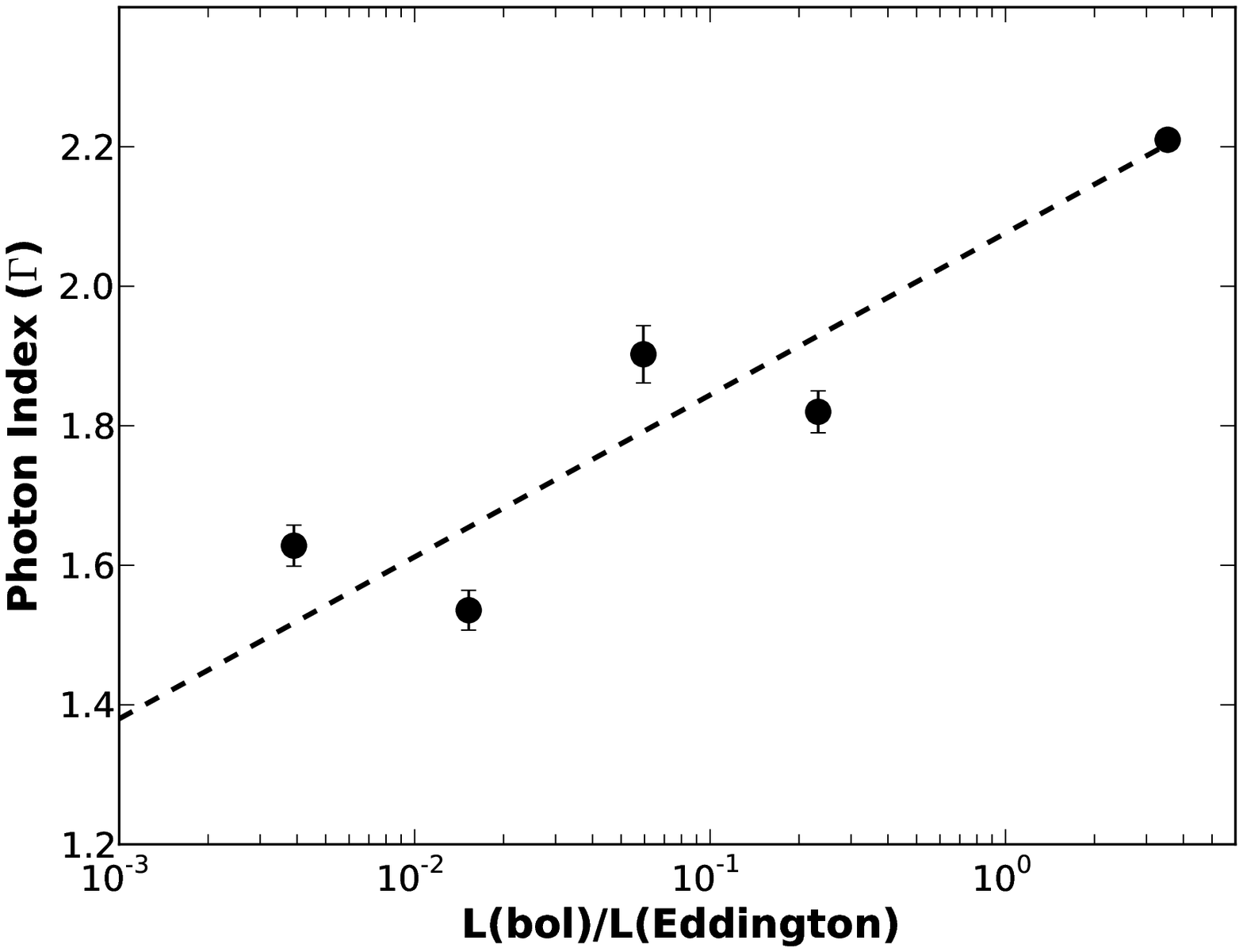}\\
\end{center}
\caption{We show the relationship between power-law index, $\Gamma$, and the bolometric luminosity (top) and Eddington ratio (bottom).  The figures at left exclude sources with $\Gamma < 1.5$, since low measured values tend to indicate difficulties modeling spectra with complex absorption.  In the plots at right, we show the binned photon index, binned by luminosity (top) and Eddington rate (bottom).  We find strong correlations between both $\Gamma$-L$_{\rm bol}$ and $\Gamma$-$L_{\rm bol}/L_{\rm Edd}$, with $R^2 = 0.82$ and $0.81$, respectively.  These values include the complex absorber sources, with measured $\Gamma < 1.5$.  The error-bars in the right panels are the logarithm of the difference between the maximum and average value, with typical values of $\Gamma = 1.07$.
\label{fig-gammalum}}
\end{figure}

\begin{figure}
\begin{center}
\includegraphics[height=8.5cm]{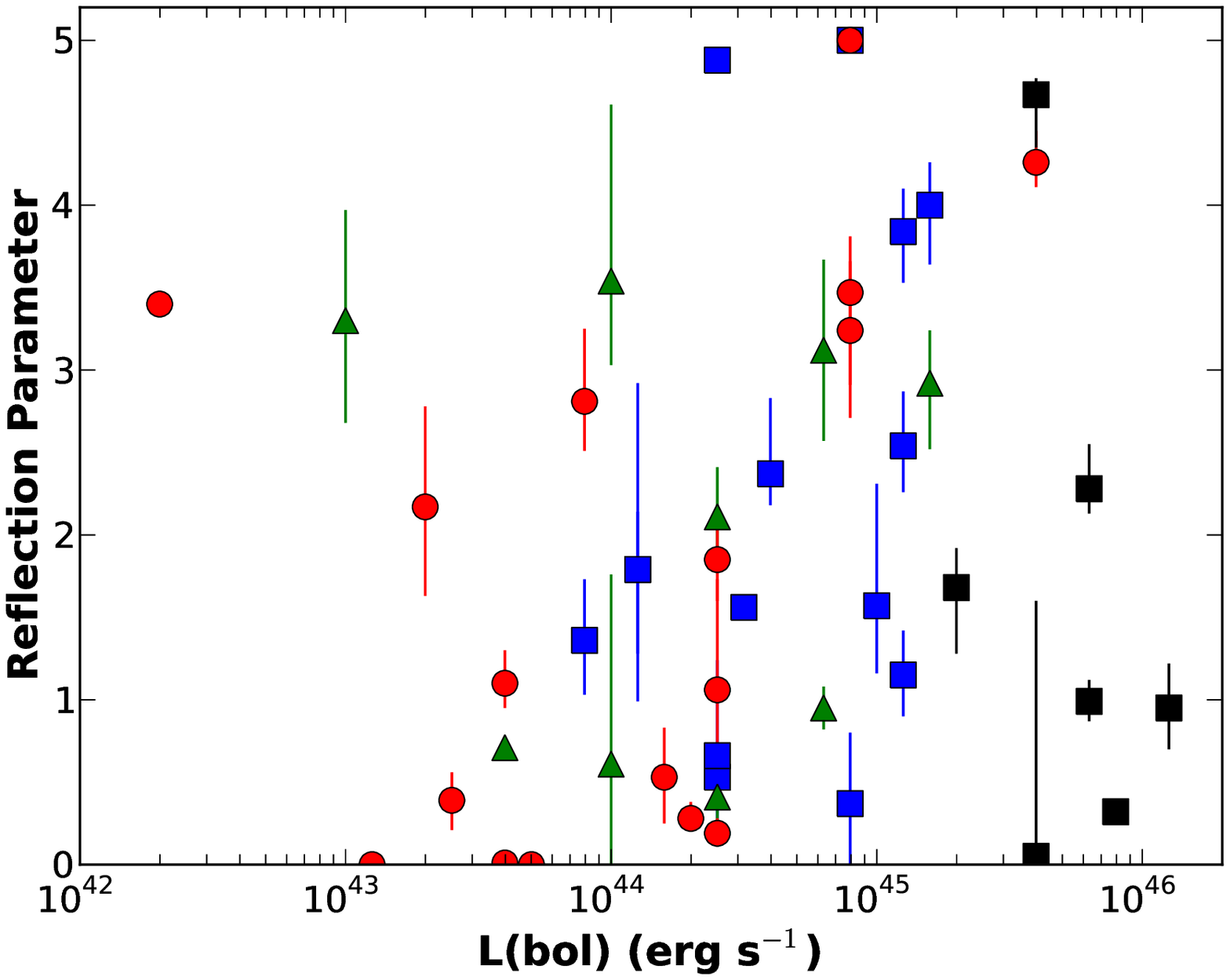}
\includegraphics[height=8.5cm]{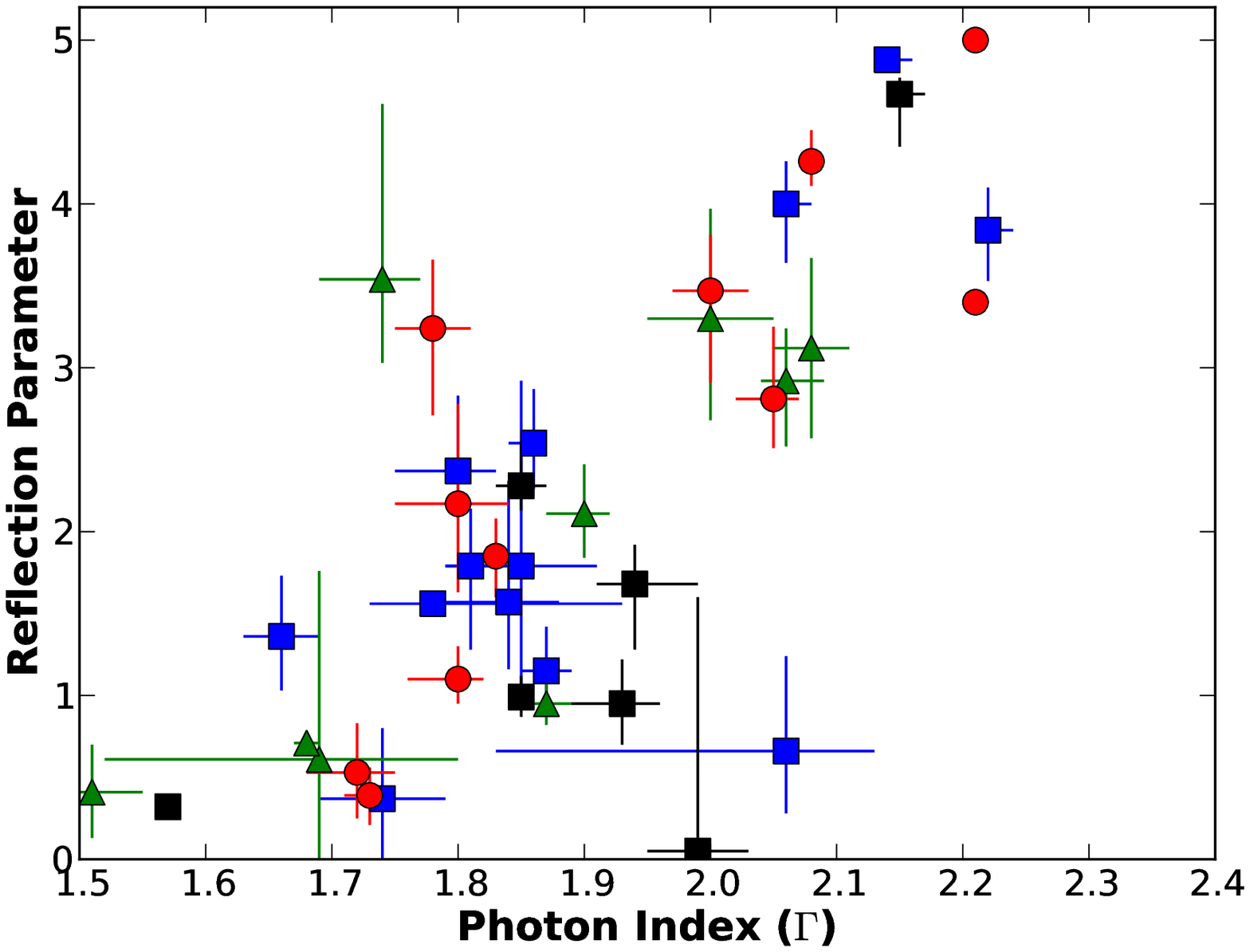}
\end{center}
\caption{The relationship between reflection and the bolometric luminosity (top) and best-fit photon index (bottom) are shown.  The reflection parameter, $R = \Omega/2\pi$, is an estimate of the reflected component to the broad-band spectral fits.  Negative values indicate fully reflected material, using the {\tt pexrav} model. While no correlation exists between reflection parameter and luminosity, there is a correlation with $\Gamma$.  Similar correlations were seen in previous samples (e.g., \citet{2007ApJ...664..101M}).  See the text for further discussion of the correlation between R and $\Gamma$.
\label{fig-reflection}}
\end{figure}

\begin{figure}
\begin{center}
\includegraphics[height=8.5cm]{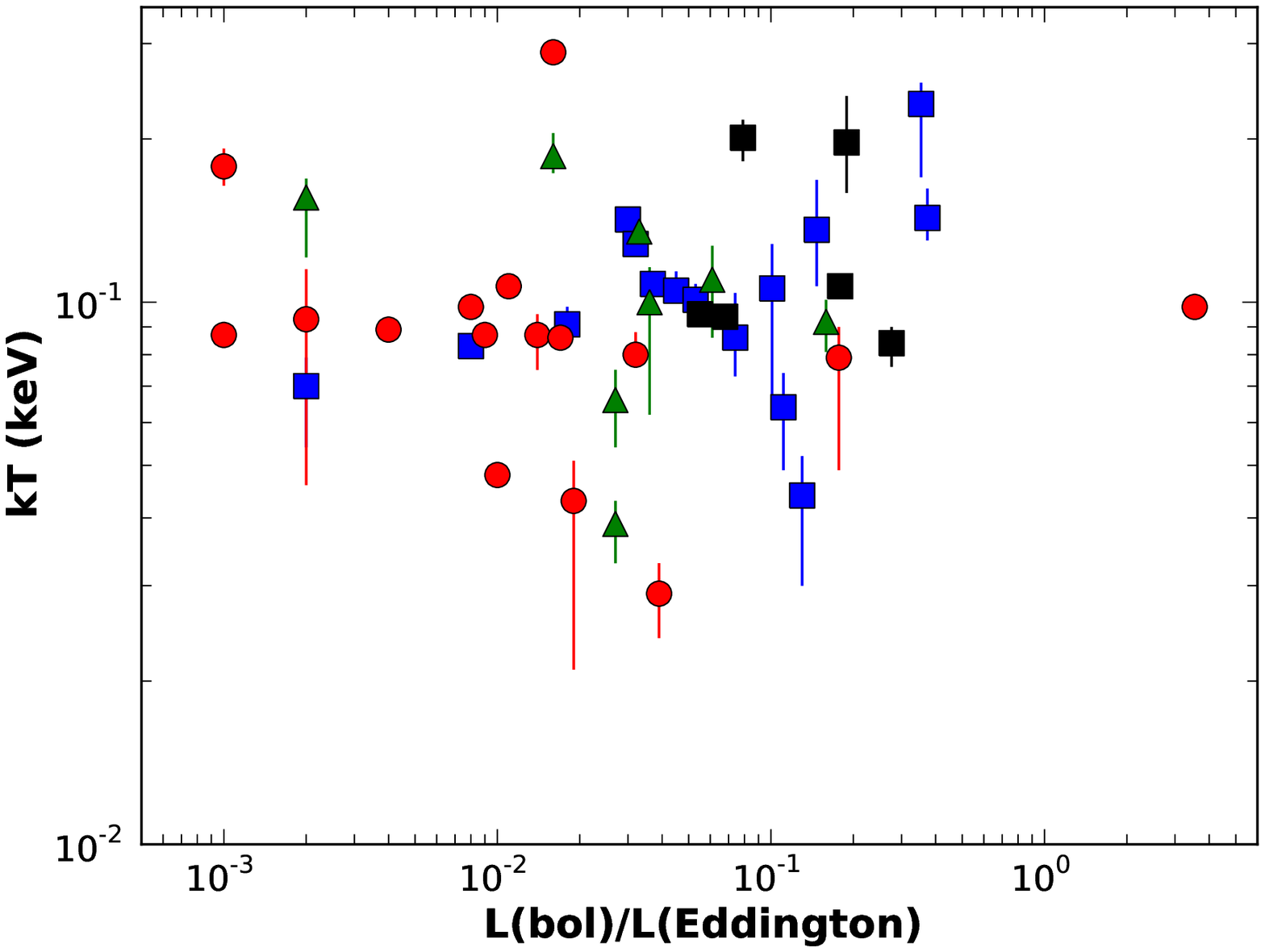}
\includegraphics[height=8.5cm]{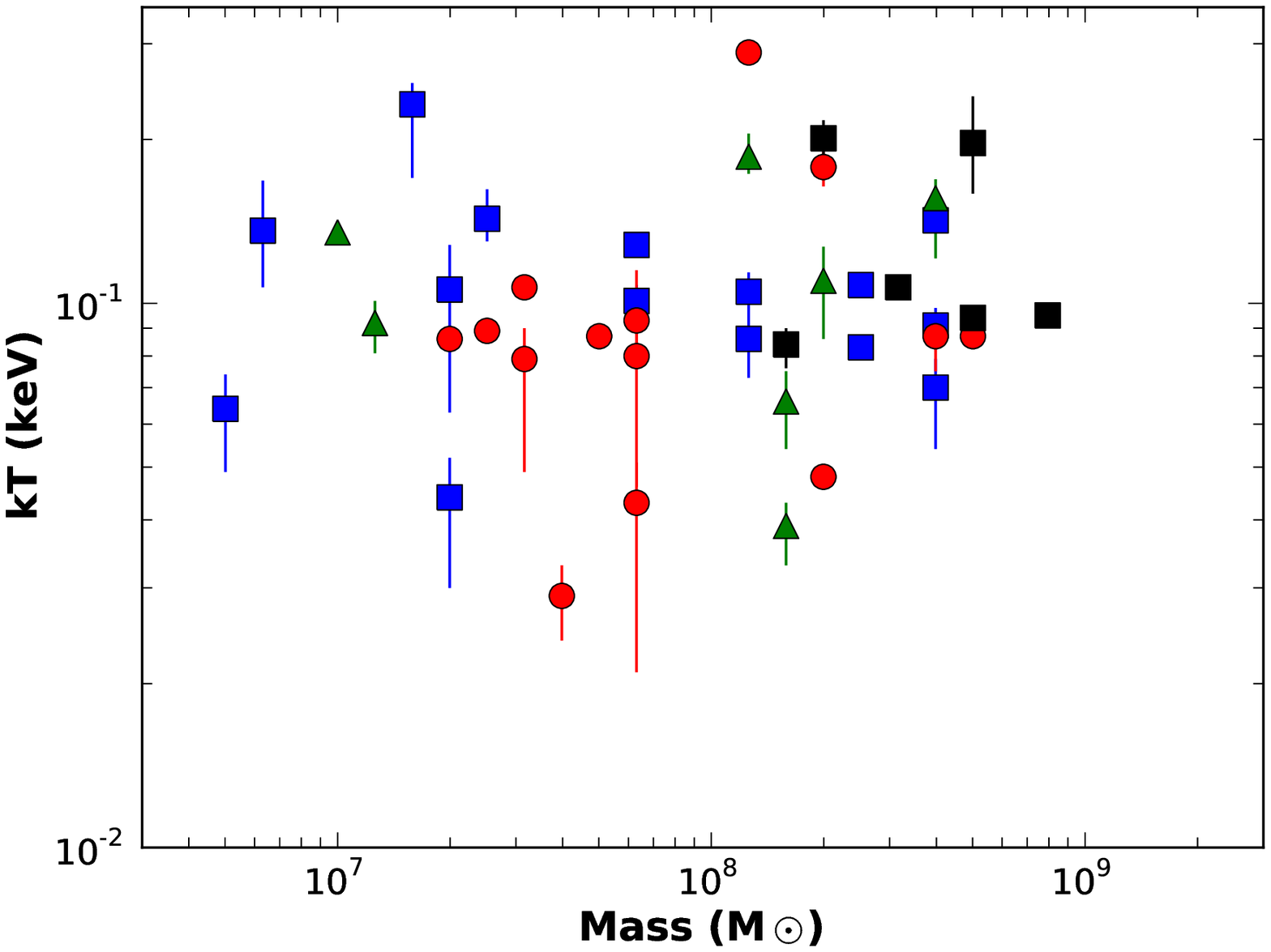}
\end{center}
\caption{As a test of a thermal origin for the soft excess, we plot the best-fit blackbody temperature ($kT$) versus the accretion rate (top) and the black hole mass (bottom).  If the soft excess were thermal, we expect a correlation between the black body temperature, such that $kT \propto M^{-1/4} L/L_{Edd}^{1/4}$.  No correlation is seen with either of these parameters.
\label{fig-bbodytemp}}
\end{figure}

\begin{figure}
\begin{center}
\includegraphics{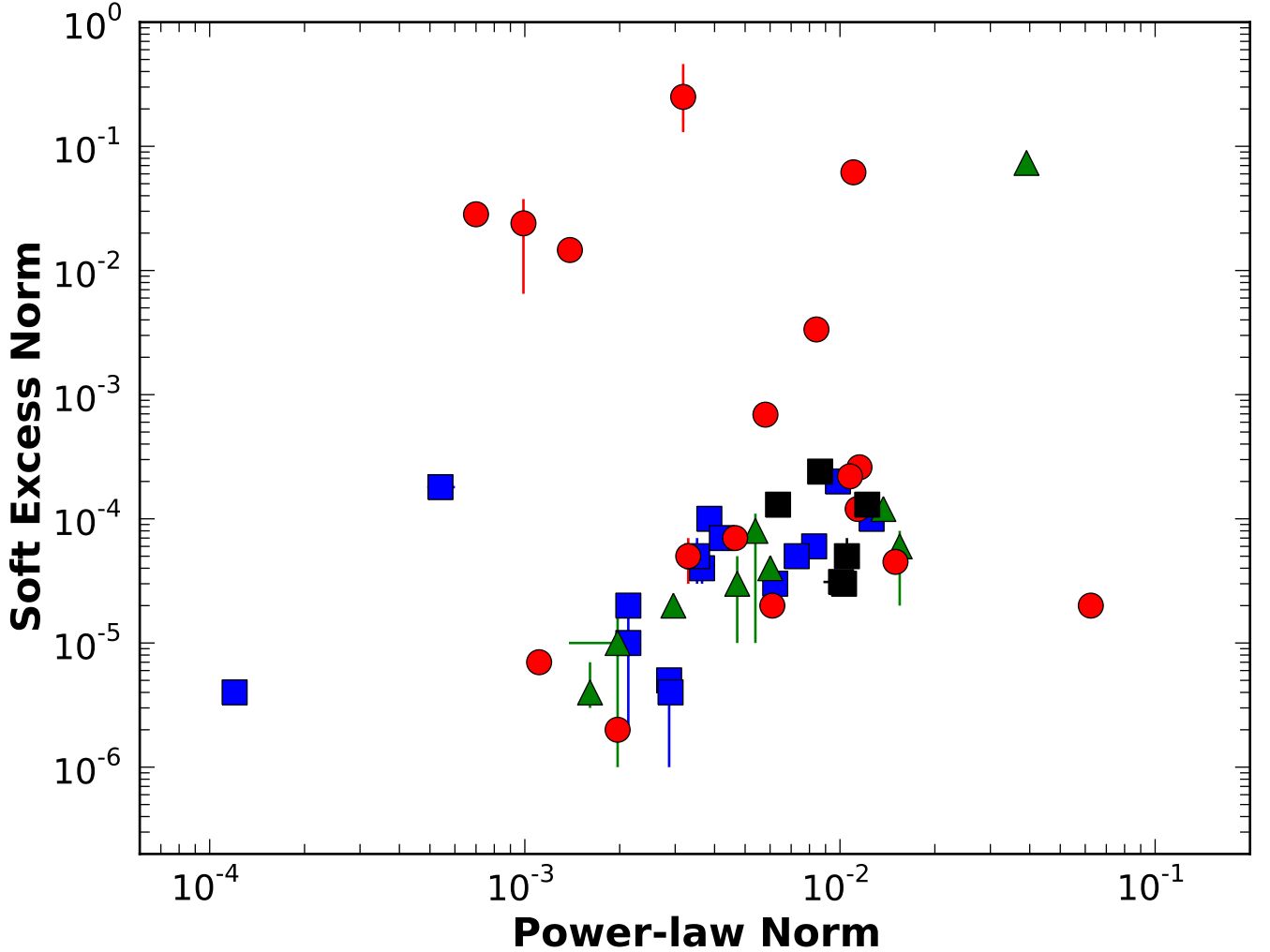}
\end{center}
\caption{The relationship between the normalization on the blackbody component and the normalization on the power-law component (see Table~\ref{tbl-fits1}) are shown.  The majority of our sources show a very tight correlation between both parameters.  This direct correlation supports our result from \citet{2009ApJ...690.1322W}, showing that the soft excess luminosity and the power-law luminosity are directly proportional.  Therefore, the soft excess is in some way created/affected by the AGN emission.
\label{fig-ktnorm}}
\end{figure}

\begin{figure}
\begin{center}
\includegraphics[height=6.25cm]{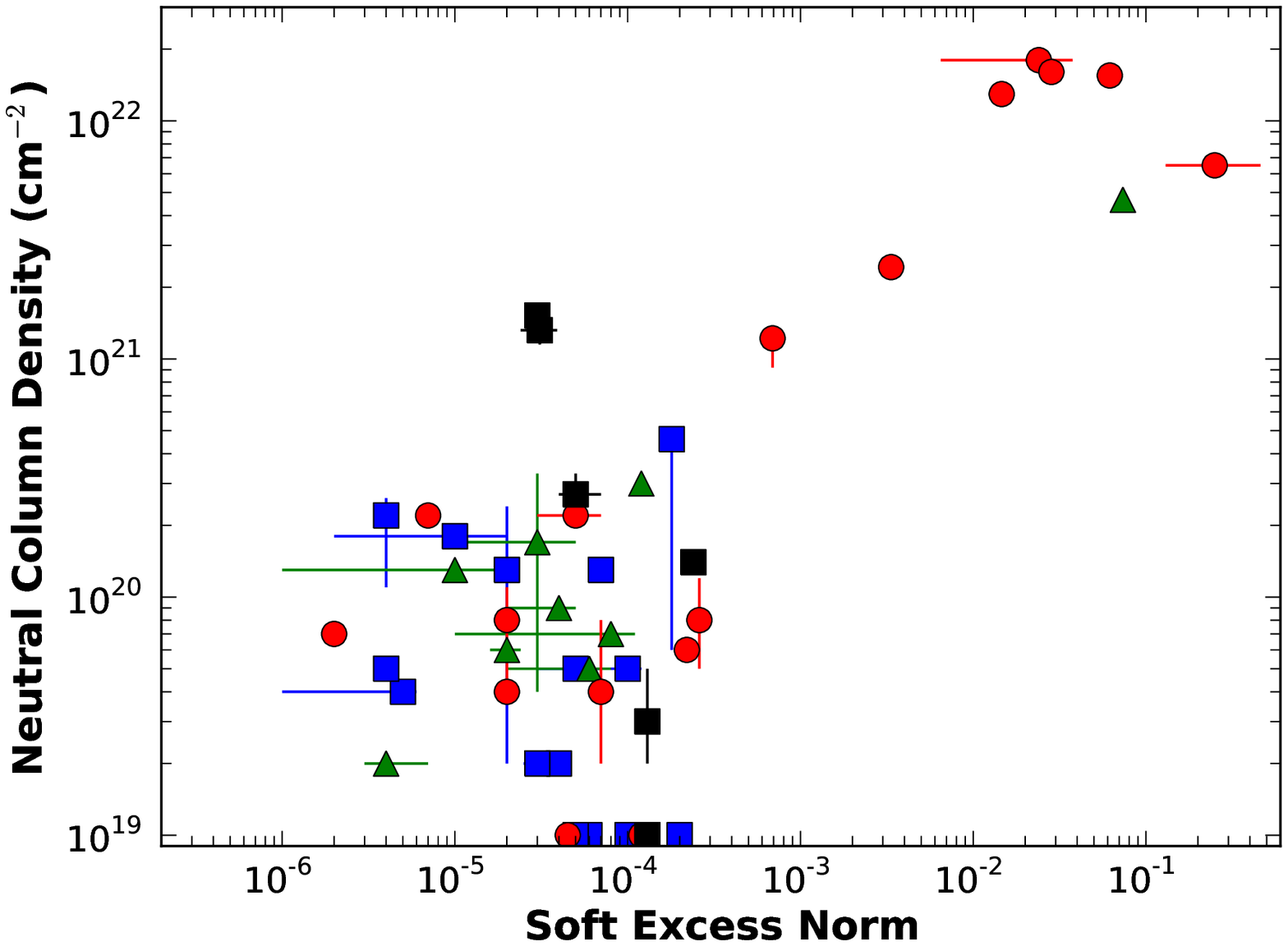}
\hspace{-1cm}
\includegraphics[height=6.25cm]{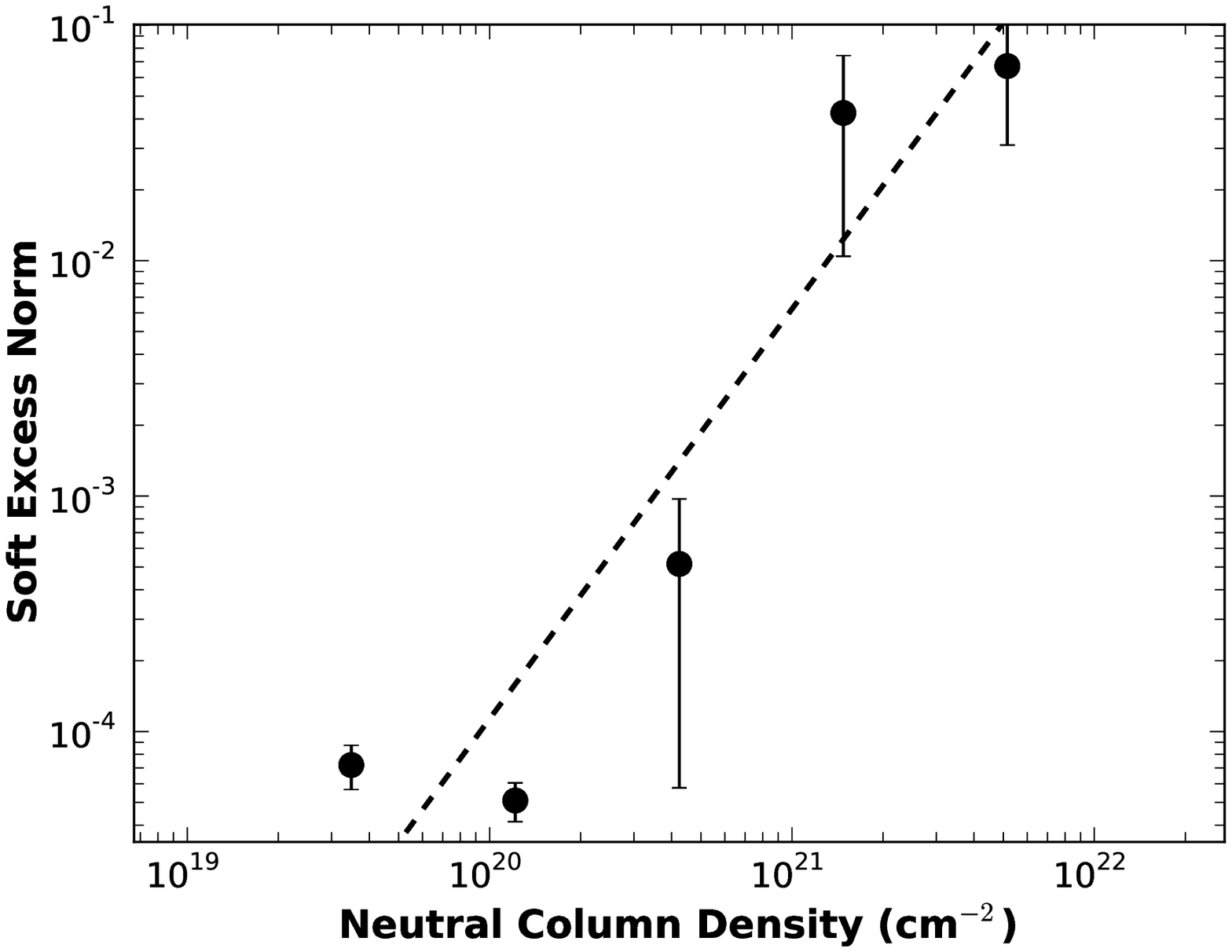} \\
\includegraphics[height=6.25cm]{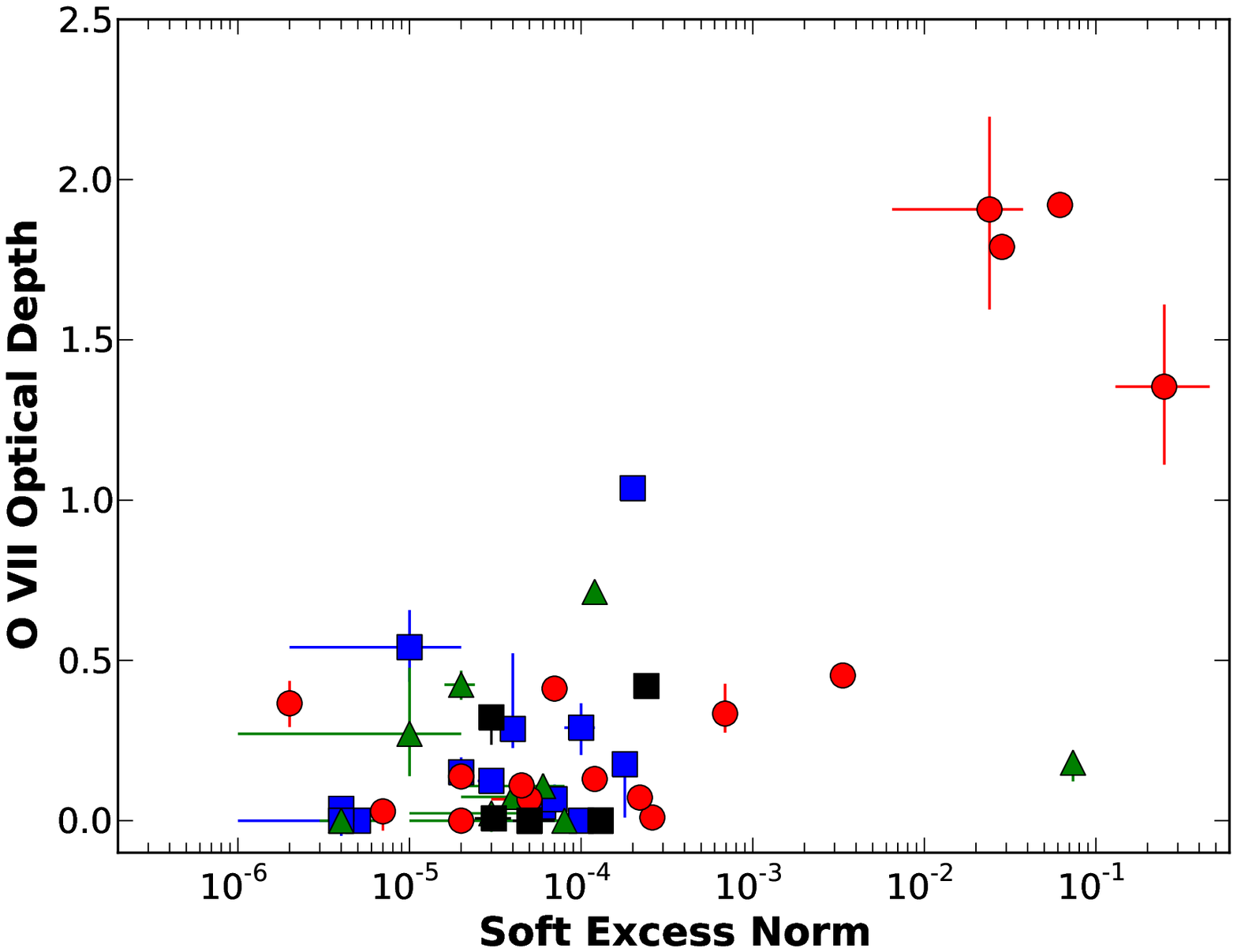}
\hspace{-1cm}
\includegraphics[height=6.25cm]{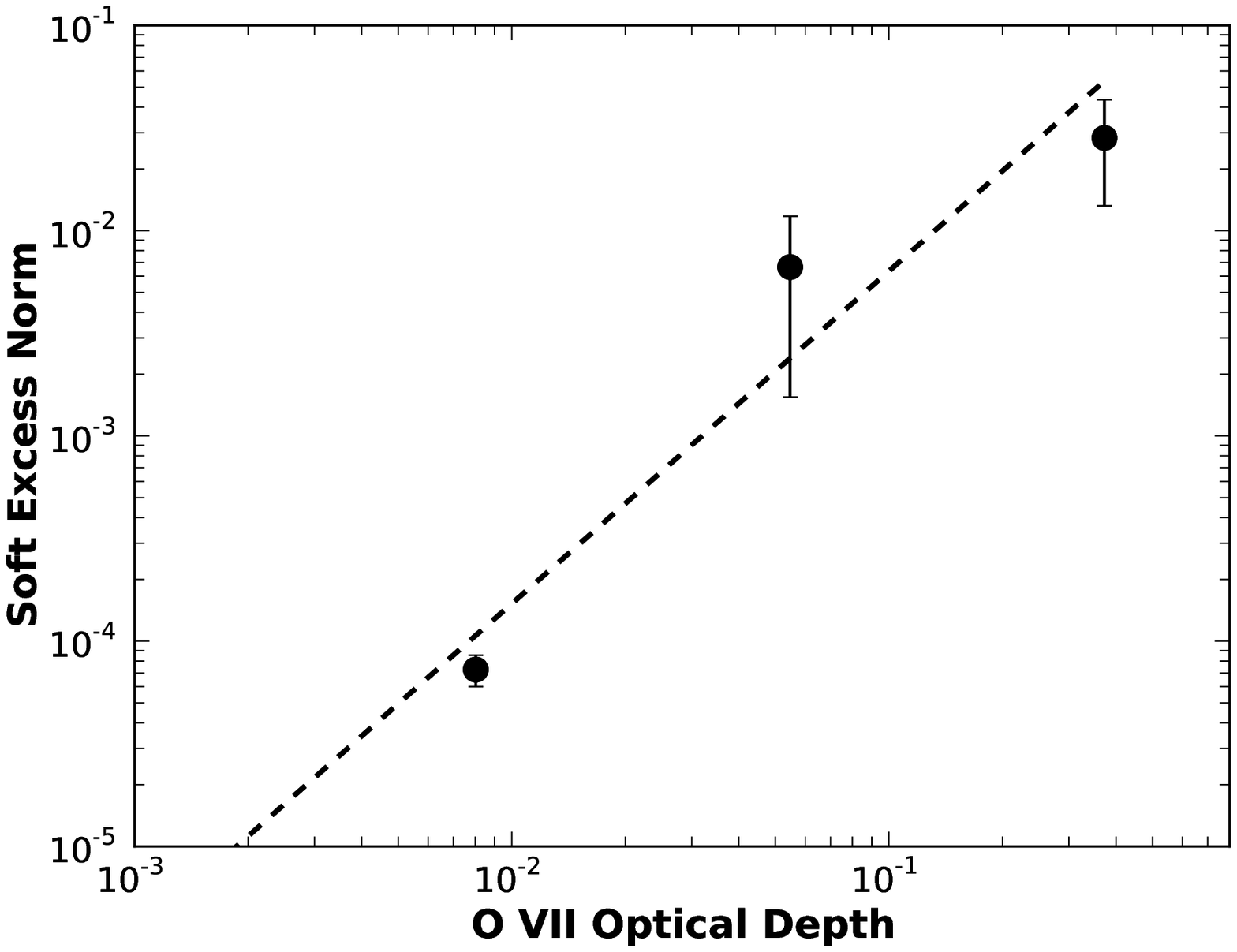}
\end{center}
\caption{The relationship between the soft excess and both the neutral column density (top) and strength of the warm absorber (bottom; through the measured optical depth in the \ion{O}{7} edge) are shown.  The right plots show the relationship, binned by column density (top) and warm absorber strength (bottom), with error-bars indicating the difference between the maximum/minimum value and the average value.  Strong correlations are found between the soft excess strength and both the neutral column density ($R^2 = 0.88$) and warm absorber strength ($R^2 = 0.92$).
\label{fig-nhoviiktnorm}}
\end{figure}

\begin{figure}
\begin{center}
\includegraphics[height=6.25cm]{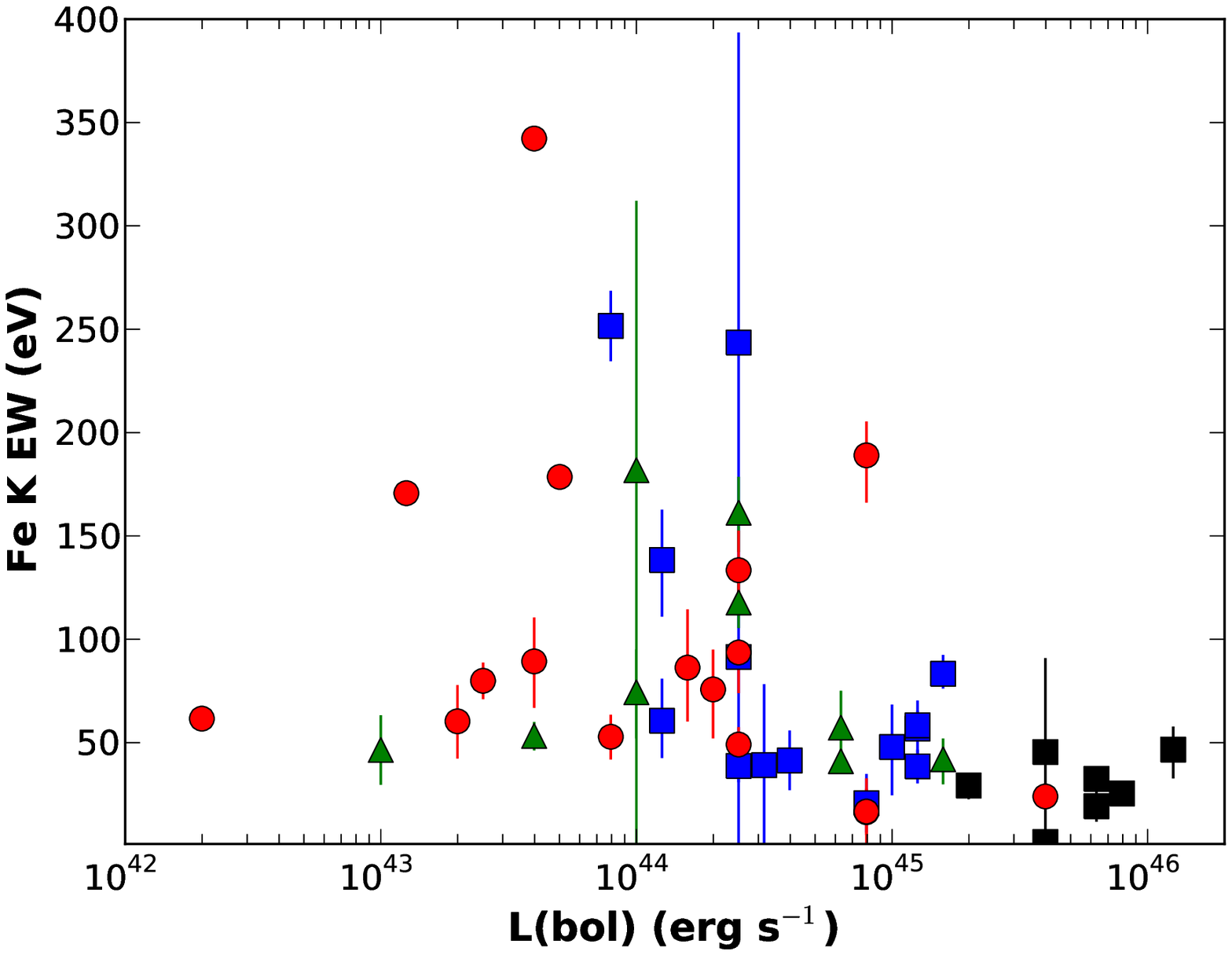}
\hspace{-1cm}
\includegraphics[height=6.25cm]{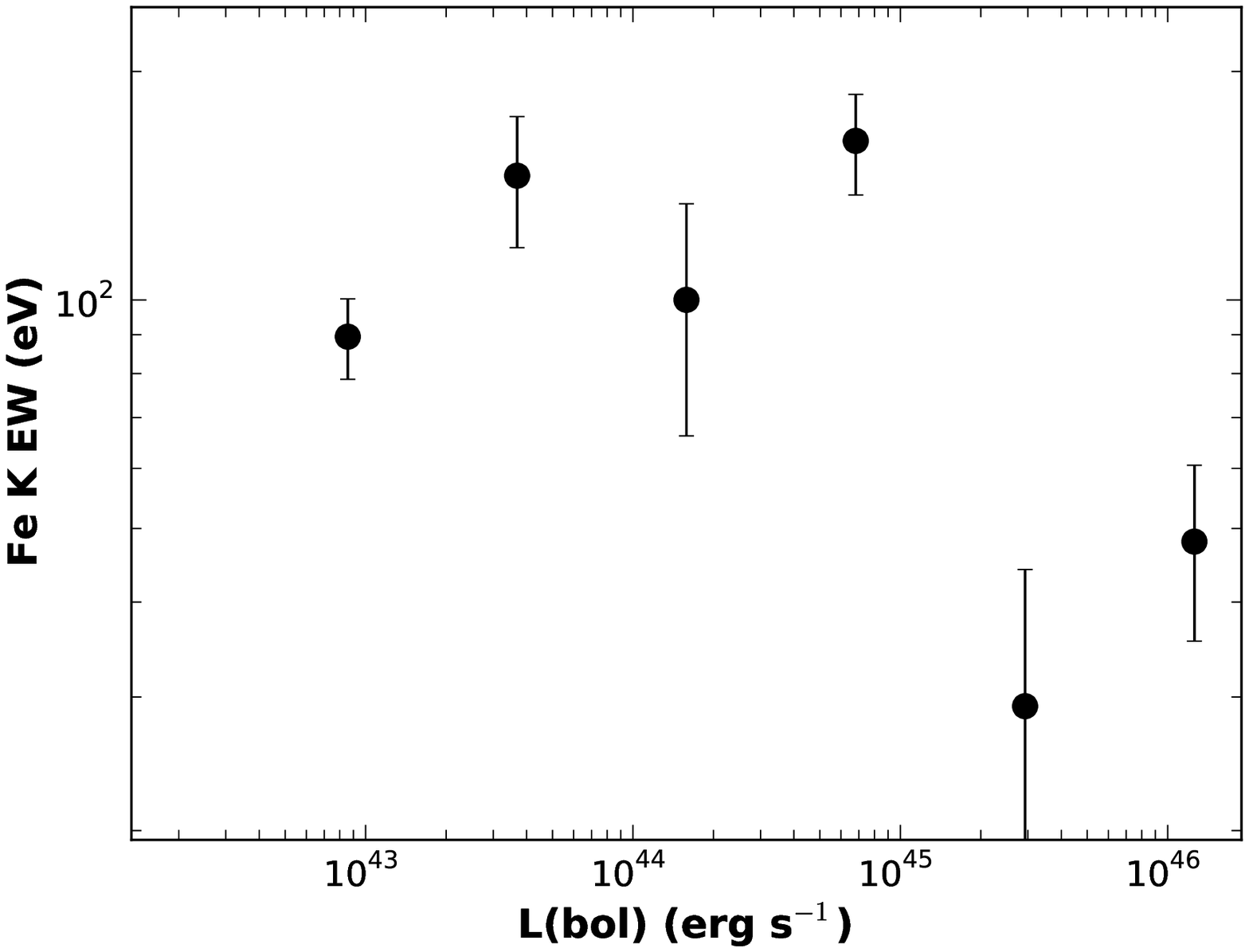}\\
\includegraphics[height=6.25cm]{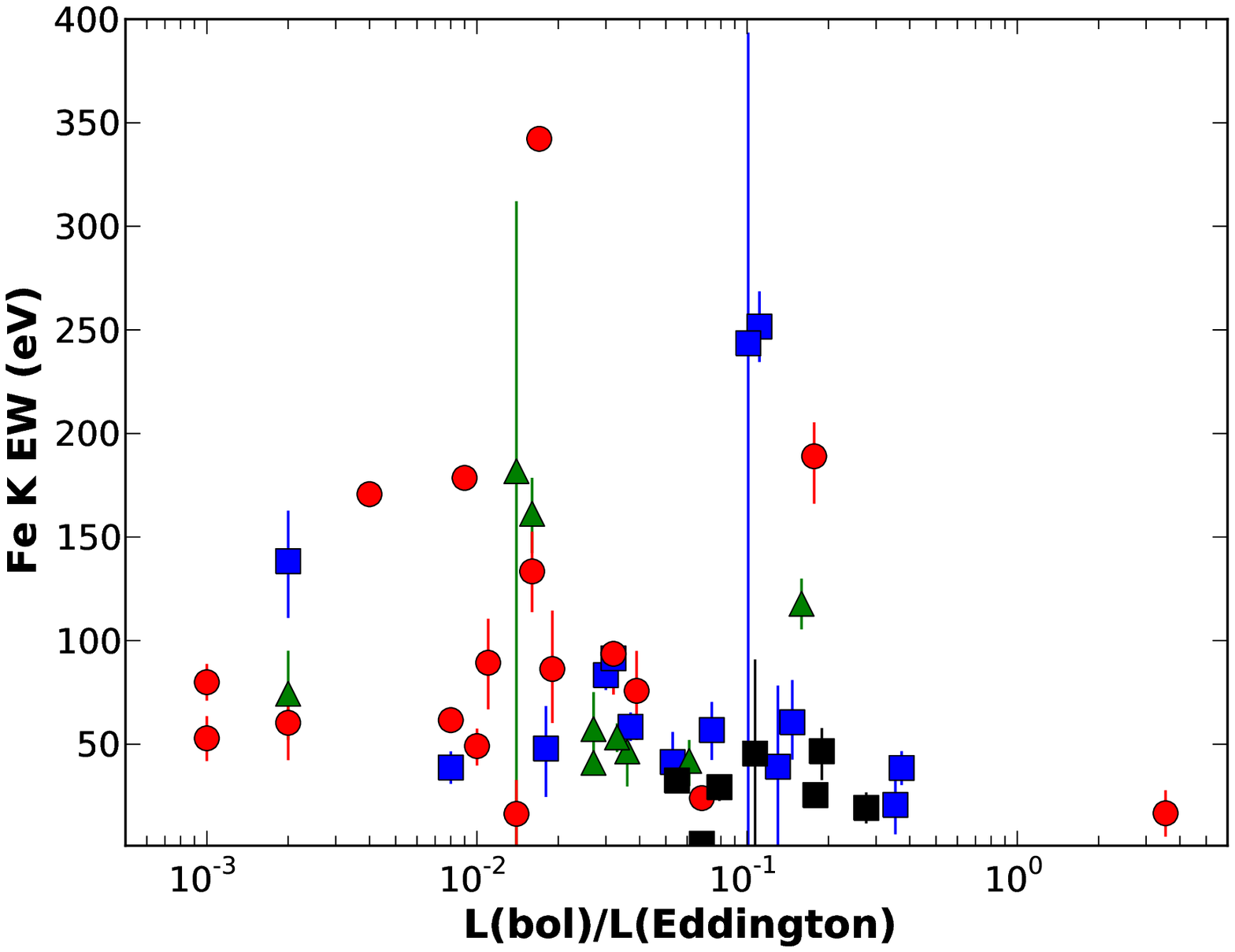}
\hspace{-1cm}
\includegraphics[height=6.25cm]{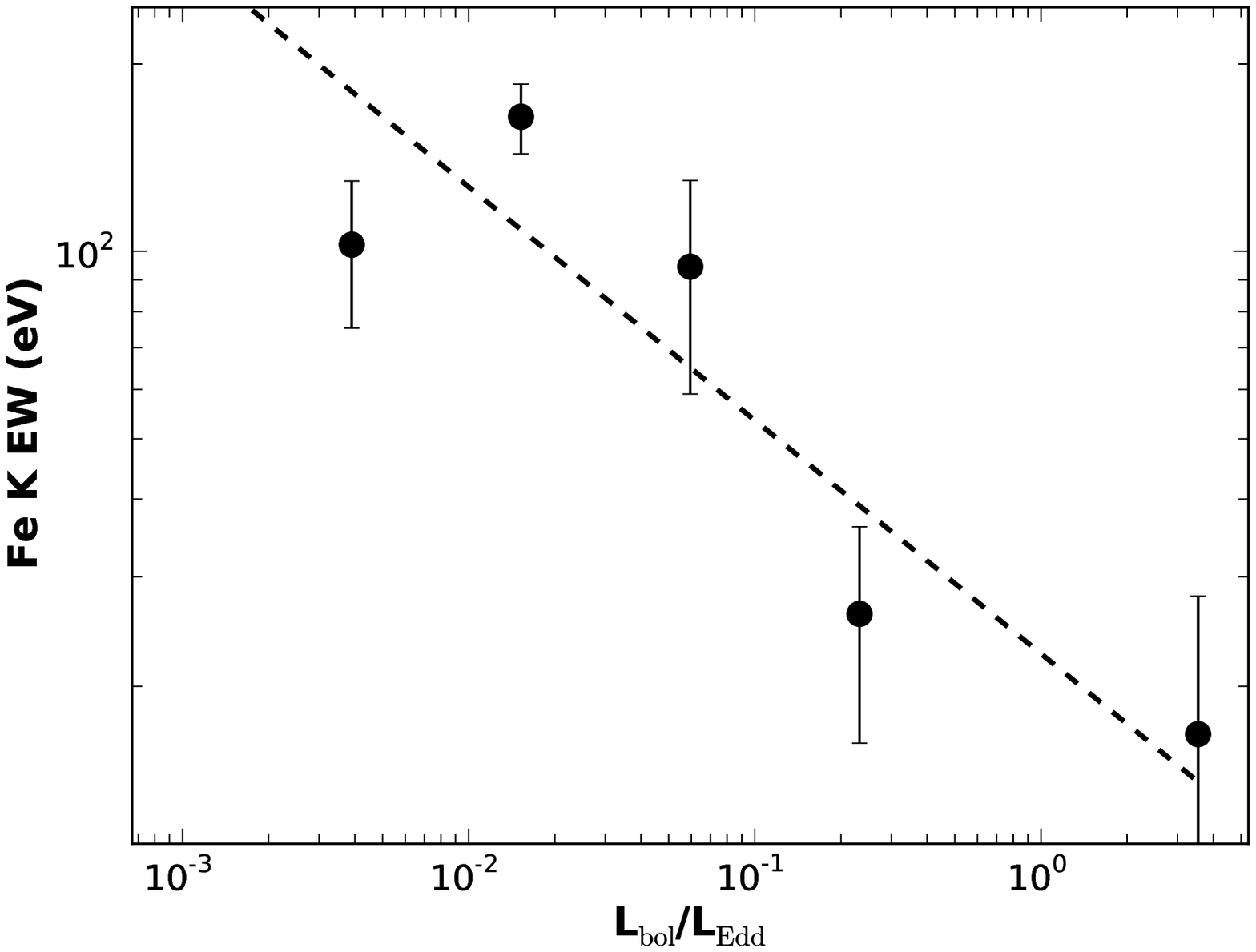}\\
\end{center}
\caption{We plot the relationship between the narrow \ion{Fe}{1} K$\alpha$ equivalent width and both the bolometric luminosity and Eddington ratio.  The right plots are binned by luminosity/accretion rate, with error-bars indicating the difference between the maximum/minimum value and the average value.  We find no strong correlation between the $EW$ and luminosity, but a strong correlation between $EW$ and accretion rate.  The linear correlation we find for the Sy 1s ($EW \propto (L_{\rm bol}/L_{\rm Edd})^{-0.38 \pm 0.07}$) is similar to that found for the entire Swift BAT-selected sample ($EW \propto(L_{2-10\,{\rm keV}}/L_{\rm Edd})^{-0.26 \pm 0.03}$; \citealt{2009ApJ...690.1322W}).
\label{fig-fekew}}
\end{figure}

\begin{figure}
\begin{center}
\includegraphics{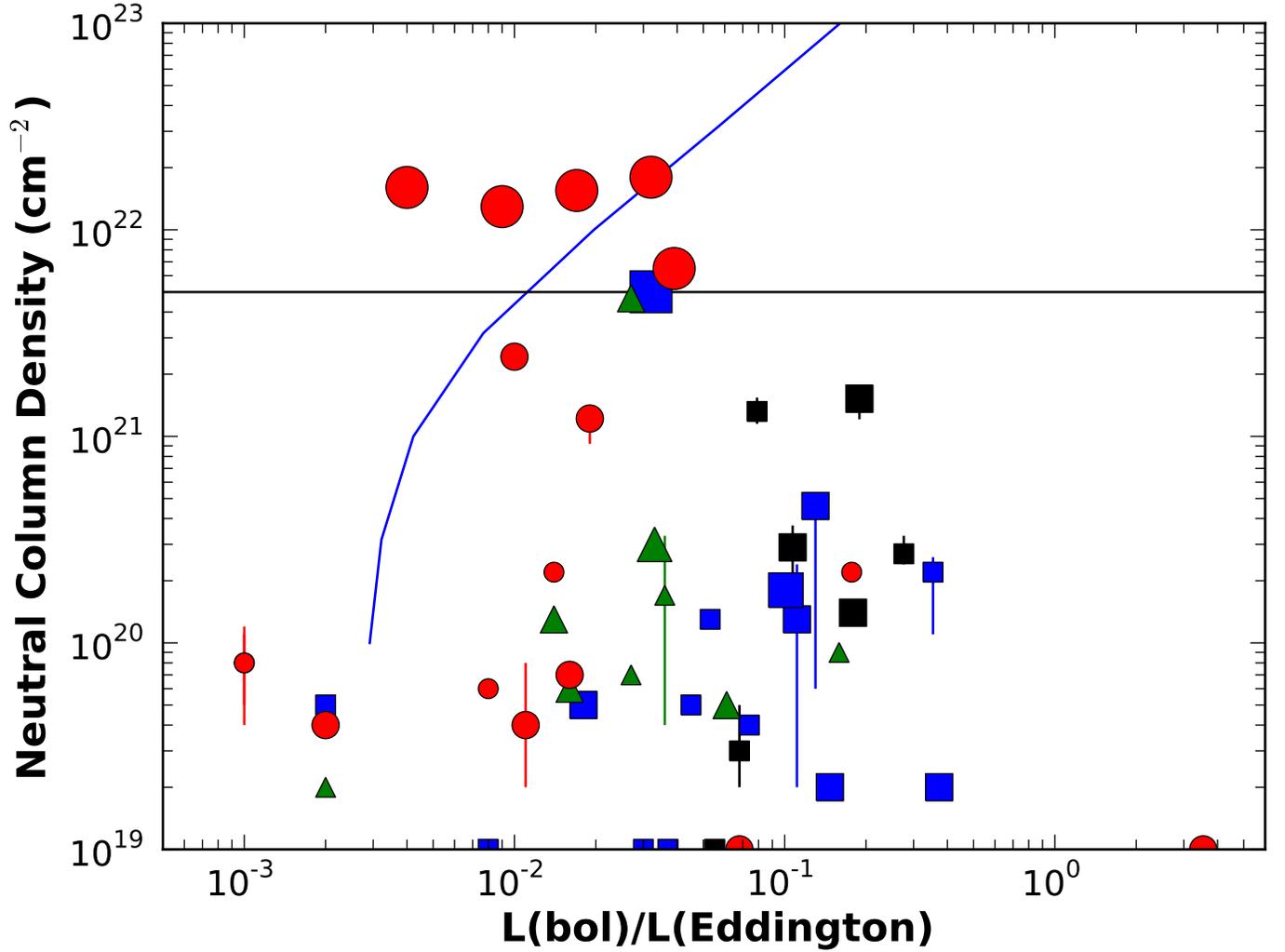}
\end{center}
\caption{We plot the neutral column density versus Eddington ratio for our sample.  The strength of the warm absorbers, from the optical depth of the \ion{O}{7} absorption edge, is correlated with the size of the data points.  The strongest warm absorber signatures are exhibited  in the sources with the highest neutral hydrogen column densities.  There is no obvious correlation with Eddington ratio.  Also, we find that the sources with the strongest detections of warm absorbers tend to lie close to the effective Eddington limit for dusty gas (blue line; assuming solar abundances).  A line at $5 \times 10^{21}$\,cm$^{-2}$ is used to denote the division between higher and lower column density sources.
\label{fig-eddingtonlimit}}
\end{figure}

\begin{figure}

\includegraphics[height=6.5cm]{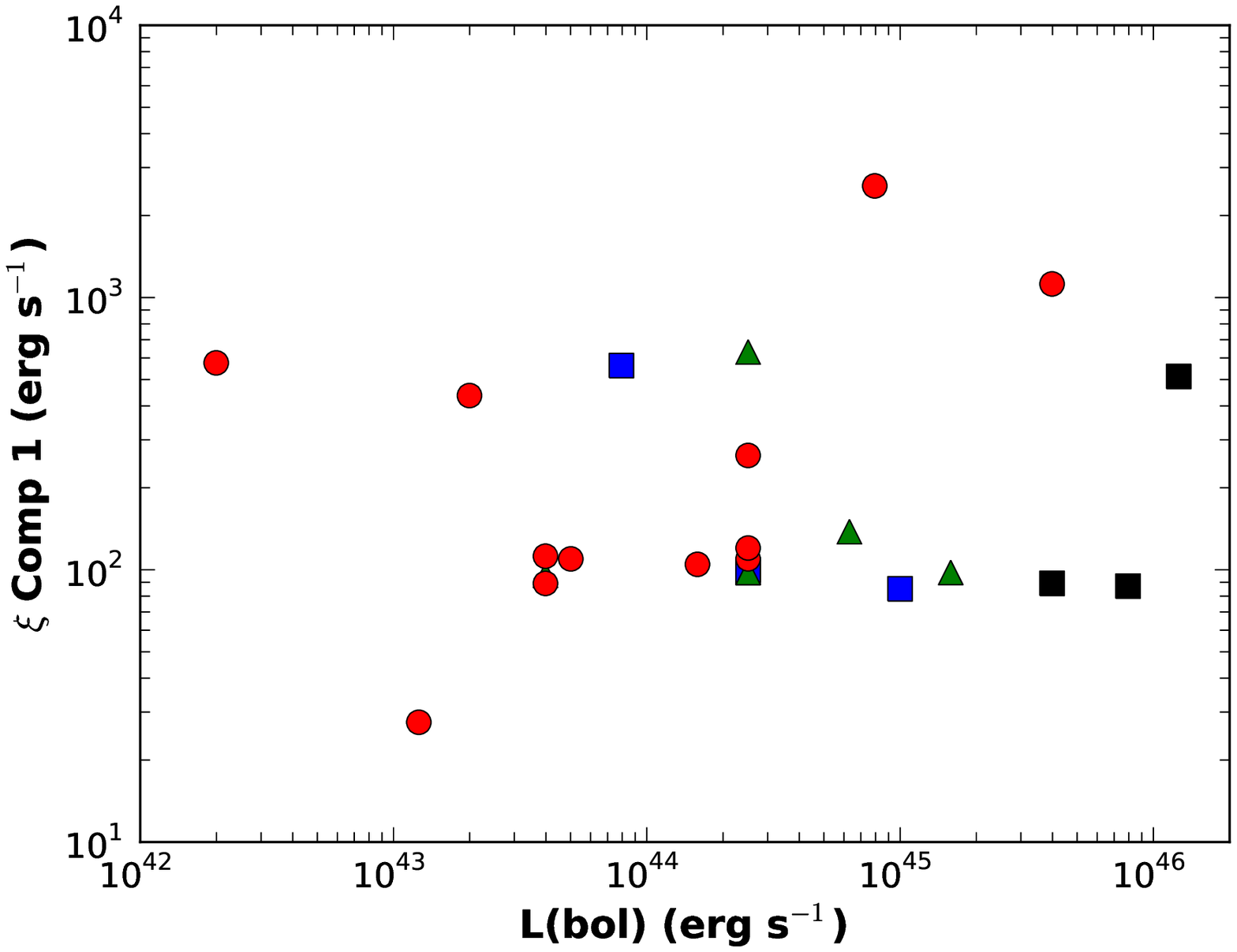}
\hspace{-1cm}
\includegraphics[height=6.5cm]{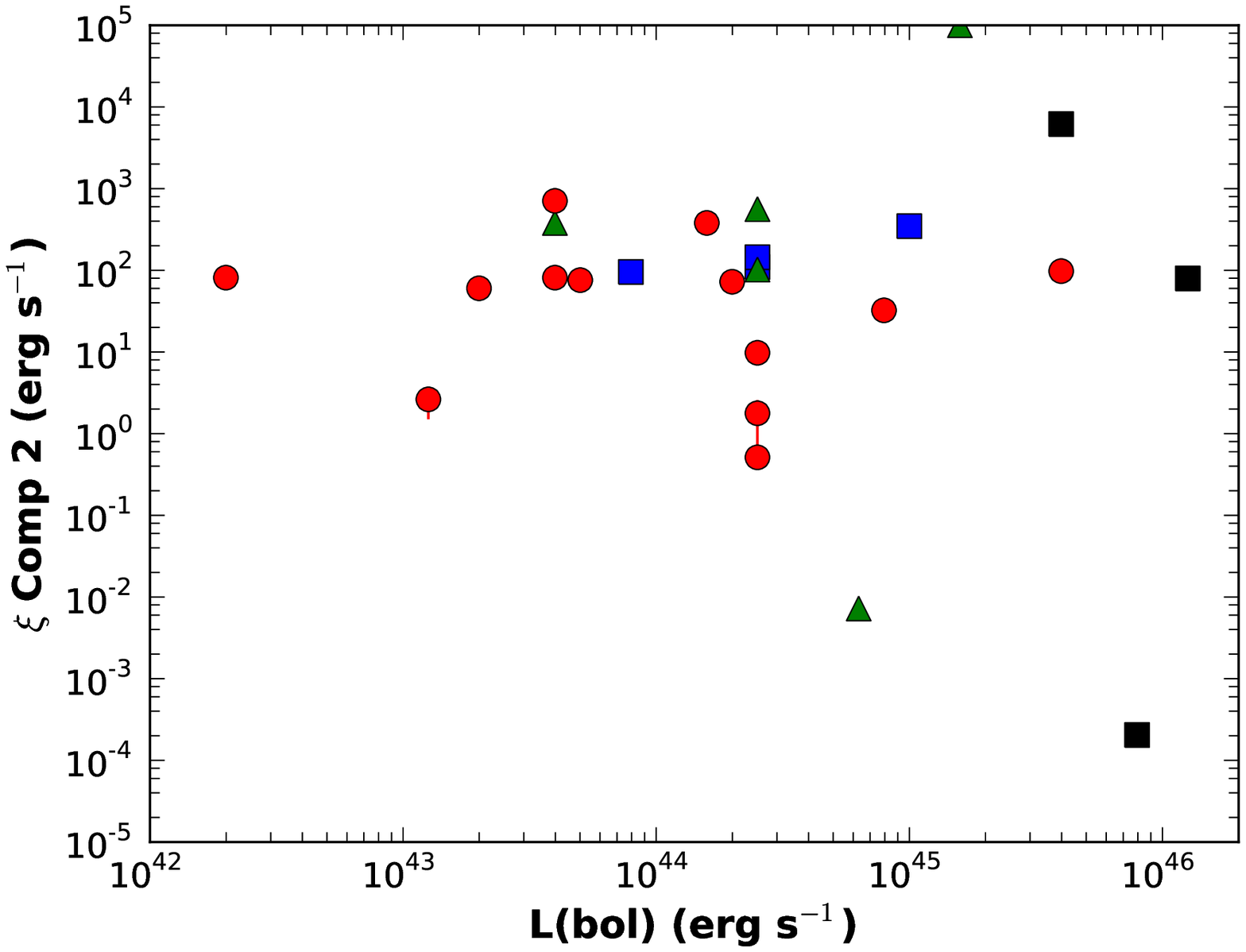}\\
\includegraphics[height=6.5cm]{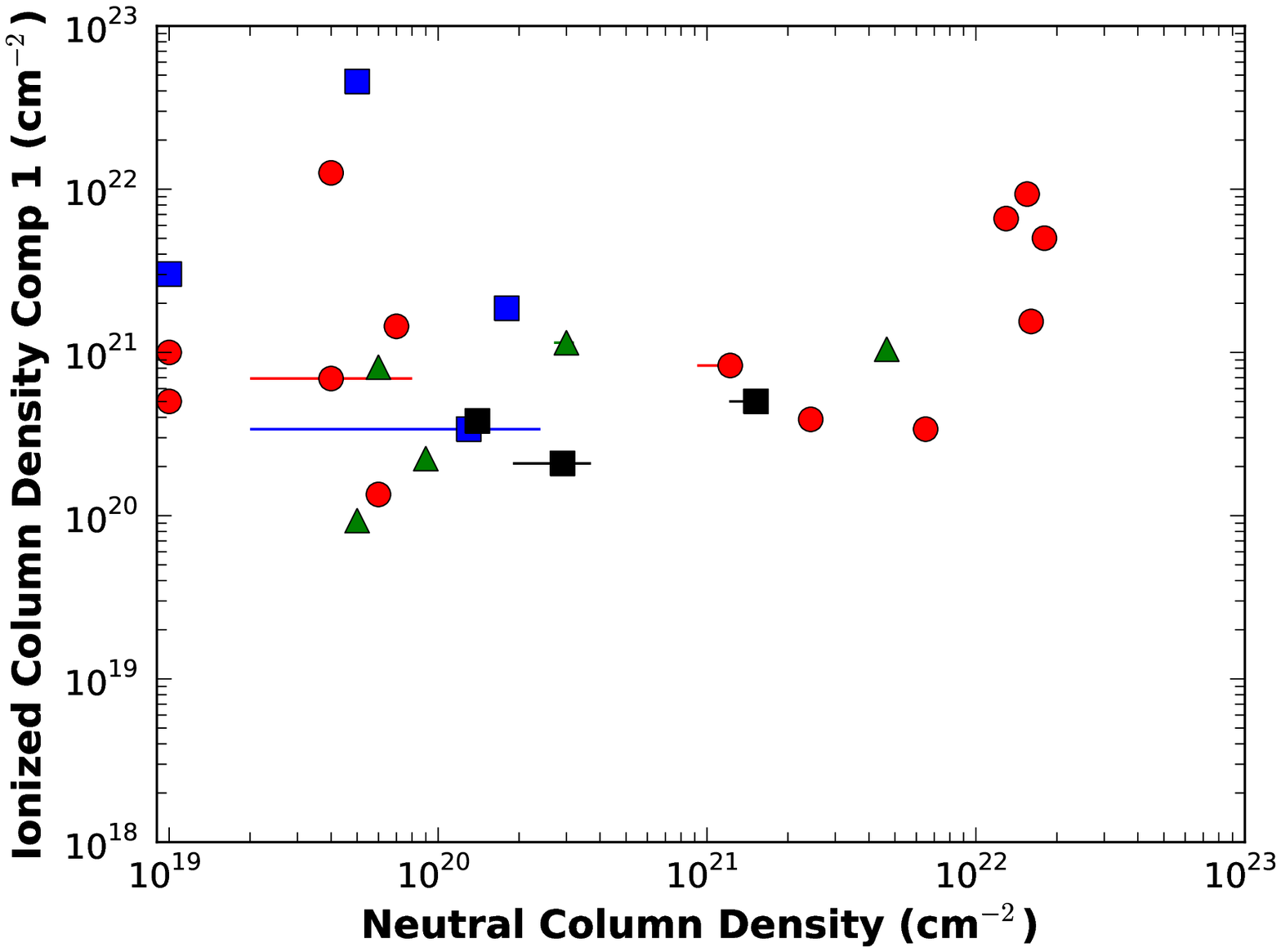}
\hspace{-1cm}
\includegraphics[height=6.5cm]{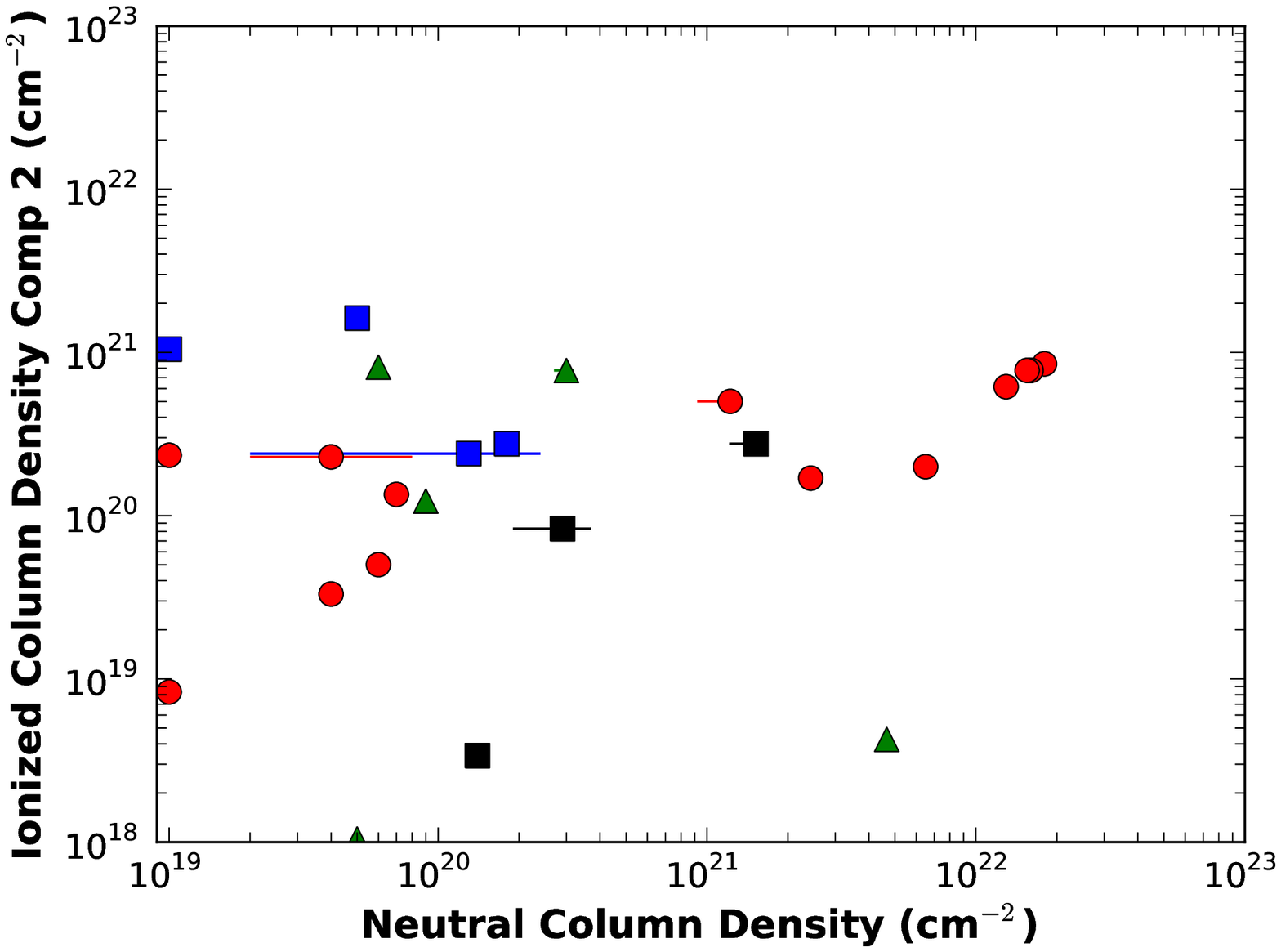}
\caption{Plotted is the ionization parameter versus the bolometric luminosity (top) and the warm ionized column density versus neutral column density (bottom) for the sources with strong detections of \ion{O}{7} and \ion{O}{8} absorption edges.  Component 1 is the component with the highest warm ionized column density, while component 2 is the second ionized component.  There is no correlation between ionization parameter and AGN luminosity or the ionized and neutral column densities.
\label{fig-nwarm1}}
\end{figure}

\end{document}